\documentclass[12pt,notitlepage,letterpaper]{article}
\usepackage{amssymb}
\usepackage{amsmath}
\usepackage{amsfonts}
\usepackage{verbatim}
\usepackage{geometry}
\usepackage[scaled]{helvet}
\usepackage{graphicx}
\usepackage{subfig}
\usepackage{amssymb}
\usepackage{amsmath}
\usepackage{amsfonts}
\usepackage{scalefnt}
\usepackage{pgflibraryarrows}
\usepackage{pgflibrarysnakes}
\usepackage{xcolor}
\usepackage{hyperref}
\usepackage{caption}
\usepackage[onehalfspacing]{setspace}
\usepackage{hyperref}
\usepackage{lipsum}
\usepackage[disable]{todonotes}
\usepackage{threeparttable}
\usepackage{longtable}
\usepackage{float}

\normalfont
\newtheorem{theorem}{Theorem}

\newtheorem{definition}{Definition}

\newtheorem{lemma}{Lemma}

\newtheorem{proposition}{Proposition}

\newenvironment{proof}[1][Proof]{\noindent\textbf{#1.} }{\ \rule{0.5em}{0.5em}}
\hypersetup{colorlinks=true,linktoc=all,linkcolor=[rgb]{0.00,0.00,0.63}}
\let\OLDthebibliography\thebibliography
\renewcommand\thebibliography[1]{
  \OLDthebibliography{#1}
  \setlength{\parskip}{0pt}
  \setlength{\itemsep}{0pt plus 0.3ex}
}

\input{tcilatex}

\begin{document}

\title{\textbf{Optimal Delegation in Markets for Matching with Signaling%
\thanks{%
This paper extends the second part of our paper, \textquotedblleft Designing
a Competitive Monotone Signaling Equilibrium.\textquotedblright\ We are very
grateful to Marzena Rostek (Editor) and two anonymous referees for their
comments and suggestions on the original paper, which allow us to greatly
improve the quality of this paper. For their comments on the original paper,
we would also like to thank Maxim Ivanov, Shuo Liu, Gabor Virag, and seminar
participants at the 2022 Australasia Meeting of the Econometric Society and
the 17th European Meeting on Game Theory. Han and Shin gratefully
acknowledge support from the Social Sciences and Humanities Research Council
of Canada, respectively.}}}
\author{Seungjin Han\thanks{%
Dept. of Economics, McMaster University, Canada. Email: hansj@mcmaster.ca}
\and Alex Sam\thanks{%
Dept. of Economics, McMaster University, Canada. Email: sama1@mcmaster.ca}
\and Youngki Shin\thanks{%
Dept. of Economics, McMaster University, Canada. Email: shiny11@mcmaster.ca}}
\date{March 14, 2023}
\maketitle

\begin{abstract}
This paper studies a delegation problem faced by the planner who wants to
regulate receivers' reaction choices in markets for matching between
receivers and senders with signaling. We provide a noble insight into the
planner's willingness to delegate and the design of optimal (reaction)
interval delegation as a solution to the planner's general mechanism design
problem. The relative heterogeneity of receiver types and the productivity
of the sender' signal are crucial in deriving optimal interval delegation in
the presence of the trade-off between matching efficiency and signaling
costs.

\medskip


\medskip

\end{abstract}









\section{Introduction\label{sec_intro}}

The design of decision rules by an uninformed principal has been extensively
studied in models where she faces an informed but biased agent. (e.g., how a
regulator regulate the prices charged by a monopolist whose cost structure
is his own private information, how the U.S. House of Representative should
regulate the drafting of bills by standing committee, how an organization
delegates the choice of decisions to a subordinate, etc.)

Holmstrom (1977, 1984) showed that the problem of finding the optimal
decision rule reduces to a delegation problem where the principal simply
decides a set of decisions and then the agent is allowed to make any
decisions from the delegated set. He characterized optimal delegated sets in
various examples. Melumad and Shibano (1991) and Alonzo and Matouschek
(2008) provide a full characterization of the optimal delegated sets in more
general environments. More recently, the studies of delegation are extended
to dynamic setting with a principal and an agent, incorporating learning
(Escobar and Zhang (2021), Guo (2016), Lipnowski and Ramos (2020)).

This paper introduces a delegation problem of the planner who wants to
regulate receivers' reaction choices in markets for matching with signaling.
In these markets, there is a continuum of heterogeneous senders and
receivers (e.g., sellers and buyers, workers and firms, and entrepreneurs
and investors) in terms of their types. Each player's type is 
their own private information.

For example, prior to entering the job market, workers (senders) often
acquire education (action) and it has a signaling effect on their type.
Signaling creates a trade-off in matching markets. It increases the
efficiency of matching between senders and receivers because separating
induces assortative matching. On the other hand, it is costly in that
senders need to choose inefficiently high levels of equilibrium actions in
order to separate themselves. Because of this trade-off, there may be
efficiency gains if the planner regulates receivers' reaction choices
(firms' wages) to prevent a separating equilibrium from happening in the
first place.

Studying the optimal delegation of the planner who wants to maximize
equilibrium aggregate net surplus in these markets, we address the following
questions. Is there no loss of generality for the planner to focus on
intervals of reactions even if she can use any mechanism?\footnote{%
We use female pronouns for the planner and senders and male pronouns for
receivers.} When is it better to impose a single reaction to all receivers
rather than the full delegation without any restrictions on reactions? When
is it optimal for the planner to impose her best single reaction to all
receivers instead of delegating reaction choices to receivers? When does the
planner provide more discretion to receivers when delegating their reaction
choices?

To address these questions, a new methodology is required because those
techniques and tools developed in the literature on optimal delegation are
largely irrelevant for the analysis of optimal delegation in markets for
matching with signaling. The first part of our paper develops a methodology.
As in most of the literature on delegation, our model assumes that the
planner cannot make transfers to receivers.

In our model, the planner moves first by publicly announcing a mechanism
that specifies a receiver's reaction as a function of his message to her.
After observing the mechanism, senders take actions. Given the announced
mechanism and the actions taken by senders, one-to-one matching occurs
between senders and receivers on the market. As matching take places, a
receiver sends a message to the planner and takes the reaction determined by
the mechanism given his message.

A key to the mechanism design is the timing of the receiver's communication.
Before entering the market and searching for their potential partners,
receivers do not have information on actions chosen by senders on the
market. This is why the planner lets receivers send messages to her at the
matching stage after they observe the distribution of sender actions, even
though the mechanism is announced before senders and receivers enter the
market. Given the announced mechanism and the distribution of sender
actions, matching and receivers' communication strategy are a \emph{stable
matching} outcome if there is no pair of a sender and a receiver who are
strictly better off by forming a new match with the receiver's alternative
message to the planner for his reaction choice.

In the example of firms (receivers) and workers (senders), a firm's reaction
is wage and a worker's action is education. Given a mechanism announced by
the planner, workers and firms have an expectation on the wage that a worker
can receive conditional on an education level chosen by any worker on the
market. This expectation is captured in a market reaction function. In
equilibrium, it crucially depends on the receiver's belief on the sender's
type conditional on her action. Given a market reaction function, each
worker chooses her education. The receiver's equilibrium communication
strategy and matching confirm the expectation in a competitive signaling
equilibrium (CSE).

We first establish the \emph{Interval Delegation Principle }when utility
functions are quasilinear with respect to the receiver's reaction: Given any
CSE that can be induced by a mechanism, there exists a corresponding
(closed) interval of reactions such that the same CSE occurs when the
planner lets receivers choose their preferred reaction directly from the
interval without communication. Therefore, there is no loss of generality
for the planner to announce an interval of reactions even if any complex
mechanism is possible. In principle, when the planner wants to elicit the
receiver's information through communication, she may need to ask him to
report not only his private type but also the whole distribution of sender
actions because he observes senders' actions after entering the market. In
this sense, the Interval Delegation Principle makes the planner's mechanism
design problem tractable particularly in a matching market.

For the implementation of an allocation, we adopt the notion of stronger
monotone CSE proposed by another paper of ours, Han, Sam, and Shin (HSS,
2023) (See Section \ref{sec_monotone_CSE}). Borrowed the terminology from
the mechanism design literature on implementation, we say that an allocation
can be \emph{fully implementable} in a stronger monotone CSE if there exists
an interval of reactions that induces it in all stronger monotone CSEs. HSS
(2023) shows that when utility functions are quasilinear, a stronger
monotone CSE is \emph{unique} and well-behaved, given any interval of
reaction and hence the full implementation in stronger monotone CSE is
possible.

Given a reaction interval, a \textquotedblleft
well-behaved\textquotedblright\ equilibrium is characterized by the two
threshold sender types. The lower threshold sender type specifies the lowest
sender type who enters the market, whereas any sender above the upper
threshold sender type pools their actions. Senders between the two threshold
types separate themselves. If the two threshold types are the same, it
becomes a pooling equilibrium. If the upper threshold type is the supremum
of the sender types and greater than the lower threshold type, it becomes a
separating equilibrium. If the upper threshold type is less than the
supremum of the sender types but greater than the lower threshold type,
separating and pooling coexist and the equilibrium is said to be strictly
well behaved. In the separating part of the equilibrium, matching is \emph{%
assortative} in terms of sender action and receiver type (and hence in terms
of sender type and receiver type), whereas in the pooling part, it is \emph{%
random}.

While the Interval Delegation Principle makes the mechanism design problem
tractable, it does not show how to derive an optimal reaction interval. The
aggregate net surplus is a function of the two threshold sender types in a
unique stronger monotone CSE. We further show that the planner's optimal
delegation problem - i.e., optimal choice of a reaction interval - comes
down to the choice of the two threshold sender types. The reason is that
choosing the two threshold sender types ($z_{\ell }$ and $z_{h}$) is
equivalent to choosing the lower and upper bounds of an interval of
reactions in the sense that we can uniquely retrieve the upper and lower
bounds, $t_{\ell }$ and $t_{h}$, of the interval of reactions from any given
two threshold sender types. If the two bounds of the interval are the same,
then the interval collapses to a singleton $\{t^{\ast }\}$. This is the case
where the planner simply imposes a single reaction $t^{\ast }$ to all
receivers on the market instead of delegating reaction decisions, inducing a
pooling equilibrium. If the upper bound of the interval is sufficiently low
but higher than the lower bound, the well-behaved equilibrium have
separating and pooling at the same time. As the upper bound increases, the
pooling part on the top shrinks. When the upper bound is sufficiently high,
the well-behaved equilibrium is separating.

For every sender type $z,$ let $n(z)$ be the type of the receiver that has
the same percentile as $z$ has. Given any sender type distribution, $n(\cdot
)$ characterizes the receiver type distribution. To solve the optimal
delegation problem, we adopt a parametrized model, proposing an approach
that approximates the distribution of receiver types by two parameters,
which are called \textquotedblleft scale\textquotedblright\ ($k$) and
\textquotedblleft relative spacing\textquotedblright\ ($q$) such that $%
n(z)=kz^{q}$. If $q=0,$ all receivers are homogeneous in terms of their
types. The gross match surplus between the receiver type $x$ and the sender
type $z$ with action $s$ is given by $v(x,s,z)=Axs^{a}z,$ where $a$ can be
interpreted as the productivity parameter of the sender's action. If $a=0$,
the sender's action only has a pure signaling effect.

We first prove that when the relative spacing parameter ($q$) and the
productivity parameter of the sender's action ($a$) are small, it is more
efficient for the planner to impose a single reaction to all receivers,
inducing a pooling CSE, instead of fully delegating reaction choices to
receivers without restriction on reactions. This results holds regardless of
the sender type distribution.

For numerical analysis, we choose various baseline distributions for sender
types with the support $[0,3]$, which is generated from the Beta
distribution multiplied by 3 with various shape parameters. We provide
comparative statics as changing various underlying parameters. In all cases,
the lower bound of the optimal reaction interval is $t_{\ell }=0$ so that it
is optimal not to exclude anyone in the market.

In Design 1, we choose $k=q=1$ so that both sides of the market have the
same type distribution. We change the upper bound of the type support.
Regardless of its value, the planner chooses the upper bound of the optimal
reaction interval $t_{h}$ in a way that it induces a strictly well-behaved
stronger monotone CSE with a pooling part on the top. This suggests that the
savings in the cost associated with the pooled action chosen by the sender
types above the upper threshold type ($z_{h}$) outweighs the inefficiency
associated with random matching in the pooling part. As the upper bound of
the type support increases, the planner gives more discretion to receivers
in that $t_{h}$ increases in $z_{h}$. While $z_{h}$ also increases in the
upper bound of the type support, the percentile of $z_{h}$ is independent of
the upper bound of the type support.

In Design 2, we choose $q=1$ given each sender type distribution, changing $%
k $. As $k$ changes, it changes every receiver type by the same scale.
Regardless of the value of $k,$ the planner again chooses $t_{h}$ in a way
that it induces a strictly well-behaved CSE with a pooling part on the top.
The planner gives more discretion to receivers in that $t_{h}$ increases in $%
k.$ However, the upper threshold type ($z_{h}$) is independent of $k,$
leaving its percentile constant with respect to $k$ due to the scaling
effect associated with $k$. In Design 3, we fix $k=1$ and change $q$ given
each sender type distribution. A change in $q$ changes every receiver type
by the same factor. As $q$ increases, the planner also increases $t_{h}$ to
raise $z_{h}$ in a way that the percentile of $z_{h}$ increases, shrinking
the pooling part on the top, due to the spacing effect associated with $q$.

Designs 2 and 3 reflect the empirical findings in Poschke (2018), which show
that the mean and variance of the firm size distribution are larger in rich
countries and have increased over time for U.S. firms. In our numerical
analysis, an increase in $k$ or $q$ each results in increases in both the
mean and variance of the receiver type distribution. The results in Designs
2 and 3 show that as the mean and variance of the firm size distribution
increase, the planner gives more discretion to receivers. However, she gives
a lot more discretion to receivers when increases in the mean and variance
of the receiver type distribution come from an increase in $q$ rather than $%
k $. The reason is that an increase in $q$ increases the percentile of the
threshold sender type that starts choosing a pooled action, whereas the
percentile is independent of $k$. This makes the solution paths of $t_{h}$
with respect to $q$ \emph{more convex} since the nonlinear transformation
moves a lot more weights on higher types.

In the first three designs, we provide comparative statics given a fixed
value for the productivity parameter $a$ of the sender's action. In Design
4, we change the value of $a$, along with the value of $q.$ Notably, when
the sender's action has only a pure signaling effect ($a=0$) and $q$ is
small, it is optimal for the planner to impose her best single reaction to
all receivers\footnote{%
Alonso and Matouschek (2007) refer this form of delegation as
\textquotedblleft centralization.\textquotedblright} and induce a pooling
CSE, instead of delegating reaction choices to receivers. However, for any
given positive value of $a$ and any given $q,$ delegation is better than
imposing a single reaction because the optimal stronger monotone CSE is
strictly well-behaved. As $a$ increases, both $t_{h}$ and $z_{h}$ increase.

Given type distributions, the utility indifference conditions, (\ref%
{jumping_sellers}) and (\ref{jumping_buyers}), play the crucial role in
determining $t_{h}$ (or equivalently $z_{h}$). Since the whole type
distributions affect the determination of $t_{h}$ ($z_{h}$) in a non-trivial
way through the indifference conditions, we cannot always provide a clear
picture about how changes in the underlying type distributions would change
the planner's delegation decision. For example, suppose that both sides of
the market have the same type distribution. As we change the type
distribution to a first order stochastically dominant one, we may expect
that it is optimal for the planner to give discretion to receiver (higher $%
t_{h}$) to push up $z_{h}$. However we do not have such a monotone
relationship (See Table \ref{tb:fsd}).

\subsection{Related literature}

Delegation problems have been studied in various problems such as tariff
caps (Amador and Bagwell (2013)), delegated project choice (Armstrong and
Vickers (2010)), consumption-saving problems (Amador, Ivan, Angeletos
(2006)), min-max optimal mechanisms (Frankel (2014)), monetary policy
(Athey, Atkeson, Kehoe (2005)), veto-based delegation (Mylovanov (2008)),
dynamic delegation (Escobar and Zhang (2021), Grenadier, Malenko, and
Malenko (2016), Guo (2016), Lipnowski and Ramos (2020)).

Most of works on optimal delegation have focused on a single principal and a
single agent. Starting from the seminal paper by Holmstrom (1977, 1984), he
asks how a principal should optimally delegate decision-making to an
informed, but biased agent. He characterizes optimal delegation sets in
various examples mostly with the constraints that only interval delegations
are allowed.

Melumad and Shinabo (1991) separately consider (i) the framework in
Holmstrom (1977, 1984) where the principal commits to take a decision rule,
and (ii) the cheap-talk framework by Crawford and Sobel (1982) where the
principal has no commitment power. They characterize the optimal delegation
sets without imposing any restrictions on the delegation sets. Their model
considers the quadratic loss function in which the preferred decisions are
linear functions of the agent's private type which is uniformly distributed.
Alonso and Matouschek (2007) also fully characterize optimal decision-making
rule for any commitment power and show conditions under which optimal
decision-making rule takes the form of interval delegation and other forms
of delegations.

Alonso and Matouschek (2008) extend the model by Melumad and Shinabo (1991)
by allowing for general distributions of the agent's private information and
more general utility functions. As assumed in all the above works, in their
model, there are no transfers between the principal and the agent. They
provide a necessary and sufficient condition for the principal to benefit
from delegating decisions to the agent. They further show that when the
agent's preferences are more aligned to the principal's, interval
(threshold) delegation is optimal. Similar to theirs, we also perform
comparative statics on the principal's willingness to delegate decisions to
agents, and when interval delegation is optimal. Unlike in Holmstrom (1977,
1984) and most of the follow-up papers, in our work, there is a continuum of
agents on both sides of a market.

To our knowledge, our paper is the first that brings a delegation problem to
markets for matching with signaling. By doing so, we provide a noble insight
into the planner's willingness to delegate and the optimal delegation.

Our methodology is new. Of course, pre-match investment competition to match
with a better partner has been studied in the literature as a potential
solution for the hold-up problem of non-contractible pre-match investment
that prevails when a match is considered in isolation (e.g.\ Grossman and
Hart (1986) and Williamson (1986)). Cole, Mailath, and Postlewaite (1995),
Rege (2008), and Hoppe, Moldovanu, and Sela (2009) consider pre-match
investment with incomplete information and non-transferable utility without
monetary transfers (i.e., no reaction choice by a receiver). Therefore, the
sender-receiver framework does not apply. Pre-match investment with
incomplete information in Hopkins (2012) includes the transferable-utility
case but with no restrictions on transfers (i.e., no restrictions on
reactions). A separating equilibrium is their focus. Furthermore, optimal
delegation has never been studied in the models for pre-match investment
competition.

While the literature has studied monotone equilibrium, exploring
complementarities between actions and types, they mostly focus on games with
simultaneous moves and no signaling (Athey (2001), McAdams (2003), Reny and
Zamir (2004), Reny (2011), Van Zandt and Vives (2007)). For the optimal
delegation in markets for matching with signaling, we adopt the solution
concept of a stronger monotone CSE (HSS (2023)), given an reaction interval
chosen by the planner. In this equilibrium, a market reaction function, a
sender action function, a matching function, and a belief function are all
monotone in the stronger set order (Shannon (1995)).\footnote{%
We allows for set-valued functions for beliefs and matching. See Section \ref%
{sec_monotone_CSE} for more details.} Given the uniqueness of a stronger
monotone CSE, we show that choosing a reaction interval is equivalent to the
two threshold sender types in a unique stronger monotone equilibrium induced
by the interval.

Liu and Pei (2020) consider a game with one sender and one receiver where
actions, reactions and types are finite. They show that the
monotone-supermodular condition of the sender's utility is sufficient for
the monotonicity of the sender's mixed strategy when the receiver has only
two feasible reactions. HSS (2023) shows that the monotone-supermodular
condition of the sender's utility is in fact sufficient for the stronger
monotonicity of a pure-strategy equilibrium in a game with one sender and
one receiver regardless of the finiteness of actions, reactions, and types.
Such a result is important in that most applications with continuous
actions, reactions, and types focus on pure-strategy equilibria.

Mensch (2020) follows Athey (2001), McAdams (2003) and Reny (2011) for the
existence of a monotone signaling equilibrium in a two-period game with
multiple principals. A problem is the potential discontinuity of the belief
function associated with a problem with off-path actions. A trick that he
employs is that beliefs are perturbed to ensure that all actions are taken
with positive probability. Taking the limit as the perturbations vanish
yields a perfect Bayesian equilibrium in the original game. One by-product
of this approach is that it uniquely pins down a stronger monotone belief.
HSS (2023) shows that the stronger monotonicity is equivalent to Criterion
D1 (Cho and Kreps (1987) and Banks and Sobel (1987)). The equivalence
between the stronger monotonicity of a belief and Criterion D1 is crucial in
deriving a unique stronger monotone CSE given any reaction interval even if
the interval induces no separating CSE.\footnote{%
Without restrictions on reactions, we only need Cho and Sobel monotonicity,
a partial implication of Criterion D1 to show that a separating equilibrium
is a unique equilibrium (See Section \ref{sec_monotone_CSE}).}

\section{Preliminaries\label{sec_model}}

There is a continuum of senders and receivers. A planner wants to regulate
receivers' reaction choices. Senders and receivers can be interpreted as
sellers and buyers, workers and firms, or entrepreneurs and investors. A
planner can be thought of as a government or policy maker. Receivers and
senders are all heterogeneous in terms of types. The sender's type set is $%
Z=[\underline{z},\overline{z}]\subset
\mathbb{R}
$ and the receiver's type set is $X=[\underline{x},\bar{x}]\subset
\mathbb{R}
.$ The set of feasible reactions for a receiver is $%
\mathbb{R}
_{+}$, whereas the set of feasible actions for a sender is $S=%
\mathbb{R}
_{+}$.

When a sender of type $z$ chooses action $s$ and is matched with a receiver
of type $x$ who takes reaction $t,$ the sender's utility is $u(t,s,z)$ and
the receiver's utility is $g(t,s,z,x)$. The interpretation of types, actions
and reactions depend on the examples considered. In the example of firms
(receivers) and workers (senders), a sender's action is education and a
receiver's reaction is wage. We assume the following quasilinear utility
functions:
\begin{eqnarray*}
u(t,s,z) &=&t-c(s,z), \\
g(t,s,z,x) &=&v(s,z,x)-t,
\end{eqnarray*}%
where $c(s,z)$ is the utility cost incurred to the sender of type $z$ when
she choose action $s$ and $v(s,z,x)$ can be thought of as match surplus
created between the receiver of type $x$ and the sender of type $z$ with
action $s$.

Assume that the measures of senders and receivers are one, respectively. Let
$G(z)$ and $H(x)$ denote cumulative distribution functions (CDFs) for sender
types and receiver types, respectively. $G$ and $H$ are public information
but each individual's type is 
their own private information. A sender (receiver) takes zero (re)action $0$
to stay out of the market.

The planner wants to regulate receivers' reaction choices. It announces a
mechanism $\psi :R\rightarrow
\mathbb{R}
_{+}$ that specifies a receiver's reaction $t\in
\mathbb{R}
_{+}$ as a function of his message $r\in R.$ We assume that $range(\psi
):=\{\psi (r):r\in R\}$ is closed. Let $\Psi $ be the set of all feasible
mechanisms.

The timing unfolds as follows.

\begin{enumerate}
\item Senders and receivers observe their own type drawn from $G$ and $H$,
respectively.

\item A planner publicly announces a mechanism $\psi :R\rightarrow
\mathbb{R}
_{+}.$

\item Each sender simultaneously chooses her action.

\item Receivers observe the whole distribution of actions chosen by senders
on the market. As one-to-one matching between senders and receivers take
places in the market, each receiver sends a message $r$ to a planner. Then,
his reaction is determined by $\psi (r)$.
\end{enumerate}

There is no restriction on the message space $R$. The receiver sends a
message to the planner at the time of matching. Therefore, a message can be
quite complex because a receiver has information not only about his type but
also about the distribution of actions chosen by senders in the market.

As described later, one-to-one matching and receivers' communication with
the planner are formulated as a stable bargaining outcome in a cooperative
game given a mechanism and senders' actions. In stable matching, there is no
pair of a sender and a receiver who are better off by forming a new match.

\section{Full Implementation through Interval Delegation}

Let us fix a mechanism $\psi :R\rightarrow
\mathbb{R}
_{+}$. Let $\sigma (z)$ be the optimal action chosen by a sender of type $z$%
. Given $\sigma :Z\rightarrow S$, let $range(\sigma) :=\{\sigma (z):z\in Z\}$
and $S^{\ast }:=range (\sigma) \backslash \{0\}$. Therefore, $S^{\ast }$
denote the set of actions chosen by senders who enter the market.

Let $co(A)$ be the convex hull of $A$ for some set $A$. When senders make
their action choices, they have a belief on the reaction chosen by a
receiver whom they can be matched with, conditional on an action $s\in
S^{\ast }.$ Receivers also have a belief on the reaction they have to take
conditional on the action of the sender whom they are matched with. This
belief is captured in a market action function $\tau :S^{\ast }\rightarrow $
$co(range(\psi )).$ A receiver's communication strategy $\rho :S^{\ast
}\times X\rightarrow \Delta (R)$ specifies a probability distribution for
his message to the planner conditional on his type $x$ and action $s\in
S^{\ast }$ of the sender whom he wants to be matched with. Let $\mathbb{E}%
_{\rho \left( s,x\right) }\left[ \cdot \right] $ be the expectation operator
over $R$ given $\rho \left( s,x\right) \in \triangle (R)$.

\begin{definition}
A receiver's communication strategy $\rho :S^{\ast }\times X\rightarrow
\Delta (R)$ is \emph{conformative} if
\begin{equation*}
\mathbb{E}_{\rho \left( s,x\right) }\left[ \psi (r)\right] =\tau (s)\text{ }%
\forall (s,x)\in S^{\ast }\times X\text{.}
\end{equation*}
\end{definition}

Let $\mu :S\rightarrow \Delta (Z)$ characterize receivers' belief on the
sender's type as a probability distribution conditional on her action. A
receiver's matching problem can be formulated as follows:
\begin{equation}
\max_{s\in S^{\ast }}\;\mathbb{E}_{\mu (s)}\left[ v(s,z,x)\right] -\tau (s)%
\text{ s.t. }\mathbb{E}_{\mu (s)}\left[ v(s,z,x)\right] -\tau (s)\geq 0,
\label{FP}
\end{equation}%
where $\mathbb{E}_{\mu (s)}\left[ \cdot \right] $ be the expectation
operator over $Z$ given $\mu (s)\in \triangle (Z)$.

Let $\xi (x)$ be the action of the sender whom the receiver of type $x$
optimally chooses as his match partner. If (\ref{FP}) has a solution for $%
x\in X,$ $\xi (x)$\ is the solution. Otherwise, $\xi (x)=0.$ Let $X^{\ast }$
denote the set of receiver types such that $\xi (x)$\ is a solution for (\ref%
{FP}). Then, $X^{\ast }$ is the set of receiver types who enter the market.

Consider a sender's action choice problem. Let $\sigma (z)\in S^{\ast }$ be
the optimal action for a sender of type $z$ if (i) it solves the following
problem,
\begin{equation}
\max_{s\in S^{\ast }}\;\tau (s)-c(s,z)\text{ s.t. }\tau (s)-c(s,z)\geq 0,
\label{WP3}
\end{equation}%
and (ii) there is no profitable sender deviation to an off-path action $%
s^{\prime }\notin range(\sigma )$ (see Definition \ref%
{def_profitable_sender_deviation} below). Note that $\sigma (z)=0$ becomes
the optimal action for a sender of type $z$ if there is no solution for (\ref%
{WP3}) and there is no profitable sender deviation to an off-path action $%
s^{\prime }\notin range(\sigma )$.

Definitions \ref{def_profitable_sender_deviation}, \ref{def_stable_matching}%
, and \ref{definition1} below define the notions of no profitable sender
deviation to an off -path action, stable matching, and competitive signaling
equilibrium (CSE), respectively. These definitions are similar to the
corresponding ones in HSS (2023) but they incorporate the receiver's
communication strategy with the planner at the time of matching.

For all $s\in S^{\ast },$ let $m(s)\in B(X^{\ast })$ be the set of receiver
types who are matched with a sender with $s$, where $B(X^{\ast })$ is the
Borel sigma-algebra on $X^{\ast }$ Therefore, $m:S^{\ast }\rightarrow
B(X^{\ast })$ is a set-valued matching function. For all $x\in X^{\ast }$, $%
m^{-1}(x)\in S^{\ast }$ denotes the action chosen by a sender with whom a
receiver of type $x$ is matched, i.e., $x\in m\left( m^{-1}(x)\right) $.

\begin{definition}
\label{def_profitable_sender_deviation}Given $\{\sigma ,\mu ,\tau ,m\}$,
there is a profitable sender deviation to an off-path action if there exists
$z$ for which there are an action $s^{\prime }\notin range(\sigma )$ and a
message distribution $\rho ^{\prime }\in \triangle (R)$ such that, for some $%
x^{\prime }$,
\begin{align}
\text{(a) }& \mathbb{E}_{\mu (s^{\prime })}\left[ v(s^{\prime },z^{\prime
},x^{\prime })\right] -\mathbb{E}_{\rho ^{\prime }}\left[ \psi (r)\right] >%
\mathbb{E}_{\mu (m^{-1}(x^{\prime }))}\left[ v(m^{-1}(x^{\prime }),z^{\prime
},x^{\prime })\right] -\tau \left( m^{-1}(x^{\prime })\right) \text{ and }
\label{receiver_matching_utility} \\
& \text{(b) }%
\begin{array}{c}
\mathbb{E}_{\rho ^{\prime }}\left[ \psi (r)\right] -c(s^{\prime },z)>\tau
(\sigma (z))-c(\sigma (z),z)\text{ if }\sigma (z)\in S^{\ast }, \\
\mathbb{E}_{\rho ^{\prime }}\left[ \psi (r)\right] -c(s^{\prime },z)>0\text{
otherwise.}%
\end{array}
\label{sender_matching_partner}
\end{align}
\end{definition}

Note that $z^{\prime }$ on each side of (\ref{receiver_matching_utility}) is
the random variable governed by $\mu (s^{\prime })$ and $\mu
(m^{-1}(x^{\prime }))$ respectively. If (\ref{receiver_matching_utility})
and (\ref{sender_matching_partner}) are not satisfied, there is no
profitable sender deviation to an off-path action. The reason is that those
are the conditions for sender $z$ to get a higher utility by forming a match
with a receiver on the market.
Let $B(S^{\ast })$ be the Borel-sigma algebra on $S^{\ast }$.

\begin{definition}
\label{def_stable_matching}Given $\{\sigma ,\mu \},$ $\{\rho ,\tau ,m\} $ is
a \emph{stable matching outcome} if

\begin{enumerate}
\item[(i)] $\rho $ is conformative,

\item[(ii)] $\tau $\emph{\ }clears the markets:
\begin{equation*}
H\left( \left\{ x|x\in m(\xi (x))\text{, }\xi (x)\in A\right\} \right)
=G\left( \left\{ z|\sigma \left( z\right) \in A\right\} \right) ~~\forall
A\in B(S^{\ast }),
\end{equation*}

\item[(iii)] $m$ is stable: there is no pair of a sender with action $s\in
S^{\ast }$ and a receiver of type $x\notin m\left( s\right) $ such that for
some $\rho ^{\prime }\in \triangle (R)$ and some $z\in $ with $\sigma (z)=s$
\begin{eqnarray*}
\text{(a) }\mathbb{E}_{\mu (s)}\left[ v(s,z^{\prime },x)\right] -\mathbb{E}%
_{\rho ^{\prime }}\left[ \psi (r)\right] &>&\mathbb{E}_{\mu (m^{-1}(x))}%
\left[ v(m^{-1}(x^{\prime }),z^{\prime },x)\right] -\tau (s) \\
\text{(b) }\mathbb{E}_{\rho ^{\prime }}\left[ \psi (r)\right] -c(s,z)
&>&\tau (s)-c(s,z)
\end{eqnarray*}
\end{enumerate}
\end{definition}

Definition \ref{def_stable_matching}.(iii) implies that the induced $m$ is
stable where no two agents would like to block the outcome after every
sender has chosen her action in the sense that they form a new match to be
better off.

\begin{definition}
\label{definition1}Fix the planner's mechanism $\psi $. $\{\sigma ,\mu ,\rho
,\tau ,m\}$ constitutes a \emph{competitive signaling equilibrium} (CSE) if

\begin{enumerate}
\item for all $z\in Z$, $\sigma (z)$ is optimal

\item $\mu $ is \emph{consistent}:

\begin{enumerate}
\item if $s\in range(\sigma) $ satisfies $G(\{z|\sigma (z)=s\})>0,$ then $%
\mu (s)$ is determined from $G$ and $\sigma ,$ using Bayes' rule.

\item if $s\in range (\sigma) $ but $G(\{z|\sigma (z)=s\})=0$, then $\mu (s)
$ is any probability distribution with $supp (\mu (s))=$ cl $\left\{
z|\sigma (z)=s\right\} $

\item if $s\notin range (\sigma) $, then $\mu (s)$ is unrestricted.
\end{enumerate}

\item given $\{\sigma ,\mu \},$ $\{\rho ,\tau ,m\}$ is a \emph{stable
matching outcome.}
\end{enumerate}
\end{definition}

\subsection{Interval delegation}

An interval menu is a special mechanism with the message space is $R=%
\mathbb{R}
_{+}$. Let $T=[t_{\ell },t_{h}]$ be an closed connected interval in $%
\mathbb{R}
_{+}.$ Let $\mathcal{T}$ be a set of all possible closed connected
intervals. For some $T\in \mathcal{T}$, let $\overline{\psi }:%
\mathbb{R}
_{+}\rightarrow T$ is called an interval menu if
\begin{equation*}
\overline{\psi }(t)=\left\{
\begin{array}{cc}
t & \text{if }t\in T \\
t^{\circ } & \text{otherwise}%
\end{array}%
\right. ,
\end{equation*}%
where $t^{\circ }$ is an arbitrary reaction in $T$. Let $\overline{\Psi }$
be the set of all possible interval menus.

Fix an interval menu $\overline{\psi }:%
\mathbb{R}
_{+}\rightarrow T.$ Given a market reaction function $\tau :S^{\ast
}\rightarrow T,$ the receiver's communication $\rho :S^{\ast }\times
X\rightarrow
\mathbb{R}
_{+}$ is \emph{pure conformative} if
\begin{equation*}
\overline{\rho }(s,x)=\tau (s)~~\forall (s,x)\in S^{\ast }\times X\text{.}
\end{equation*}

\begin{proposition}[Interval Delegation Principle]
\label{prop_1}Let $\{\sigma ,\mu ,\rho ,\tau ,m\}$ be a CSE given a
mechanism $\psi :R\rightarrow
\mathbb{R}
_{+}$ announced by the planner. Then, $\{\sigma ,\mu ,\overline{\rho },\tau
,m\}$ is a CSE given an interval menu $\overline{\psi }:%
\mathbb{R}
_{+}\rightarrow T$ with $T=co($range $\psi )$.
\end{proposition}

\begin{proof}
Given a mechanism $\psi :R\rightarrow
\mathbb{R}
_{+}$, we have that
\begin{equation}
\left\{ \mathbb{E}_{\rho ^{\prime }}[\psi (r)]\in
\mathbb{R}
_{+}:\rho ^{\prime }\in \triangle (R)\right\} =co(\text{range }\psi ).
\label{prop1_1}
\end{equation}%
Suppose that the planner announces an interval menu $\overline{\psi }:%
\mathbb{R}
_{+}\rightarrow T$ with $T=co($range $\psi )$. Because $\{\sigma ,\mu ,\rho
,\tau ,m\}$ is a CSE given $\psi :R\rightarrow
\mathbb{R}
_{+},$ (\ref{prop1_1}) implies that there is no profitable sender deviation
to an off-path action given $\{\sigma ,\mu ,\tau ,m\}$, that is, there is no
sender type $z$ for which there are an action $s^{\prime }\notin $ range $%
\sigma $ and and $t^{\prime }\in T$ such that, for some $x^{\prime }$,
\begin{gather*}
\text{\textrm{(a)} }\mathbb{E}_{\mu (s^{\prime })}\left[ v(s^{\prime
},z^{\prime },x^{\prime })\right] -t^{\prime }>\mathbb{E}_{\mu
(m^{-1}(x^{\prime }))}\left[ v\left( m^{-1}(x^{\prime }),z^{\prime
},x^{\prime }\right) \right] -\tau \left( m^{-1}(x^{\prime })\right) \text{,
} \\
\text{\textrm{(b)} }%
\begin{split}
t^{\prime }-c(s^{\prime },z)& >\tau (\sigma (z))-c(\sigma (z),z)\hskip15pt%
\mbox{ if }\sigma (z)\in S^{\ast }\text{, } \\
t^{\prime }-c(s^{\prime },z)& >0,\hskip95pt\mbox{
otherwise}.
\end{split}%
\end{gather*}%
Furthermore, (\ref{prop1_1}) implies that given $(\sigma ,\mu ),$ $\{%
\overline{\rho },\tau ,m\}$ is a \emph{stable matching outcome} because (i) $%
\overline{\rho }$ is (pure) conformative, (ii) $\tau $\emph{\ }clears the
markets, i.e., for all $A\in P(S^{\ast })$ such that $H\left( \left\{ x|x\in
m(\xi (x))\text{, }\xi (x)\in A\right\} \right) =G\left( \left\{ z|\sigma
\left( z\right) \in A\right\} \right) $, (iii) $m$ is stable, i.e., there is
no pair of a sender with action $s$ and a receiver of type $x\notin m\left(
s\right) $ such that, for some $t^{\prime }\in T$ , some $z$ with $\sigma
(z)=s\in S^{\ast }$,
\begin{align*}
\mbox{(a) }& \mathbb{E}_{\mu (s)}\left[ v(s,z^{\prime },x)\right] -t^{\prime
}>\mathbb{E}_{\mu (m^{-1}(x))}\left[ v(m^{-1}(x),z^{\prime },x)\right] -\tau
(s), \\
\mbox{(b) }& t^{\prime }-c(s,z)>\tau (s)-c(s,z).
\end{align*}%
Therefore, $\{\sigma ,\mu ,\overline{\rho },\tau ,m\}$ is a CSE given an
interval menu $\overline{\psi }:%
\mathbb{R}
_{+}\rightarrow T$ with $T=co($range $\psi )$.
\end{proof}

\bigskip

In principle, when the planner wants to elicit the receiver's information
through communication, she may need to ask him to report not only his
private type but also the whole distribution of sender actions because he
observes senders' actions after entering the market. In this sense, the
Interval Delegation Principle in Proposition \ref{prop_1} makes the
planner's mechanism design problem tractable particularly in matching
markets.

From now on, we use an interval $T$ and an interval menu $\overline{\psi }:%
\mathbb{R}
_{+}\rightarrow T$ interchangeably. Further, we drop the pure 
conformative communication strategy $\overline{\rho }$ when we refer a CSE. $%
\{\sigma ,\mu ,\tau ,m\}$ is a CSE given an interval $T$ with the
understanding that a receiver directly chooses his reaction from $T$ when
the planner announces $T\subset \mathcal{T}$. We say that $T$ provides \emph{%
more discretion} to receivers than $T^{\prime }$ does if $T\supset T^{\prime
}$ and $T\neq T^{\prime }$.

\subsection{Full implementation}

\label{sec_monotone_CSE}

The allocation is $\{\sigma ,\tau ,m\}$ given a CSE, $\{\sigma ,\mu ,\tau
,m\}$. Borrowing the terminology from the mechanism design literature on
implementation, we say that an allocation $\{\sigma ,\tau ,m\}$ is \emph{%
fully implemented} in stronger monotone CSE if every stronger monotone CSE
given some $T\in \mathcal{T}$ induces $\{\sigma ,\tau ,m\}$.

HSS (2023) formulates a stronger monotone CSE. Consider two sets $A$ and $B$
in the power set $P(Y)$ for $Y$ a lattice with a given relation $\geq $. We
say that $A\leq _{c}B$, read \textquotedblleft $A$ is \emph{completely lower}
than $B$\textquotedblright\ in the stronger set order if for every $a\in A$
and $b\in B,$ $a\leq b.$

\begin{definition}[Stronger set order]
Given a partially ordered set $K$ with the given relation $\geq $, a
set-valued function $M:K\rightarrow P(Y)$ is monotone non-decreasing in the
stronger set order if $k^{\prime }\leq k$ implies that $M(k^{\prime })\leq
_{c}M(k)$.
\end{definition}

Fix a CSE $\{\sigma ,\mu ,\tau ,m\}$ given an interval of reactions $T$
chosen by the planner. $\mu :S\rightarrow \triangle (Z)$ is said to be
non-decreasing in the stronger set order if $s^{\prime }<s$ implies supp $%
\mu (s^{\prime })\leq _{c}$supp $\mu (s)$ for all $s,s^{\prime }\in
\mathbb{R}
_{+}.$

A (set-valued) matching function $m:S^{\ast }\rightarrow P(X^{\ast })$ is
said to be non-decreasing in the stronger set order if $s^{\prime }<s$
implies $m(s^{\prime })\leq _{c}m(s)$ for $s,s^{\prime }\in
\mathbb{R}
_{+}.$ Note that if $\mu $ is monotone in the stronger set order, then the
union of supp $\mu (s^{\prime })$ and supp $\mu (s)$ is either the empty set
or a singleton for any $s$ and $s^{\prime }$ such that $s\neq s^{\prime }$.
The same property holds for $\sigma ,\tau ,$and $m$ if they are monotone in
the stronger set order.

\begin{definition}[Stronger Monotone CSE]
\label{definition_monotone_eq}A CSE $\left\{ \sigma ,\mu ,\tau ,m\right\} $
is stronger monotone if $\sigma ,$ $\mu ,\tau ,$ and $m$ are non-decreasing
in the stronger set order.
\end{definition}

A stronger monotone CSE is called well-behaved if it is characterized by two
threshold sender types, $z_{\ell }$ and $z_{h}$ such that every sender of
type below $z_{\ell }$ stays out of the market, every sender in $[z_{\ell
},z_{h})$ differentiates herself with her unique action choice, and all
senders in $[z_{h},\overline{z}]$ pool themselves with the same action. If $%
z_{\ell }<z_{h}=\bar{z},$ then a well-behaved CSE is separating. If $z_{\ell
}=z_{h},$ then a well-behaved CSE is pooling. If $z_{\ell }<z_{h}<\bar{z},$
then it is strictly well-behaved with both separating and pooling parts in
the equilibrium.

Given technical assumptions, HSS (2023) shows that a stronger monotone CSE $%
\left\{ \sigma ,\mu ,\tau ,m\right\} $ is \emph{unique} and \emph{%
well-behaved} given any interval of reactions $T$ chosen by the planner
(Theorems 5, 6, and 7) and that if a well-behaved equilibrium includes a
separating part, it is differentiable (Theorem 4). Because a stronger
monotone CSE $\left\{ \sigma ,\mu ,\tau ,m\right\} $ is unique given any
interval of reactions $T$, the planner can fully implement an allocation $%
\{\sigma ,\tau ,m\}$ given $T$.

\section{Optimal Delegation}

This section develops a method for solving the planner's optimal delegation
problem based on the characterization of CSE given any reaction interval in
HSS (2023). We start with a stronger monotone separating CSE $\{\tilde{\sigma%
},\tilde{\mu},\tilde{\tau},\tilde{m}\}$ given an interval of reaction $%
T=[t_{\ell },t_{h}]\subset
\mathbb{R}
_{+},$ assuming that $t_{h}$ is sufficiently high. A bilaterally efficient
action $\zeta (x,z)$ for type $z$ given $x$ maximizes $v(x,s,z)-c(s,z)$ and
we assume that
\begin{equation}
v(\underline{x},\zeta (\underline{x},\underline{z}),z)-c(\zeta (\underline{x}%
,\underline{z}),\underline{z})\geq 0.  \label{constrained_eff_b}
\end{equation}%
We normalize $\zeta (\underline{x},\underline{z})$ to $0.$ Let $z_{\ell }$
be the lowest sender type who is matched in equilibrium and $s_{\ell }$ her
action. Then, the following two inequalities must be satisfied at $(s_{\ell
},z_{\ell })$:
\begin{align}
v\left( n\left( z\right) ,s,z\right) -t_{\ell }& \geq 0,  \label{lem1} \\
t_{\ell }-c\left( s,z\right) & \geq 0.  \label{lem2}
\end{align}%
If $z_{\ell }=\underline{z},$ then all types are matched in equilibrium and
there is no information rent in the bottom match between type $\underline{z}$
and type $\underline{x}$. Therefore, the equilibrium action $s_{\ell }$ in
the lowest match is bilaterally efficient and it is $\zeta (\underline{x},%
\underline{z})=0$. In this case, we normalize $t_{\ell }=c\left( 0,%
\underline{z}\right) =0$. If $t_{\ell }$ is so high that type $\underline{x}$
cannot achieve a non-negative value of $v-t_{\ell }$ in a match with type $%
\underline{z}$ who takes an action that costs her $t_{\ell }$, then the
lowest match must be between types $z_{\ell }$ and $x_{\ell }:=n(z_{\ell })$
in the interior of both type distributions. (\ref{lem1}) and (\ref{lem2})
must be satisfied with equality at $(s_{\ell },z_{\ell })$.

We define the function $n$\ as $n\equiv H^{-1}\circ G$\ so that $%
H(n(z))=G(z) $ for all $z\in \lbrack \underline{z},\overline{z}]$. In a
separating equilibrium, matching is assortative and hence the matching
function $m$ satisfies $m(s)=n(\mu \left( s\right) )$. The first-order
conditions for senders and receivers induce the first-order ordinary
differential equation for $\mu $:
\begin{equation}
v_{s}(n(\mu (s)),s,\mu (s))+v_{z}(n(\mu (s)),s,\mu (s))\mu ^{\prime
}(s)-c_{s}(s,\mu (s))=0  \label{diff_eq}
\end{equation}%
with the initial condition $\mu (s_{\ell })=z_{\ell }$.

Once we drive a solution $\tilde{\mu}$, we can construct the functions $%
\tilde{\sigma},$ $\tilde{\tau},$ and $\tilde{m}$ implied by $\tilde{\mu}$ as
follows. $\tilde{\sigma}(z)$ for all $z\in \lbrack z_{\ell },\overline{z}]$
is determined by $\tilde{\sigma}(z)=\tilde{\mu}^{-1}(z)$ for all $z\in
\lbrack z_{\ell },\overline{z}],$ where $\tilde{\mu}^{-1}(z)$ is the type of
a sender that satisfies $z=\tilde{\mu}\left( \tilde{\mu}^{-1}(z)\right) $
for all $z\in \lbrack z_{\ell },\overline{z}].$ For $s\in \lbrack s_{\ell },%
\tilde{\sigma}(\overline{z})],$ we can derive the matching function $\tilde{m%
}$ according to $\tilde{m}(s)=n\left( \tilde{\mu}(s)\right) $. The market
reaction function $\tilde{\tau}$ becomes%
\begin{equation}
\tilde{\tau}(s)=\int_{s_{\ell }}^{s}\left[ v_{s}(\tilde{m}(y),y,\tilde{\mu}%
\left( y\right) )+v_{z}(\tilde{m}(y),y,\tilde{\mu}\left( y\right) )\tilde{\mu%
}^{\prime }\left( y\right) \right] dy+t_{\ell }.  \label{equilibrium_wage}
\end{equation}

If $t_{h}<\tilde{\tau}(\tilde{\sigma}(\bar{z}))$, then we have no separating
CSE. In this case, there are two possible stronger monotone equilibria: (i)
a stronger monotone pooling CSE and (ii) a strictly well-behaved stronger
monotone CSE with $z_{\ell }<z_{h}<\bar{z}$ (i.e., both separating and
pooling parts).

If $T$ is a singleton (i.e., $t_{\ell }=t_{h}=t^{\ast }$), a stronger
monotone is pooling in the sense that every sender type above $z^{\ast }$
enters the market, choosing a pooled action $s^{\ast }$. The following
conditions are satisfied at $(z^{\ast },s^{\ast })$:%
\begin{gather}
t^{\ast }-c(s^{\ast },z^{\ast })\geq 0,  \label{pooling_sender1} \\
\mathbb{E}\left[ v\left( n\left( z^{\ast }\right) ,s^{\ast },z^{\prime
}\right) |z^{\prime }\geq z^{\ast }\right] -t^{\ast }\geq 0,\text{ }
\label{pooling_receiver1}
\end{gather}%
where each condition holds with equality if $z^{\ast }>\underline{z}.$

A strictly well-behaved stronger monotone CSE with both separating and
pooling parts emerges when $t_{\ell }<t_{h}<\tilde{\tau}(\tilde{\sigma}(\bar{%
z}))$. Any sender type above $z_{h}$ choose the same action $s_{h}.$ The
following two equations are satisfied at $(s_{h},z_{h})$ and they are the
key to understanding jumping and pooling in the upper tail of the match
distributions:
\begin{gather}
t_{h}-c\left( s,z\right) =\tilde{\tau}\left( \tilde{\sigma}\left( z\right)
\right) -c\left( \tilde{\sigma}\left( z\right) ,z\right) ,
\label{jumping_sellers} \\
\mathbb{E}[v(n\left( z\right) ,s,z^{\prime })|z^{\prime }\geq
z]-t_{h}=v\left( n\left( z\right) ,\tilde{\sigma}\left( z\right) ,z\right) -%
\tilde{\tau}\left( \tilde{\sigma}\left( z\right) \right) .
\label{jumping_buyers}
\end{gather}

Let $x_{h}:=n(z_{h})$. Theorem 6 in HSS (2023) establishes the existence of
a unique well-behaved stronger monotone CSE $\left\{ \hat{\sigma},\hat{\mu},%
\hat{\tau},\hat{m}\right\} $ given $T=[t_{\ell },t_{h}]$ with $0\leq t_{\ell
}<\tilde{\tau}(\tilde{\sigma}(\bar{z}))<t_{h}$. It is strictly well behaved.

\begin{figure}[tbp]
\centering\includegraphics[scale=0.4]{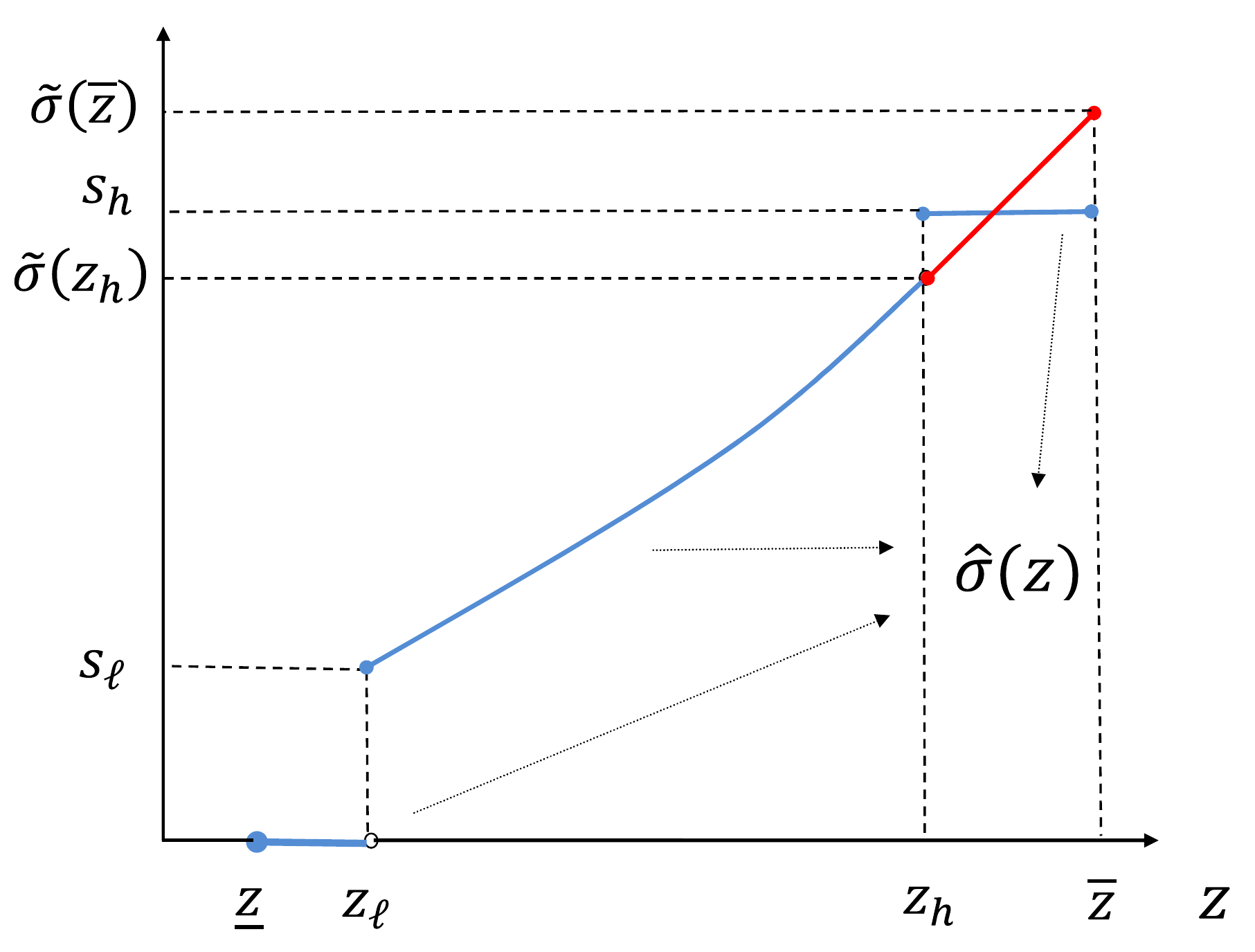}
\caption{Senders' equilibrium actions}
\label{fig:figure1}
\end{figure}

Figure \ref{fig:figure1} (borrowed from HSS (2023)) shows the equilibrium
sender actions consist of the three different blue parts in a strictly
well-behaved stronger monotone CSE $\left\{ \hat{\sigma},\hat{\mu},\hat{\tau}%
,\hat{m}\right\} $.\footnote{%
Note that $\lim_{z\nearrow z_{h}}\hat{\sigma}(z)=\tilde{\sigma}(z_{h})$ in
Figure \ref{fig:figure1}. If $0\leq t_{\ell }<\tilde{\tau}(\tilde{\sigma}(%
\bar{z}))\leq t_{h}$ so that the separting part of the well-behaved CSE can
be extended to the red part, the well-behaved equilibrium is separating.}
Note that equilibrium matching is assortative in terms of sender action and
receiver type (and therefore in terms of sender type and receiver type) in
the separating part of the CSE but it is random in the pooling part of the
CSE. Therefore, there is matching inefficiency in the pooling part but there
may be potential savings in the signaling cost associated with the pooled
action choice by senders above $z_{h}$.

Theorems 5, 6, and 8 in HSS (2023) establish the unique stronger monotone
CSE given each type of the feasible reaction intervals: (i) If $0\leq
t_{\ell }<\tilde{\tau}(\tilde{\sigma}(\bar{z}))\leq t_{h}$, then a unique
stronger monotone CSE is well-behaved and separating, (ii) If there is only
a single feasible reaction, then it is pooling, (iii) If $0\leq t_{\ell
}<t_{h}<\tilde{\tau}(\tilde{\sigma}\left( \overline{z}\right) ),$ then it is
strictly well-behaved with both separating and pooling.

Lemma \ref{lem_unified_well_behaved} shows that the planner only needs to
consider a strictly well-behaved stronger monotone CSE because it also
covers a stronger monotone separating CSE and a stronger monotone pooling
CSE.

\begin{lemma}
\label{lem_unified_well_behaved}As $z_{h}\rightarrow \bar{z},$ a strictly
well-behaved stronger monotone CSE $\left\{ \hat{\sigma},\hat{\mu},\hat{\tau}%
,\hat{m}\right\} $ converges to the stronger monotone separating CSE with
the same lower threshold sender type $z_{\ell }.$ As $z_{h}\rightarrow
z_{\ell },$ $\left\{ \hat{\sigma},\hat{\mu},\hat{\tau},\hat{m}\right\} $
converges to the stronger monotone pooling CSE in which $t_{\ell }$ is a
single feasible reaction, $z_{\ell }$ is the threshold sender type for
market entry, and $s_{\ell }$ is the pooled action for senders in the market.%
\footnote{%
As $z_{h}\rightarrow z_{\ell }$, (\ref{jumping_sellers}) and (\ref%
{jumping_buyers}) become (\ref{pooling_sender1}) and (\ref{pooling_receiver1}%
), each with equality if $z^{\ast }>$\underline{$z$}. From (\ref%
{pooling_sender1}) and (\ref{pooling_receiver1}), we can also directly
derive $(t^{\ast },s^{\ast })$ given each $z^{\ast }$ or $(z^{\ast },s^{\ast
})$ given each $t^{\ast }$ for a pooling equilibrium.}
\end{lemma}

Combining with Lemma \ref{lem_unified_well_behaved}, the next two
propositions reduce the planner's optimal delegation problem to the choice
of $z_{\ell }$ and $z_{h}$ subject to $z_{\ell }\in Z$ and $z_{h}\geq
z_{\ell }$ given that a unique stronger monotone CSE is well-behaved. In
other words, for the planner's optimal delegation, choosing the lower and
upper bounds of the feasible reaction interval $T$ is equivalent to choosing
the two threshold sender types $z_{\ell }$ and $z_{h}$, one for market entry
and the other for pooling on the top.

\begin{proposition}
\label{prop_unbounded_design}(i) For any given $z_{\ell }\in \lbrack
\underline{z},\overline{z}),$ there exists a unique solution $(s_{\ell
},t_{\ell })$ of (\ref{lem1}) and (\ref{lem2}). (ii) Suppose that the
planner chooses $t_{\ell }$ in (i) above Then, $(z_{\ell },s_{\ell })$
solves (\ref{lem1}) and (\ref{lem2}) given $t_{\ell }$ and it is unique.
\end{proposition}

Note that if $z_{\ell }=\underline{z}$, then $s_{\ell }=\zeta (\underline{x},%
\underline{z})=0$. If $z_{\ell }>\underline{z},$ then $s_{\ell }$ is
determined uniquely by $z_{\ell }$ because it solves
\begin{equation}
v(n(z_{\ell }),s,z_{\ell })-c(s,z_{\ell })=0,  \label{s_l_determination}
\end{equation}%
which is the sum of (\ref{lem1}) and (\ref{lem2}) with equality. Therefore,
Proposition \ref{prop_unbounded_design} implies that we can retrieve $%
t_{\ell }$ that induces $(s_{\ell },z_{\ell })$ from (\ref{lem1}) with
equality when $z_{\ell }=\underline{z}$, or either (\ref{lem1}) or (\ref%
{lem2}), each with equality when $z_{\ell }>\underline{z}$. Therefore, the
planner's point of view, choosing $z_{\ell }$ is equivalent to choosing $%
t_{\ell }$.

\begin{proposition}
\label{prop_bounded_design}(i) For any given $z_{h}\in (z_{\ell },\overline{z%
}),$ there exists a unique $(s_{h},t_{h})$ of (\ref{jumping_sellers}) and (%
\ref{jumping_buyers}). (ii) Suppose that the planner chooses $t_{h}$ in (i)
above. Then, $(z_{h},s_{h})$ solves (\ref{jumping_sellers}) and (\ref%
{jumping_buyers}) given $t_{h}$ and it is unique.
\end{proposition}

Note that $s_{h}$ is determined solely by $z_{h}$ because it solves
\begin{equation}
\mathbb{E}[v(n\left( z_{h}\right) ,s,z^{\prime })|z^{\prime }\geq
z_{h}]-c\left( s,z_{h}\right) =v\left( n\left( z_{h}\right) ,\tilde{\sigma}%
\left( z_{h}\right) ,z_{h}\right) -c\left( \tilde{\sigma}\left( z_{h}\right)
,z_{h}\right) ,  \label{s_h_determination}
\end{equation}%
which is the sum of (\ref{jumping_sellers}) and (\ref{jumping_buyers}) at $%
z_{h}$. Therefore, Proposition \ref{prop_bounded_design} implies that we can
retrieve $t_{h}$ that induces $(s_{h},z_{h})$ from either (\ref%
{jumping_sellers}) or (\ref{jumping_buyers}). The planner can first choose
the threshold sender type $z_{h}$ and retrieve the upper bound of feasible
reactions $t_{h}$ that induces $z_{h}$ in a well-behaved equilibrium.

Given a well-behaved stronger monotone CSE $\left\{ \hat{\sigma},\hat{\mu},%
\hat{\tau},\hat{m}\right\} $ with the lower and upper threshold sender
types, $z_{\ell }$ and $z_{h}$, the aggregate net surplus is
\begin{multline*}
\Pi (z_{\ell },z_{h}):=\int_{z_{\ell }}^{z_{h}}v(n(z),\hat{\sigma}%
(z),z)dG\left( z\right) -\int_{z_{\ell }}^{z_{h}}c(\hat{\sigma}%
(z),z)dG\left( z\right) \\
+\int_{z_{h}}^{\bar{z}}\mathbb{E}\left[ v(n(z),s_{h}(z_{h}),z^{\prime
})|z^{\prime }\geq z_{h}\right] dG\left( z\right) -\int_{z_{h}}^{\bar{z}%
}c(s_{h}(z_{h}),z)dG\left( z\right) ,
\end{multline*}%
where $s_{h}(z_{h})$ is the pooled action chosen by all sender types above $%
z_{h}$ and it is unique given any $z_{h}\in \lbrack z_{\ell },\bar{z}]$
because of Proposition \ref{prop_bounded_design}(i). Note that the first
line in $\Pi (z_{\ell },z_{h})$ is the aggregate net surplus in the
separating part of the equilibrium where matching is \emph{assortative} in
terms of types. The second line is the aggregate net surplus in the pooling
part of the equilibrium with \emph{random} matching and hence matching
efficiency is lower in this pooling part but there is potential savings in
the cost due to the pooled action chosen by all senders above $z_{h}$.

\begin{theorem}
\label{thm_optimal_design}Suppose that the planner wants to maximize the
aggregate net surplus. The solution for the planner's optimal delegation
problem is the same as the solution for the planner's design following
problem of the optimal stronger monotone equilibrium:%
\begin{equation*}
\max_{\overline{z}>z_{\ell }\geq \underline{z},z_{h}\geq z_{\ell }}\Pi
(z_{\ell },z_{h})
\end{equation*}
\end{theorem}

If $z_{\ell }<z_{h}<\bar{z},$ then the stronger monotone equilibrium is
strictly well-behaved. If $z_{\ell }<z_{h}=\bar{z},$ it is separating. If $%
z_{\ell }=z_{h}<\bar{z},$ it is pooling.

If $z_{\ell }=z_{h}<\bar{z}$, then the planner simply imposes her best
uninformed reaction choice $t^{\ast }$ to all receivers, inducing a pooling
equilibrium. When is imposing a single reaction to all receivers better than
the full delegation ($T=%
\mathbb{R}
_{+}$)? When is it optimal for the planner to simply impose her best single
reaction to all receivers instead of delegating the reaction choice to
receivers? Putting it differently, when does the planner benefit from
delegation? When does the planner provide a more discretion to receivers
when delegating their reaction choices? We address these questions in a
parametrized model.

\subsection{Parametrized optimal delegation problem}

Generally, the aggregate equilibrium surplus depends on $v,$ $c,$ $G$, and $%
H.$ For the optimal equilibrium design, we propose an approach that
approximates the distribution of receiver types with the \textquotedblleft
shift\textquotedblright\ and \textquotedblleft relative
spacing\textquotedblright\ parameters given an arbitrary distribution of
sender types. Consider a gross match surplus function that follows the form
of $v(x,s,z)=As^{a}xz$ with $0\leq a<1.$ The cost of choosing an action $s$
is $c(s,z)=\beta \frac{s^{2}}{z}$ for the sender of type $z,$ where $\beta
>0 $. The lowest sender type is $\underline{z}=0.$

A sender's type follows a probability distribution $G$, whereas a receiver's
type follows $H$. Recall that $n$ is defined as $H^{-1}\circ G$\ so that $%
H(n(z))=G(z)$ for all $z.$ We assume that $n$ takes the following form:
\begin{equation}
n(z)=kz^{q},  \label{matching_function}
\end{equation}%
where $k>0$ and $q\geq 0.$ Note that $k$ is the \textquotedblleft
scale\textquotedblright\ parameter and $q$ is the \textquotedblleft relative
spacing\textquotedblright\ parameter. The relative spacing parameter $q$
shows the relative heterogeneity of receiver types to sender types. Recall
that $n(z)$ denotes the type of a receiver who is matched with the sender of
type $z$ in the stronger monotone separating equilibrium. This approach is
general in the sense that it approximates the distribution of receiver types
with the \textquotedblleft scale\textquotedblright\ and \textquotedblleft
relative spacing\textquotedblright\ parameters for any arbitrary
distribution of sender types.

To derive a well-behaved stronger monotone equilibrium, we first need to
solve the first-order differential equation (\ref{diff_eq}) for $\mu $ with
the initial condition $(z_{\ell },s_{\ell }).$ The value of $s_{\ell }$ only
depends on $z_{\ell }.$ If $z_{\ell }=0,$ then $s_{\ell }(z_{\ell })=\zeta
(0,0)=0.$ If $z_{\ell }>0,$ then $s_{\ell }(z_{\ell })$ is determined by (%
\ref{s_l_determination}) and it is $s_{\ell }(z_{\ell })=\left( \frac{Ak}{%
\beta }z_{\ell }^{q+2}\right) ^{\frac{1}{2-a}}$. Note that $s_{\ell
}(z_{\ell })$ is continuous everywhere including at $z_{\ell }=0.$

\begin{proposition}
\label{prop_diff_eq}Given any initial condition $(z_{\ell },s_{\ell }\left(
z_{\ell }\right) ),$ the solution for first-order differential equation $\mu
^{\prime }(s)=\phi (s,\mu (s))$ is
\begin{equation*}
\tilde{\mu}(s)=\left[
\begin{array}{c}
\left( \dfrac{2\beta (2+q)}{Ak}\right) \dfrac{s^{2-a}}{2+a+aq} \\
+\left( \dfrac{s_{\ell }(z_{\ell })}{s}\right) ^{a(2+q)}\dfrac{\left[
Ak(2+a+aq)z_{\ell }^{2+q}-2\beta (2+q)s_{\ell }(z_{\ell })^{(2-a)}\right] }{%
Ak(2+a+aq)}%
\end{array}%
\right] ^{\dfrac{1}{2+q}}.
\end{equation*}
\end{proposition}

Note that $\tilde{\sigma}(z)$ is the inverse of $\tilde{\mu}(s)$, which is
derived numerically as $\tilde{\mu}(s)$ does not allow for a closed-form
solution for its inverse. Given $z_{h},$ $s_{h}(z_{h})$ is unique and it
solves (\ref{s_h_determination}), which is%
\begin{equation}
Aks_{h}^{a}z_{h}^{q}\mathbb{E}[z^{\prime }|z^{\prime }\geq z_{h}]-\beta
\frac{s_{h}^{2}}{z_{h}}=Ak\tilde{\sigma}\left( z_{h}\right)
^{a}z_{h}^{1+q}-\beta \frac{\tilde{\sigma}\left( z_{h}\right) ^{2}}{z_{h}}.
\label{s_h_choice}
\end{equation}%
We need to numerically derive $s_{h}(z_{h})$ as it does not allow for a
closed form solution. Given a choice of $z_{\ell }$ and $z_{h}$, the
aggregate net surplus is
\begin{multline*}
\Pi _{w}(z_{\ell },z_{h},q,a,G)=\int_{z_{\ell }}^{z_{h}}\left[ v(n(z),\hat{%
\sigma}(z),z)-c(\hat{\sigma}(z),z)\right] g(z)dz \\
+\int_{z_{h}}^{\bar{z}}\left[ \mathbb{E}[v(n\left( z\right) ,s_{h}\left(
z_{h}\right) ,z^{\prime })|z^{\prime }\geq z_{h}]-c(s_{h}\left( z_{h}\right)
,z)\right] g(z)dz \\
=\int_{z_{\ell }}^{z_{h}}\left( Akz^{q+1}\hat{\sigma}(z)^{a}-\beta \frac{%
\hat{\sigma}(z)^{2}}{z}\right) g(z)dz \\
+Aks_{h}(z_{h})^{a}\mathbb{E}[z^{\prime }|z^{\prime }\geq
z_{h}]\int_{z_{h}}^{\bar{z}}z^{q}g(z)dz-\beta s_{h}(z_{h})^{2}\int_{z_{h}}^{%
\bar{z}}\frac{1}{z}g(z)dz,
\end{multline*}%
where $\hat{\sigma}(z)=\tilde{\sigma}\left( z\right) $ for $z\in \lbrack
z_{\ell },z_{h}].$

Given $(a,q,G),$ we can find the best well-behaved equilibrium through the
following maximization problem:%
\begin{eqnarray*}
&&\max_{(z_{\ell },z_{h})}\Pi _{w}(z_{\ell },z_{h},q,a,G) \\
&&\text{subject to }0\leq z_{\ell }\leq z_{h}\leq \bar{z}.
\end{eqnarray*}

\subsection{More efficient non-separating equilibria}

Suppose that the planner fixes the lower bound of feasible reactions such
that $z_{\ell }=0$ and hence $s_{\ell }(z_{\ell })=0$ (no sender stays out
of the market). In this case, the belief function $\tilde{\mu}(s)$ allows
for the closed-form expression of its inverse, which is the sender's
equilibrium action function in the separating equilibrium:%
\begin{equation}
\tilde{\sigma}(z)=\left( \frac{Ak}{2\beta }\frac{aq+a+2}{q+2}\right) ^{\frac{%
1}{2-a}}z^{\frac{q+2}{2-a}}\text{ for }z<z_{h}.  \notag
\end{equation}%
The aggregate net surplus in the well-behaved stronger monotone equilibrium
is
\begin{multline*}
\Pi _{w}(0,z_{h},q,a,G)= \\
\left( \left( Ak\right) ^{\frac{2}{2-a}}\left( \frac{aq+a+2}{2\beta \left(
q+2\right) }\right) ^{\frac{a}{2-a}}-\beta \left( \frac{Ak}{2\beta }\frac{%
aq+a+2}{q+2}\right) ^{\frac{2}{2-a}}\right) \int_{0}^{z_{h}}z^{\frac{2q+2+a}{%
2-a}}dG(z) \\
+Aks(z_{h})^{a}\mathbb{E}[z|z\geq z_{h}]\int_{z_{h}}^{\bar{z}%
}z^{q}g(z)dz-\beta s_{h}(z_{h})^{2}\int_{z_{h}}^{\bar{z}}\frac{1}{z}g(z)dz.
\end{multline*}%
As $z_{h}$ approaches $\bar{z}$, the maximum of the support of $G,$ $\Pi
_{w}(0,z_{h},q,a,G)$ becomes the aggregate net surplus $\Pi _{s}(a,q,G)=$ $%
\Pi _{w}(0,\bar{z},q,a,G)$ without any restrictions on feasible reactions
(i.e., $\Pi _{s}(a,q,G)$ is the aggregate net surplus in the baseline
separating equilibrium). We show that, when the relative heterogeneity of
receiver types ($q$) and the productivity of the sender action ($a$) are not
too large, there is an interval of reactions that induces a strictly
well-behaved equilibrium that is more efficient than the separating
equilibrium with the full delegation (i.e., $T=%
\mathbb{R}
_{+}$) regardless of $G.$

\begin{theorem}
\label{thm_well_behaved_design}There are $\hat{q},\hat{a}>0$ such that for
any given $(q,a)\in \lbrack 0,\hat{q}]\times \lbrack 0,\hat{a}]$, we have an
interval of feasible reactions $[0,\hat{t}]$ that induces a unique strictly
well-behaved stronger monotone equilibrium, which is more efficient than the
stronger monotone separating equilibrium with the full delegation. Given $[0,%
\hat{t}]$, it is a unique stronger monotone equilibrium.
\end{theorem}

\noindent \textbf{Proof}. First, we construct a (unique) strictly
well-behaved stronger monotone equilibrium with $0=z_{\ell }<z_{h}<$
supremum of the support of $G.$ Let $s_{h}(z_{h},a,q)$ be the value of $%
s_{h} $ that solves (\ref{s_h_choice}) at every $a$ and $q.$ Because
functions in (\ref{s_h_choice}) are continuous in $a$ and $q$, $%
s_{h}(z_{h},a,q)$ is continuous in $a$ and $q.$ Given (\ref{s_h_choice}), we
have
\begin{multline*}
\lim_{q,a\rightarrow 0}\left( As_{h}(z_{h},q,a)^{a}kz_{h}{}^{q}\mathbb{E}%
\left[ z|z\geq z_{h}\right] -\beta \frac{s_{h}(z_{h},q,a)^{2}}{z_{h}}\right)
\\
= \lim_{q,a\rightarrow 0}\left( Ak\tilde{\sigma}\left( z_{h}\right)
^{a}z_{h}{}^{q+1}-\beta \frac{\tilde{\sigma}\left( z_{h}\right) ^{2}}{z_{h}}%
\right).
\end{multline*}%
Therefore, we have that $\lim_{q,a\rightarrow 0}s_{h}(z_{h},q,a)=\sqrt{%
z_{h}{}^{2}Ak\left( \mathbb{E}\left[ z|z\geq z_{h}\right] -1\right) /\beta }%
. $ This implies that
\begin{multline*}
\lim_{q,a\rightarrow 0}\Pi _{w}(0,z_{h},q,a,G)=\int_{0}^{z_{h}}\frac{Akz}{2}%
dG(z)+\int_{z_{h}}^{\bar{z}}Ak\mathbb{E}\left[ z|z\geq z_{h}\right] dG(z) \\
-z_{h}{}^{2}Ak\left( \mathbb{E}\left[ z|z\geq z_{h}\right] -1\right)
\int_{z_{h}}^{\bar{z}}\frac{1}{z}dG(z).
\end{multline*}%
Taking the limit of $\lim_{q,a\rightarrow 0}\Pi _{w}(0,z_{h},q,a,G)$ with
respect to $z_{h}$ yields
\begin{equation*}
\lim_{z_{h}\rightarrow 0}\left[ \lim_{q,a\rightarrow 0}\Pi
_{w}(0,z_{h},q,a,G)\right] =\int_{0}^{\bar{z}}AkzdG(z)=Ak\mu _{z},
\end{equation*}%
where $\mu _{z}$ is the unconditional mean of the sender type $z.$

On the other hand, the limit of the aggregate net surplus in the stronger
monotone separating equilibrium is
\begin{equation*}
\lim_{q,a\rightarrow 0}\Pi _{s}(q,a,G)=\int_{0}^{\bar{z}}AkzdG(z)-\int_{0}^{%
\bar{z}}\frac{Akz}{2}dG(z)=\frac{Ak\mu _{z}}{2}.
\end{equation*}%
Therefore, we have that
\begin{equation}
\lim_{z_{h}\rightarrow 0}\left[ \lim_{q,a\rightarrow 0}\Pi
_{w}(0,z_{h},q,a,G)\right] -\lim_{q,a\rightarrow 0}\Pi _{s}(q,a,G)=\frac{%
Ak\mu _{z}}{2}>0.  \label{well_behaved_zero}
\end{equation}

Because $\Pi _{w}(0,z_{h},q,a,G)$ and $\Pi _{s}(q,a,G)$ are continuous,
there exists $\hat{q}>0$, and $\hat{a}>0$ and $\hat{z}_{h}(\hat{q},\hat{a}%
)\in $ Int $Z$ such that for every $(q,a)\in \lbrack 0,\hat{q}]\times
\lbrack 0,\hat{a}]$ and every $z_{h}\in (0,\hat{z}_{h}(\hat{q},\hat{a})]$, $%
\Pi _{w}(0,z_{h},q,a,G)>\Pi _{s}(q,a,G)$. We can retrieve $t_{h}$ given $%
z_{h}\in (0,\hat{z}_{h}(\hat{q},\hat{a})]$. $\blacksquare $

In the well-behaved stronger monotone equilibrium constructed above, a small
fraction of senders and receivers on the low end of the type distribution
follow their equilibrium sender actions, reactions, and assortative matching
that would have occurred in the stronger monotone separating equilibrium.
The rest of senders and receivers are matched randomly because the rest of
senders all choose the same action. We can also establish that when $a$ and $%
q$ are small, no delegation is better than the full delegation.

\begin{theorem}
\label{thm_eq_design}There are $\tilde{q},\tilde{a}>0$ such that, for any
given $(q,a)\in \lbrack 0,\tilde{q}]\times \lbrack 0,\tilde{a}]$, we have $%
t^{\ast }>0$ such that it induces a unique stronger monotone pooling
equilibrium that is more efficient than the stronger monotone separating
equilibrium with the full delegation.
\end{theorem}

The intuition behind Theorem \ref{thm_eq_design} is the same as that behind
Theorem \ref{thm_well_behaved_design}. A major difference is that a pooling
equilibrium forces a small fraction of senders and receivers on the low end
of type distribution to stay out of the market even though they can produce
positive net surplus, whereas everyone is matched in the strictly
well-behaved equilibrium identified in Theorem \ref{thm_well_behaved_design}%
. Theorems \ref{thm_well_behaved_design} and \ref{thm_eq_design} in fact
show that a separating equilibrium with the full delegation is not optimal
in the classical Spencian model (Spence 1973) of pure signaling with no
heterogeneity of receivers (i.e., $a=q=0$).


\section{Numerical analysis}

We turn our attention to numerical analysis. For the concreteness of our
numerical analysis, one can think of senders as workers and receivers as
firms. A sender's type can be then viewed as unobservable skill and her
action as observable skill. A firm's type can be viewed as its size. In the
literature of empirical macro/development economics, firm size is measured
by the amount of labour it employees; In finance, it is measured by the
firm's market value if it is publicly traded. In an entry-level job market,
a worker's unobservable skill could be her ability to understand a task
given to her and to figure out how to complete it. In a managerial job
market, a worker's unobservable skill could be her ability to come up with
new business idea or innovation.%

\begin{figure}[h]
\begin{threeparttable}
\caption{Probability Density Functions of the Beta Distribution}
\label{fg:beta-pdf}\centering
\begin{tabular}{cccc}
~~Beta(1,1) & ~~Beta(5,5) & ~~Beta(3,5) & ~~Beta(5,3) \\
\includegraphics[scale=0.19]{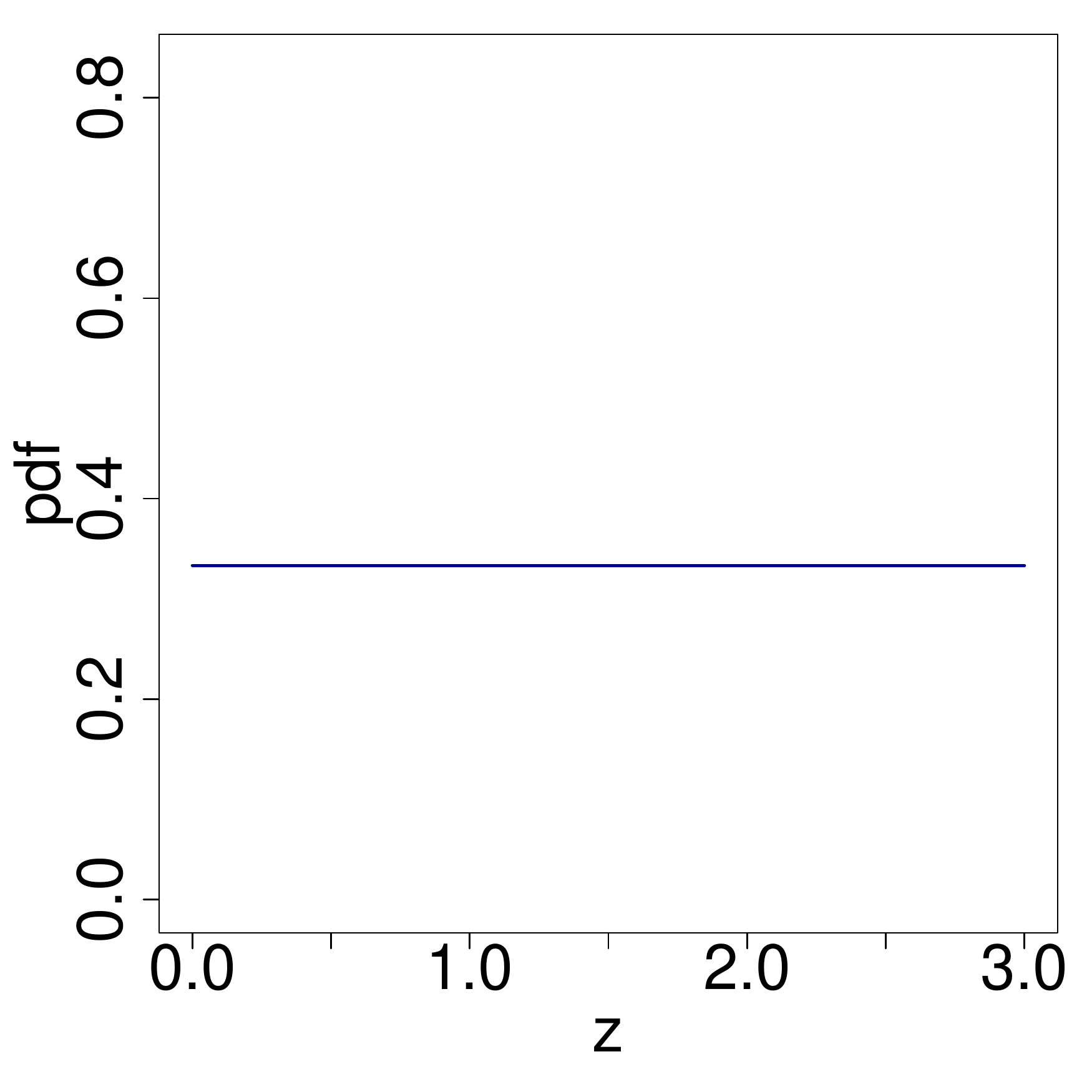} & %
\includegraphics[scale=0.19]{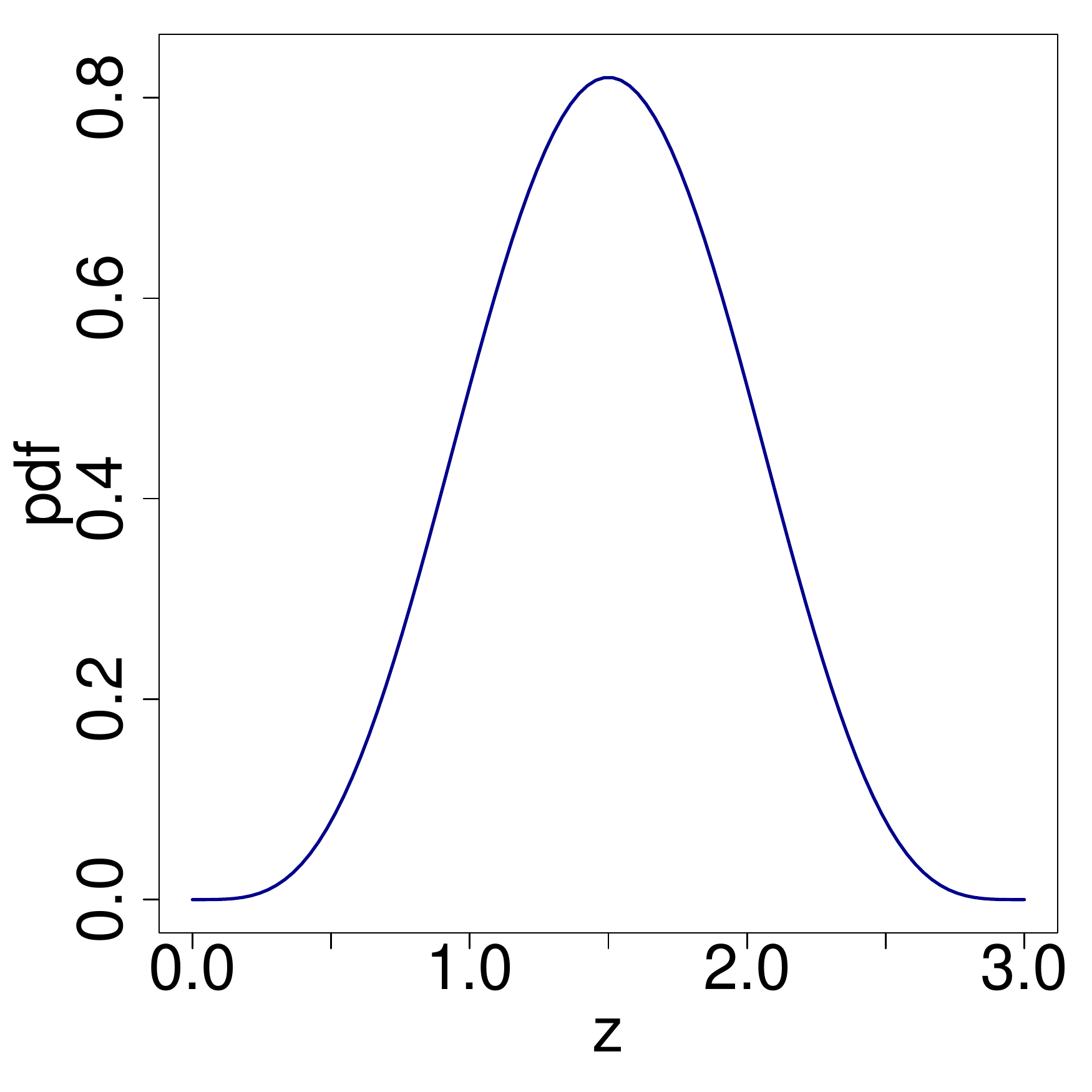} & %
\includegraphics[scale=0.19]{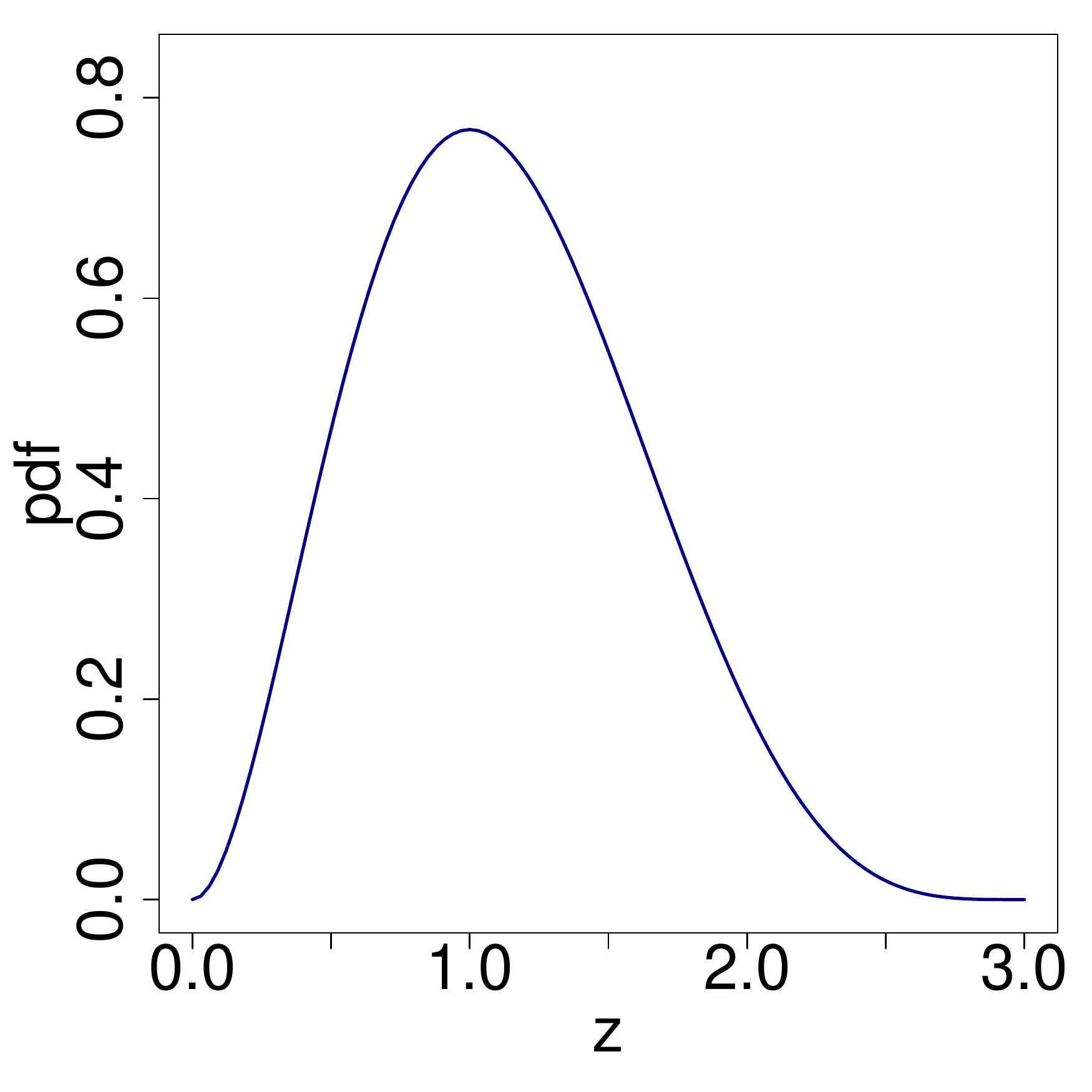} & %
\includegraphics[scale=0.19]{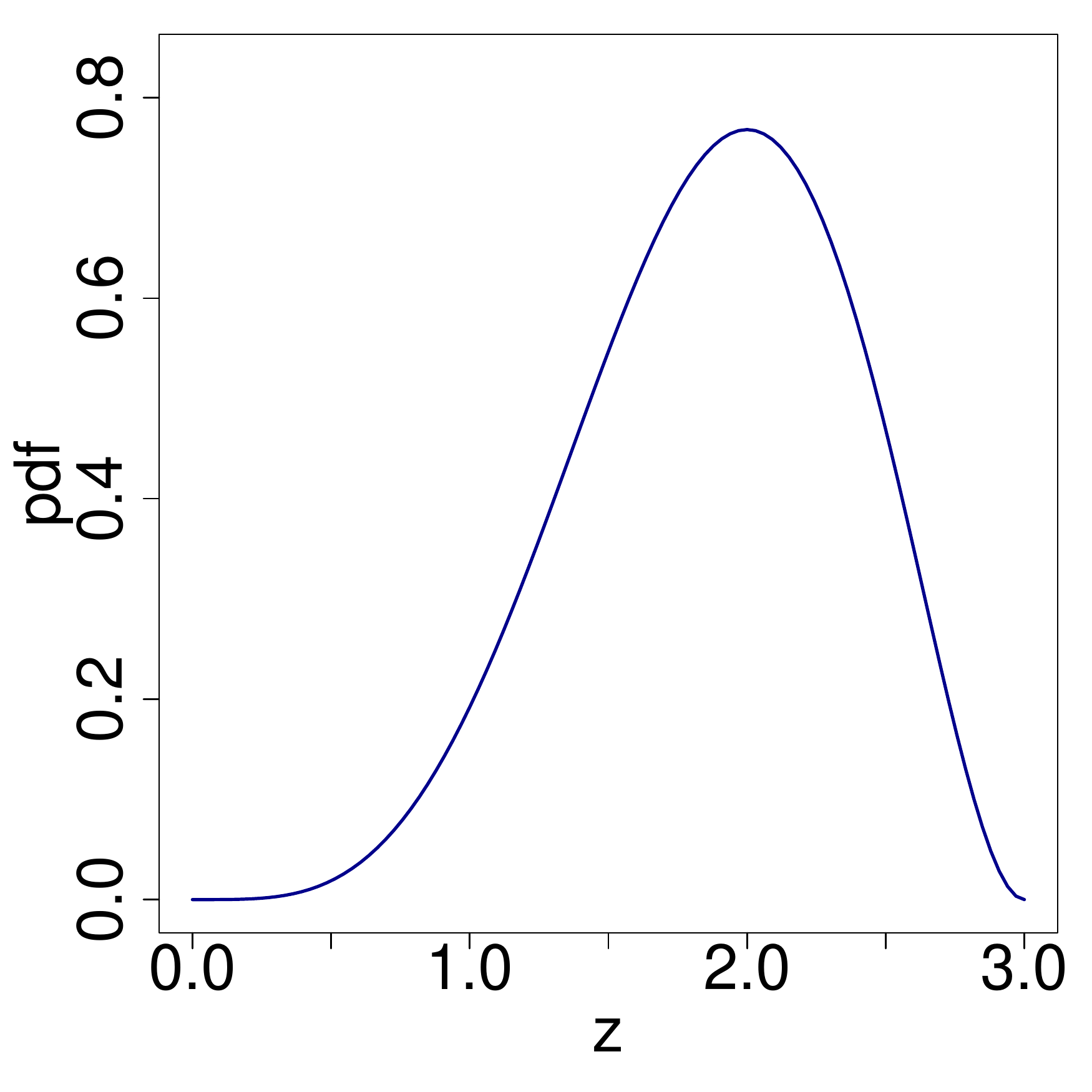}
\end{tabular}
    \begin{tablenotes}
      \footnotesize
      \item Notes. The send type variable $z$ is generated by $3\cdot Beta(\cdot,\cdot).$
    \end{tablenotes}
\end{threeparttable}
\end{figure}

For numerical analysis, we consider a specific distribution of $G$ with
various combinations of underlying parameters. The support of the sender's
type $z$ is set to be $[0,3]$ in the baseline design and is generated from
the Beta distribution multiplied by 3 with the following shape parameters: $%
\{(1,1),(5,5),(3,5),(5,3)\}.$ Figure~\ref{fg:beta-pdf} shows the probability
density functions with different shape parameter values. Note that Beta(1,1)
corresponds to the uniform distribution and Beta(5,5) to a symmetric
bell-shaped distribution. Beta(3,5) and Beta(5,3) correspond to right-skewed
and left skewed sender type distributions, respectively. We set the
effective zero as $10^{-6}$.

\begin{figure}[p]
\begin{threeparttable}
\caption{Design 1. Support Change with Symmetric Distributions}\label{fg:D1}
\centering
\begin{tabular}{c c}
\multicolumn{2}{c}{\underline{Beta(1,1)}}   \\
\includegraphics[scale=0.25]{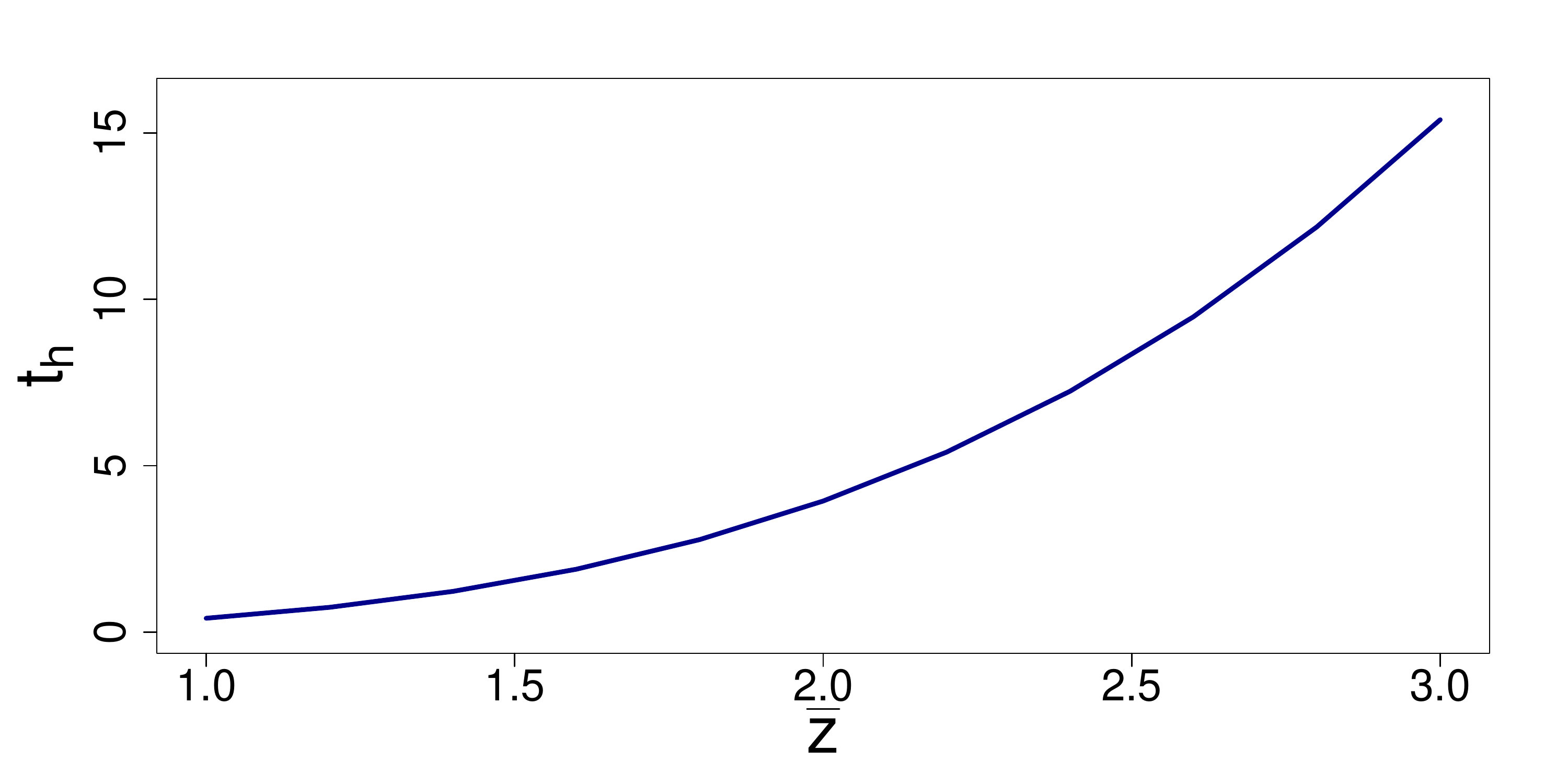} &
\includegraphics[scale=0.25]{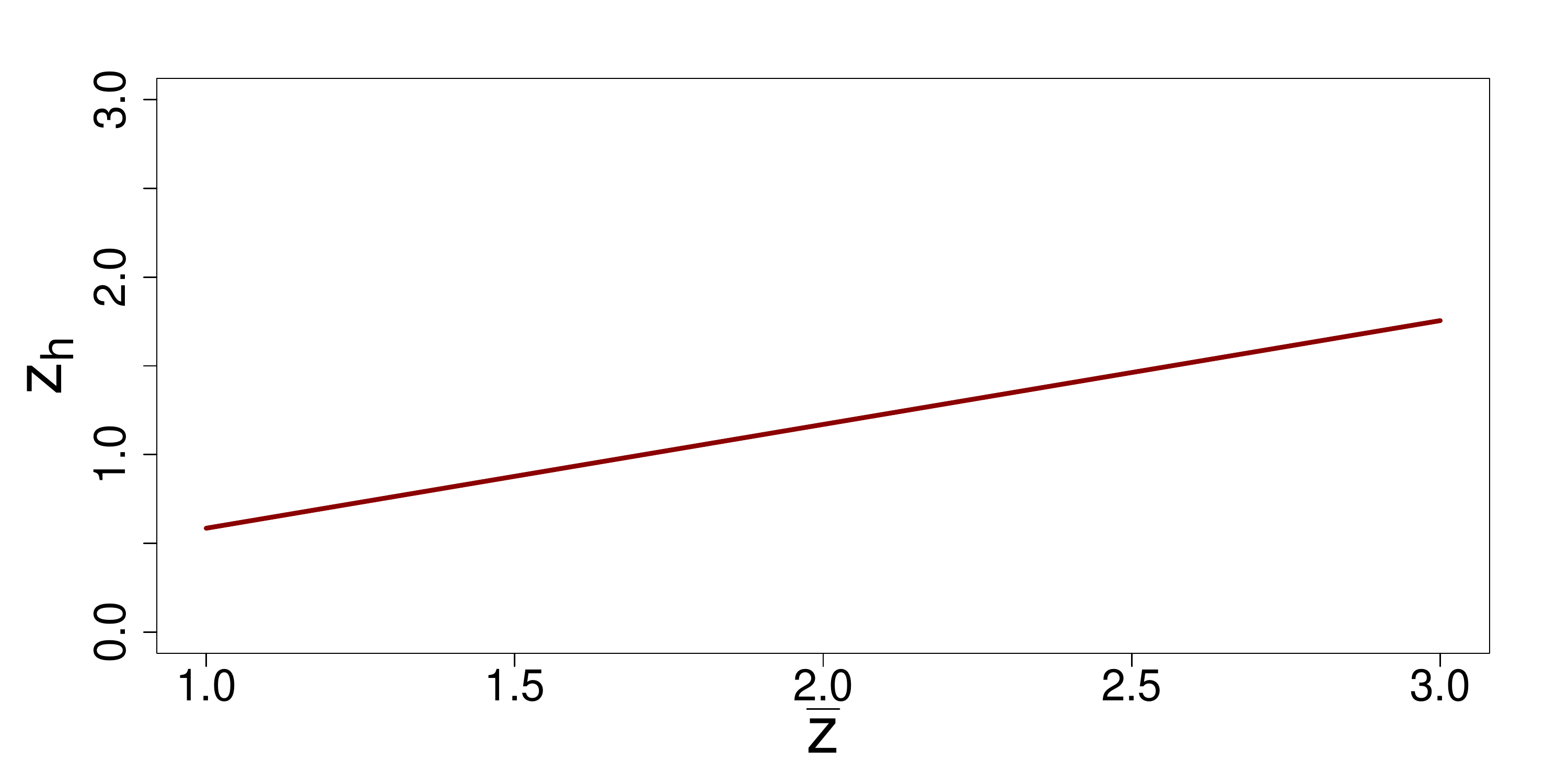} \\
\multicolumn{2}{c}{\underline{Beta(3,5)}}   \\
\includegraphics[scale=0.25]{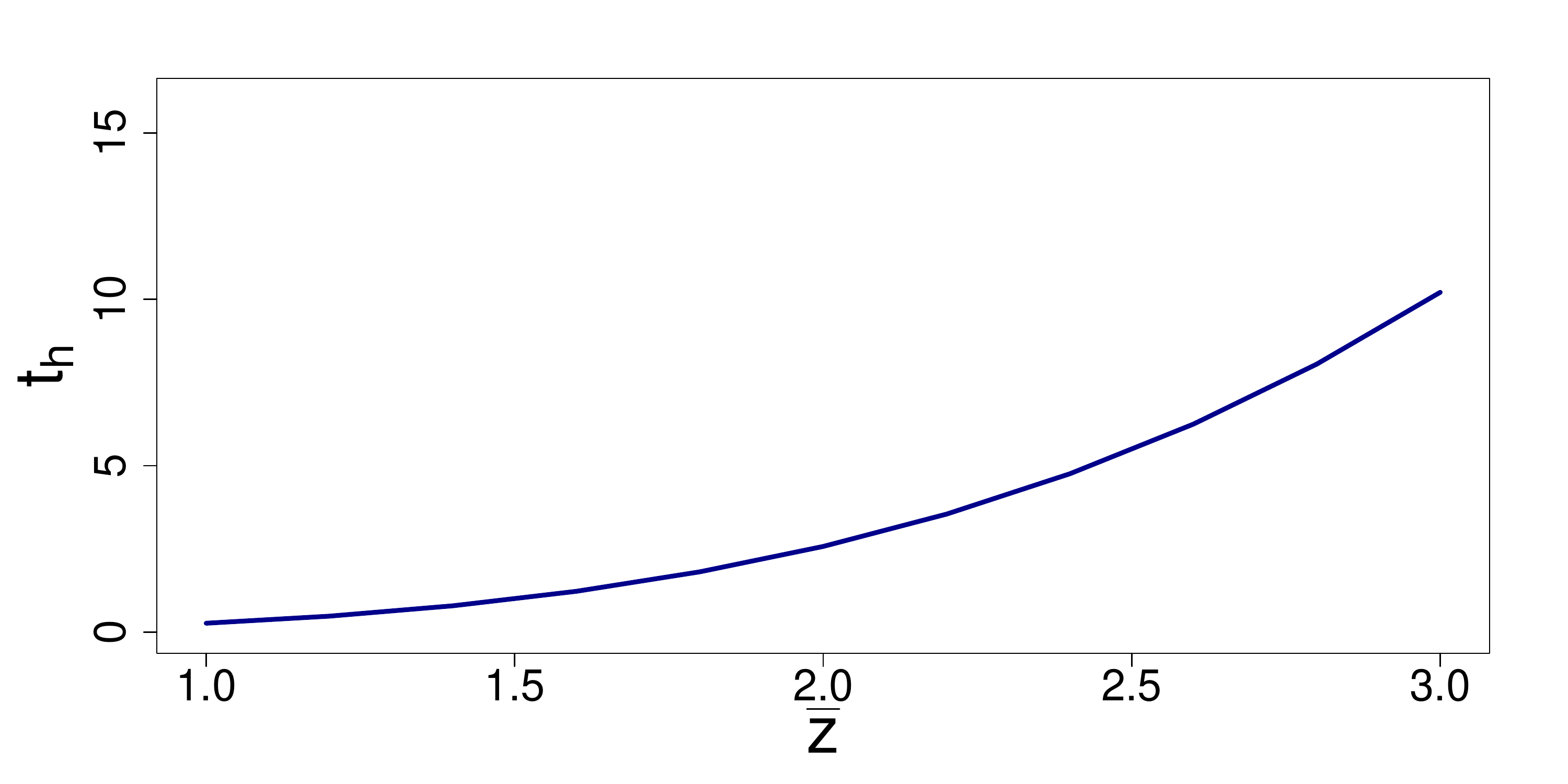} &
\includegraphics[scale=0.25]{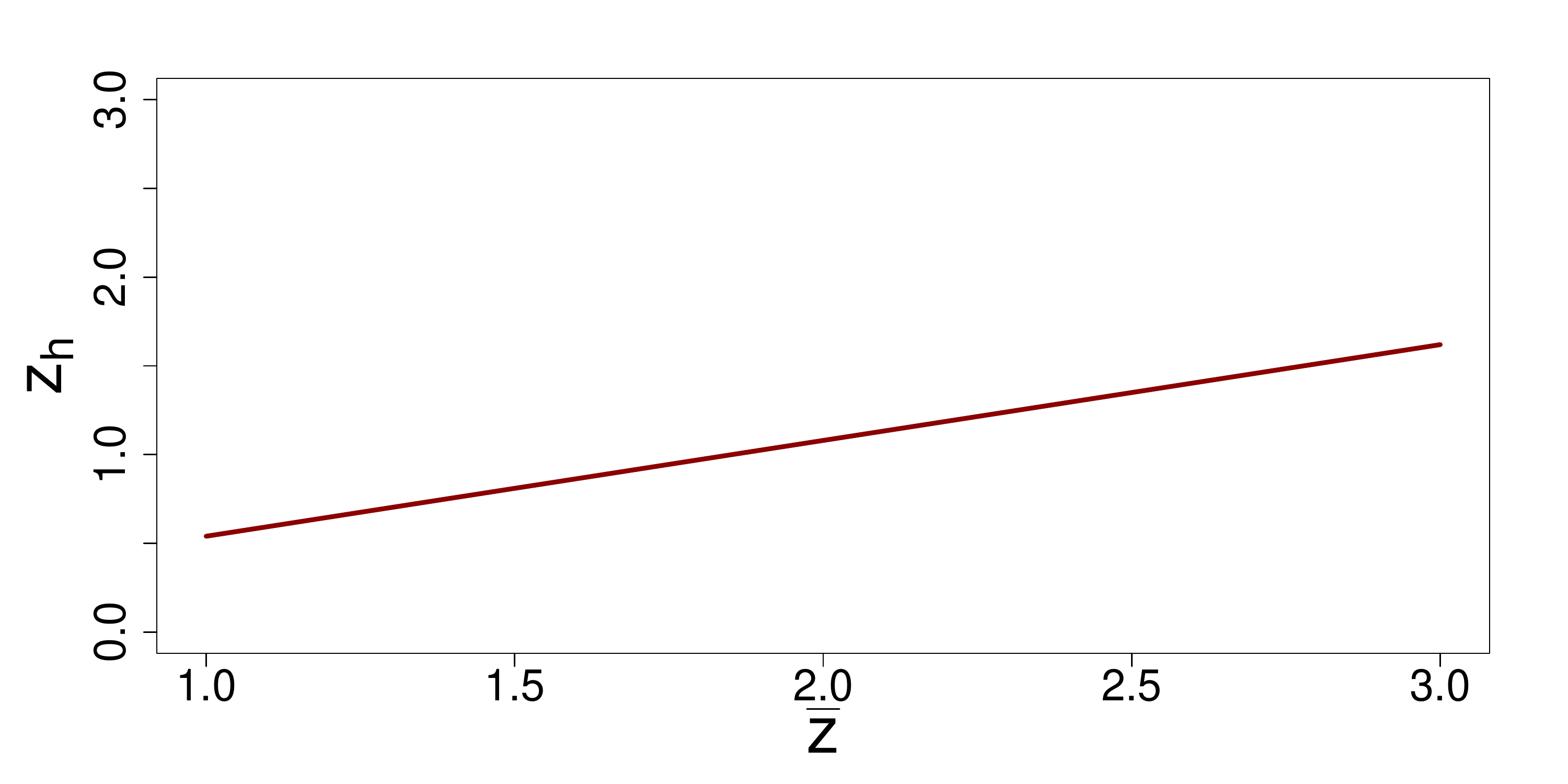} \\
\multicolumn{2}{c}{\underline{Beta(5,5)}}   \\
\includegraphics[scale=0.25]{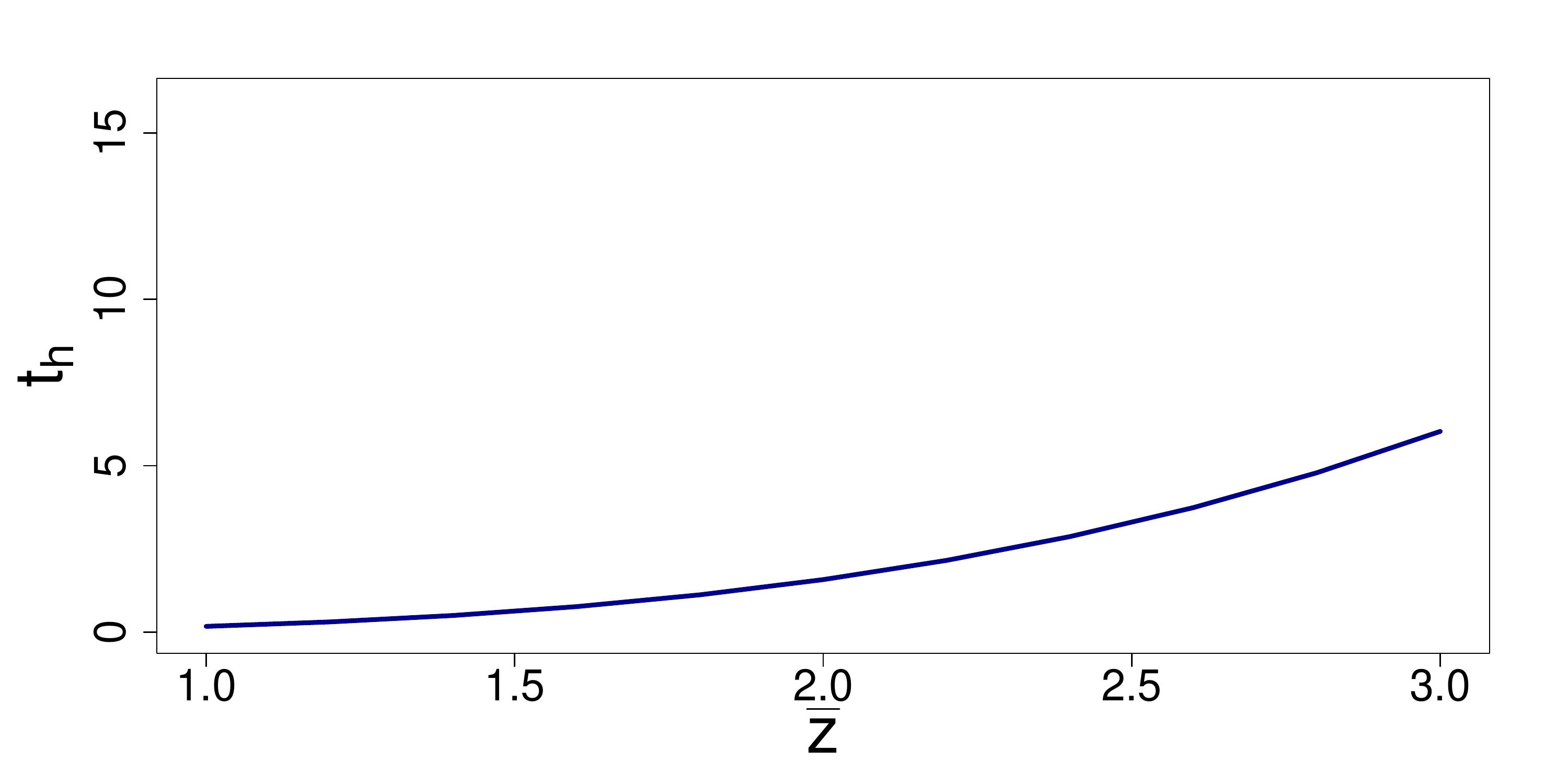} &
\includegraphics[scale=0.25]{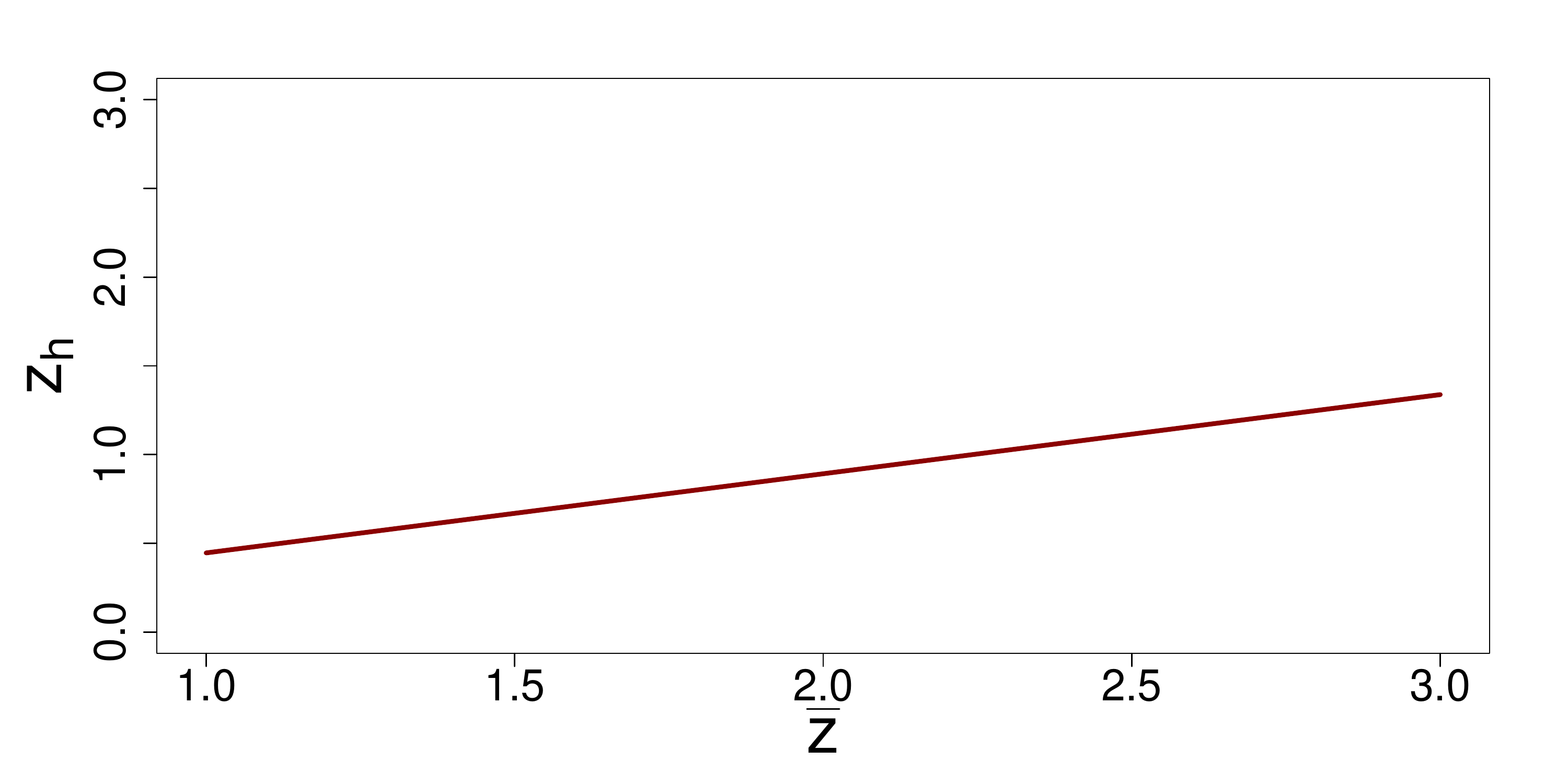} \\
\multicolumn{2}{c}{\underline{Beta(5,3)}}   \\
\includegraphics[scale=0.25]{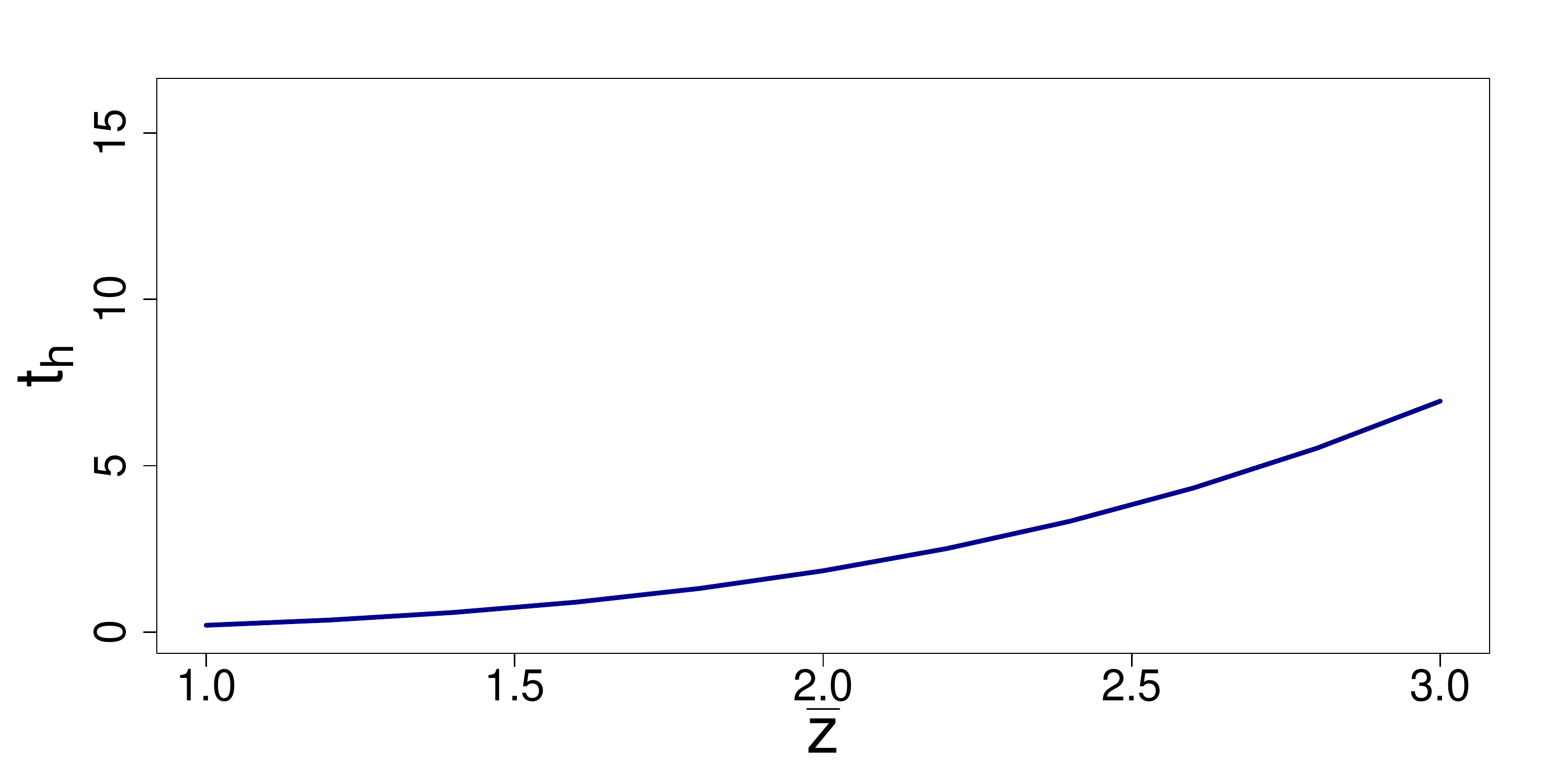} &
\includegraphics[scale=0.25]{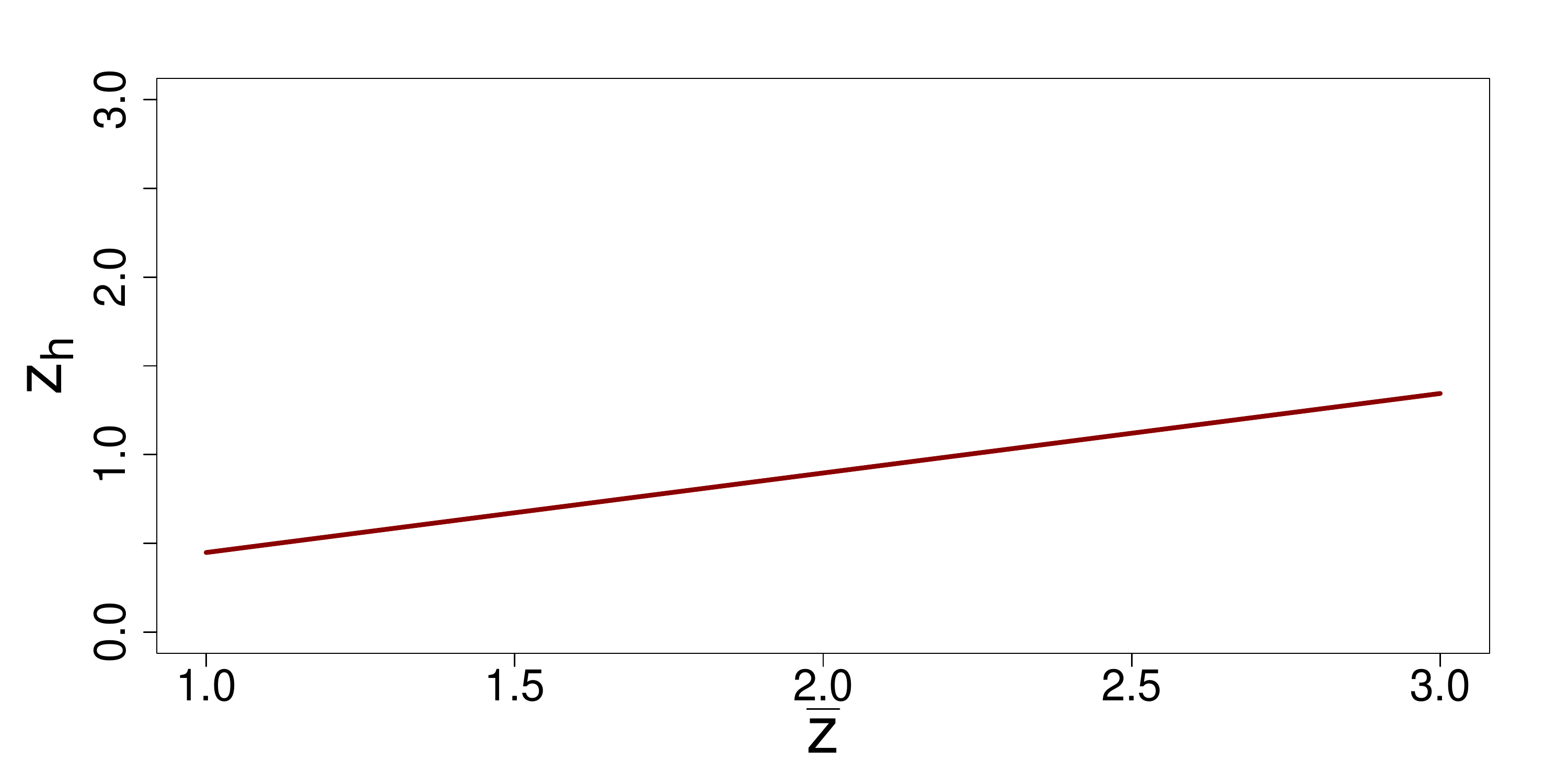} \\
\end{tabular}
    \begin{tablenotes}
      \footnotesize
      \item Notes. This design imposes the same distribution on senders and receivers. We change the upper bound of their supports ($\bar{z} = \bar{x}$). The lower bound of an interval delegation is always zero ($t_l = 0$). The lower bound of a well-behaved equilibrium is also zero ($z_l=0$).
    \end{tablenotes}
\end{threeparttable}
\end{figure}

\begin{figure}[p]
\begin{threeparttable}
\caption{Design 2. Receiver Distribution Change by $k$}\label{fg:D2}
\centering
\begin{tabular}{c c}
\multicolumn{2}{c}{\underline{Beta(1,1)}}   \\
\includegraphics[scale=0.25]{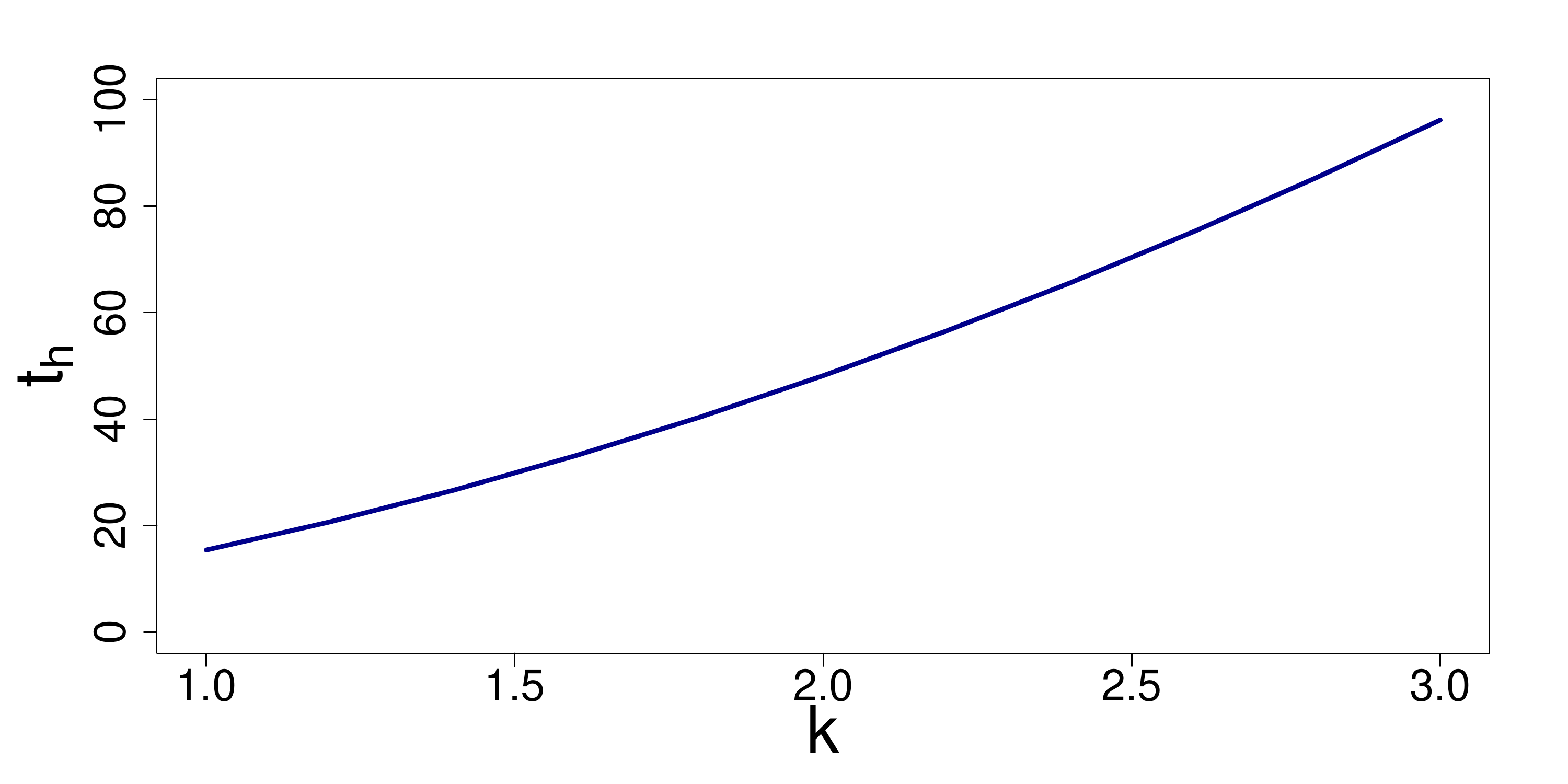} &
\includegraphics[scale=0.25]{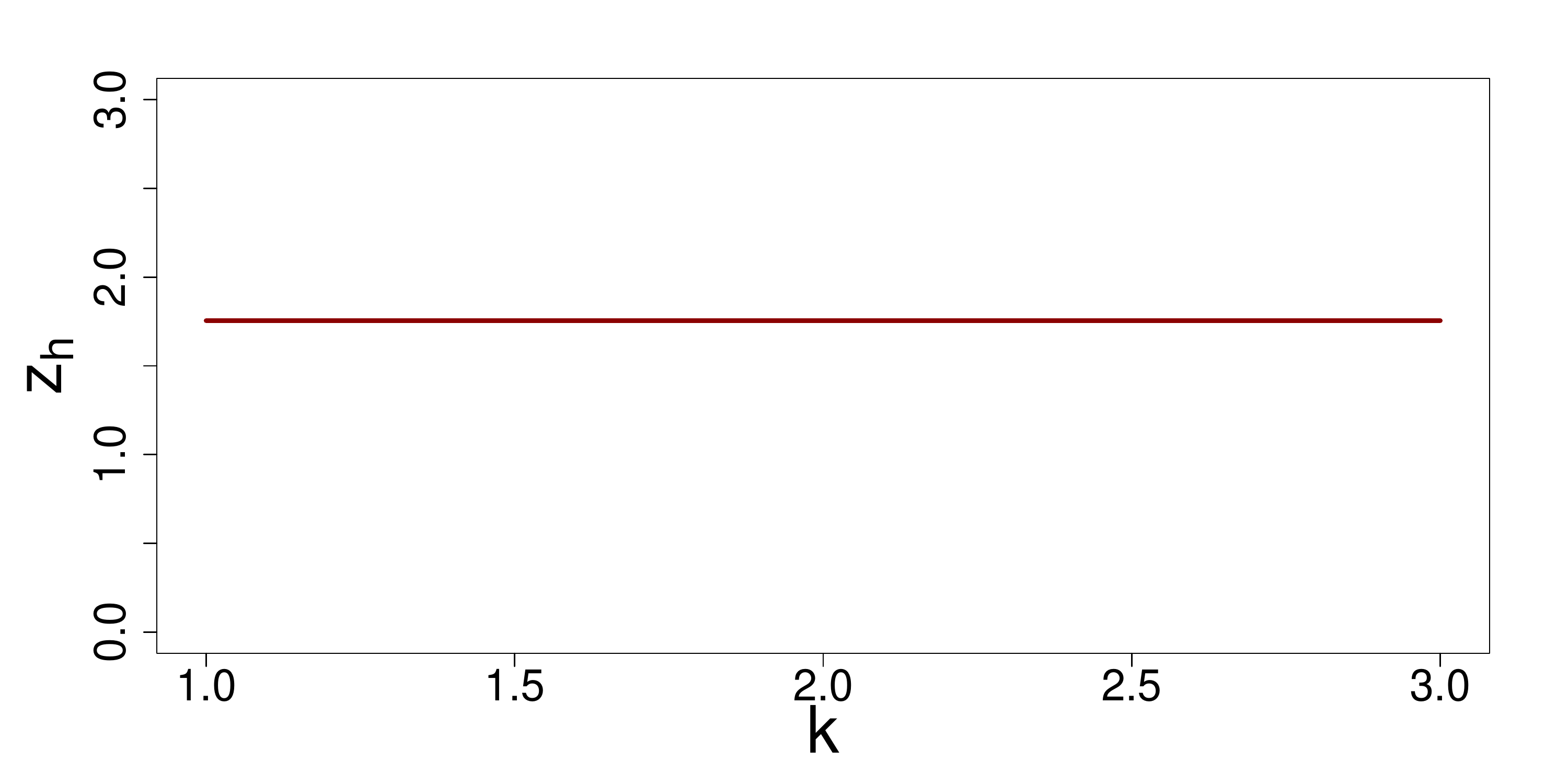} \\
\multicolumn{2}{c}{\underline{Beta(3,5)}}   \\
\includegraphics[scale=0.25]{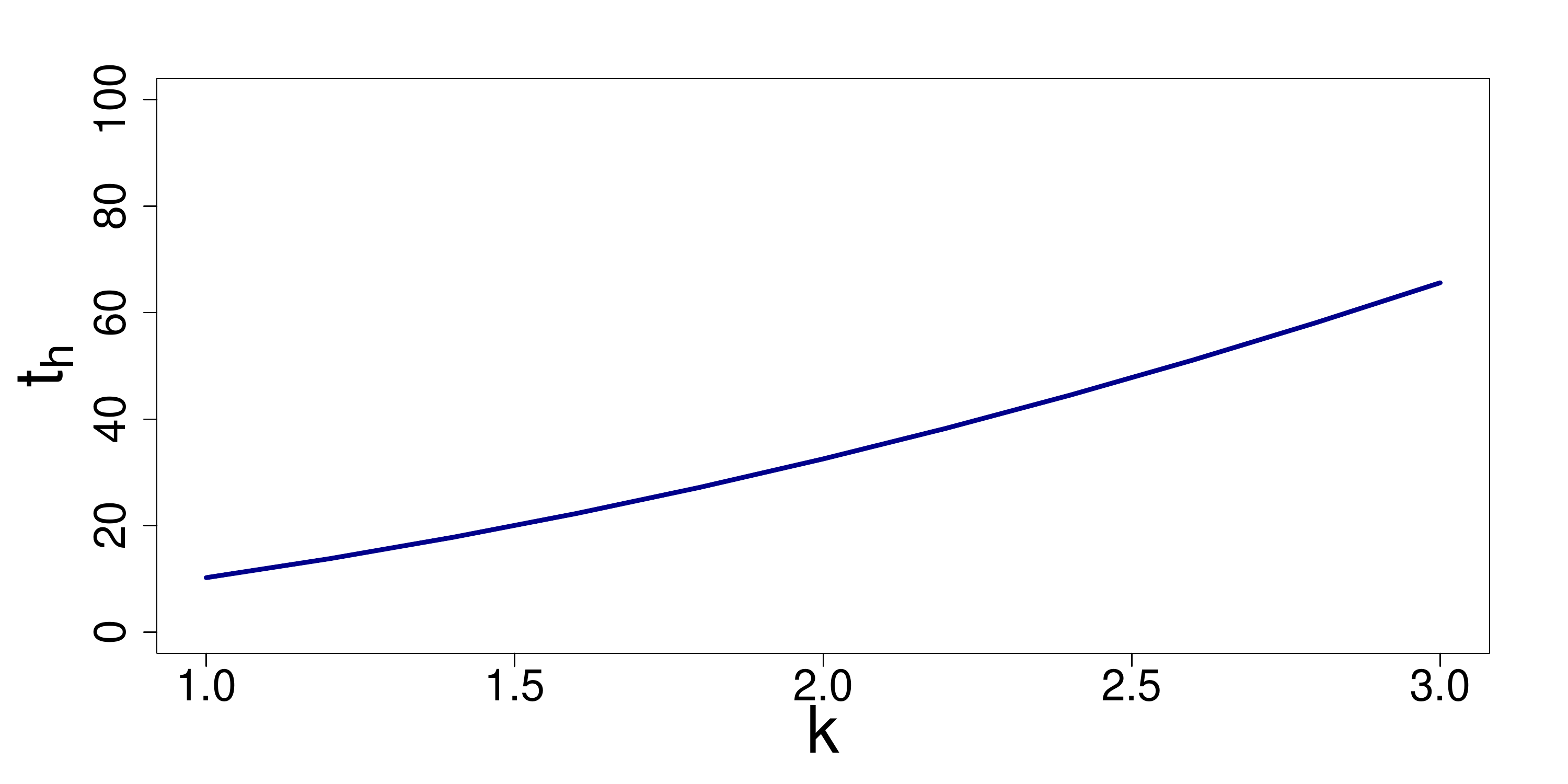} &
\includegraphics[scale=0.25]{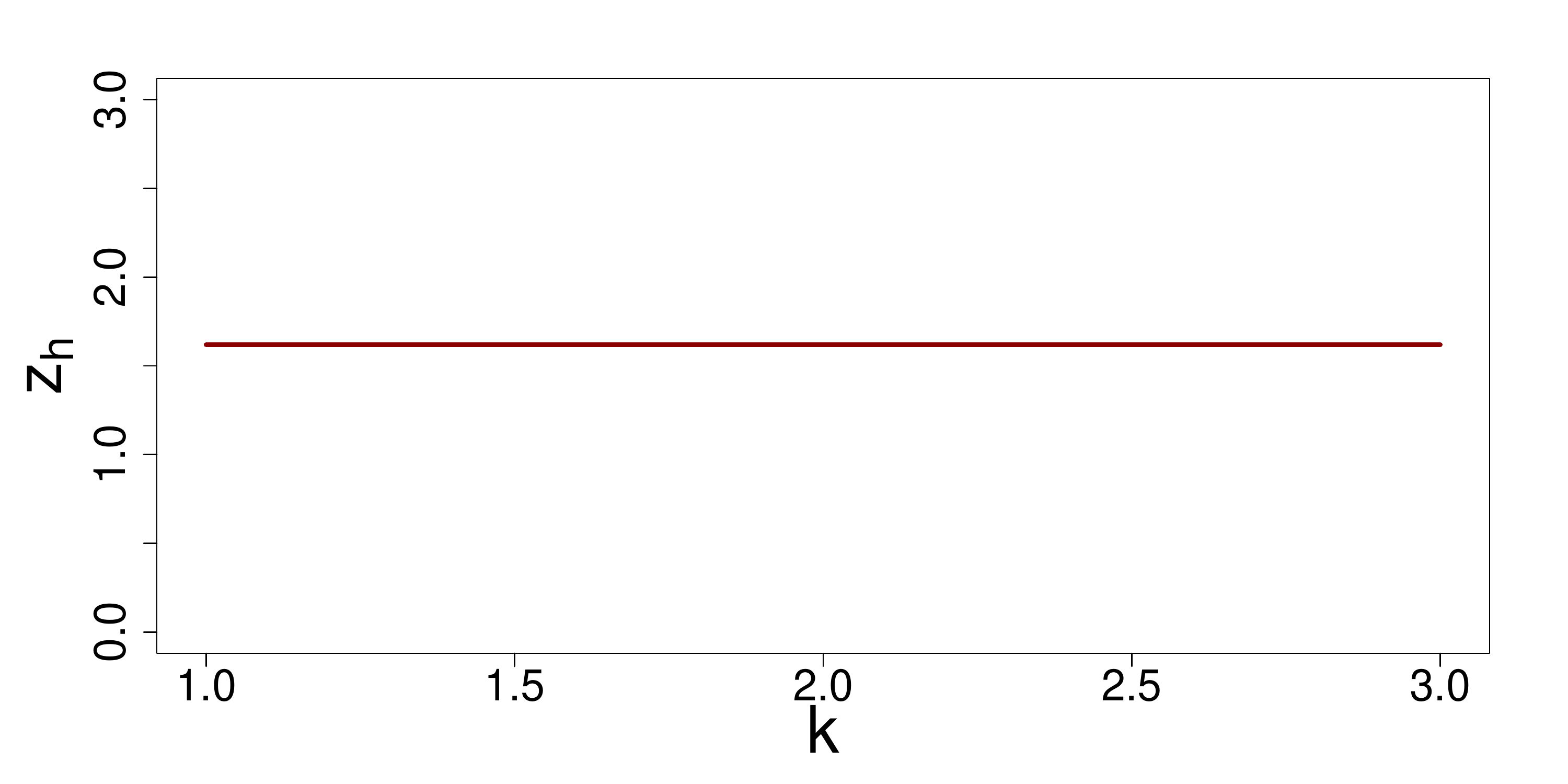} \\
\multicolumn{2}{c}{\underline{Beta(5,5)}}   \\
\includegraphics[scale=0.25]{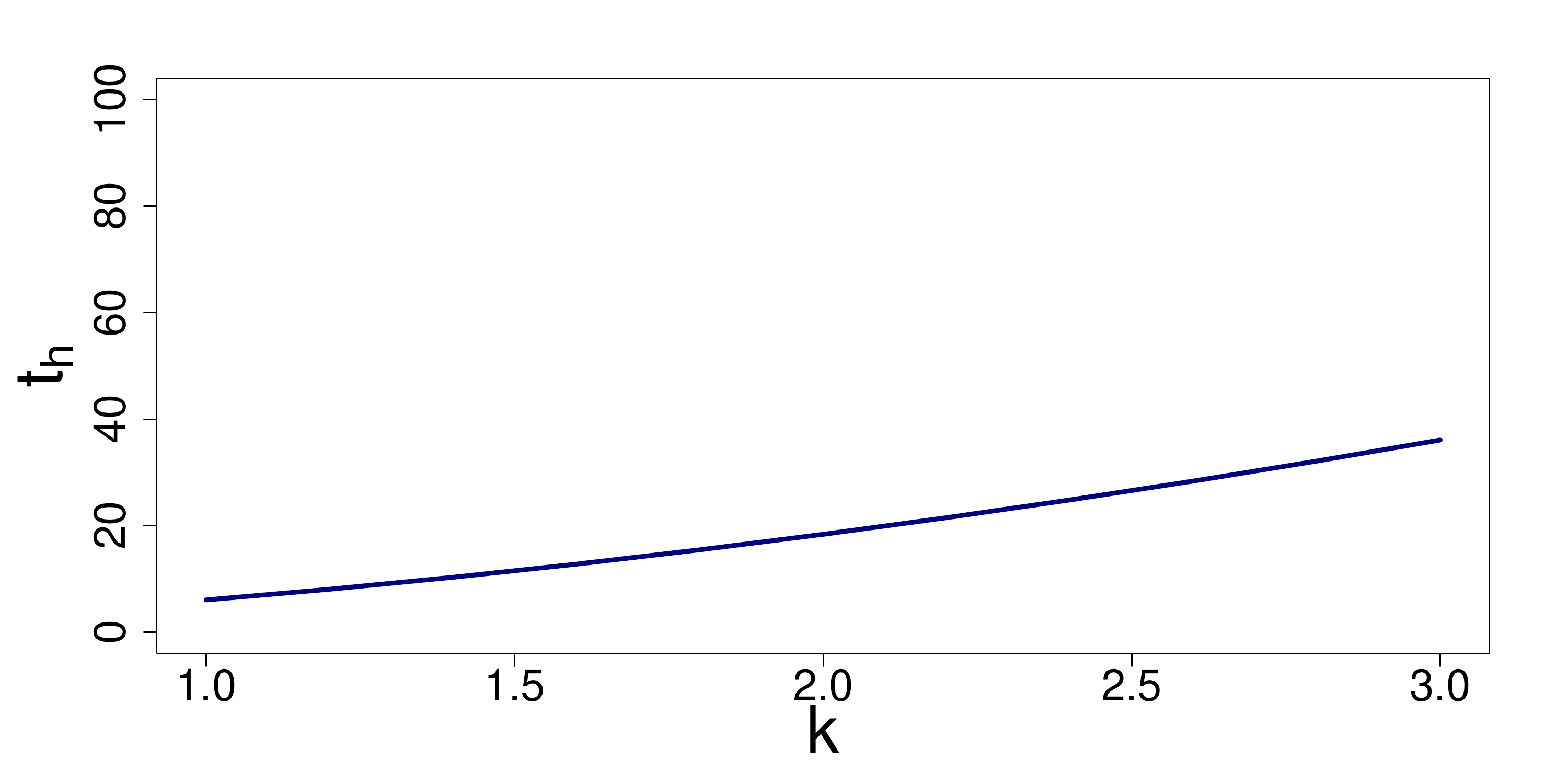} &
\includegraphics[scale=0.25]{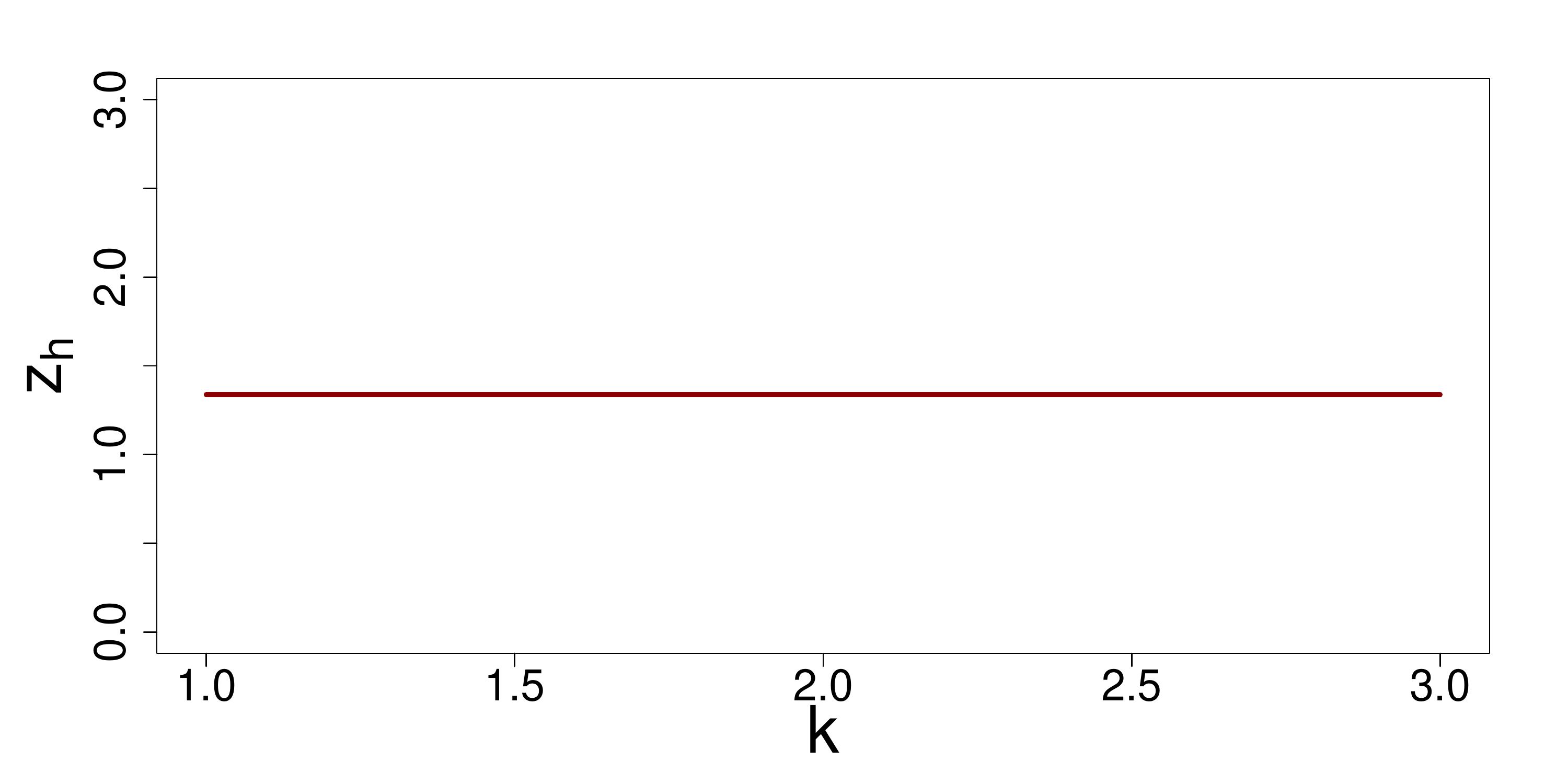} \\
\multicolumn{2}{c}{\underline{Beta(5,3)}}   \\
\includegraphics[scale=0.25]{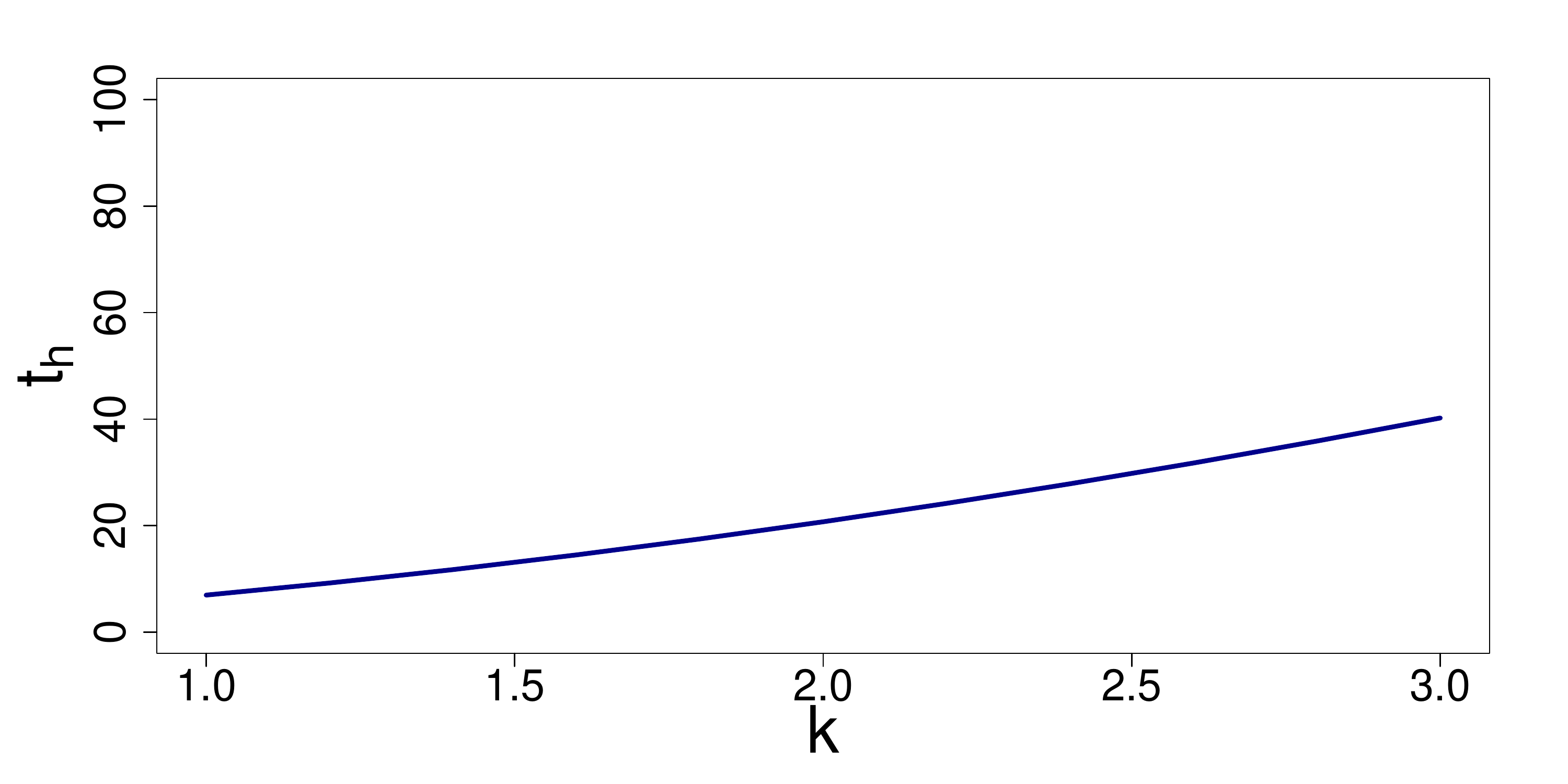} &
\includegraphics[scale=0.25]{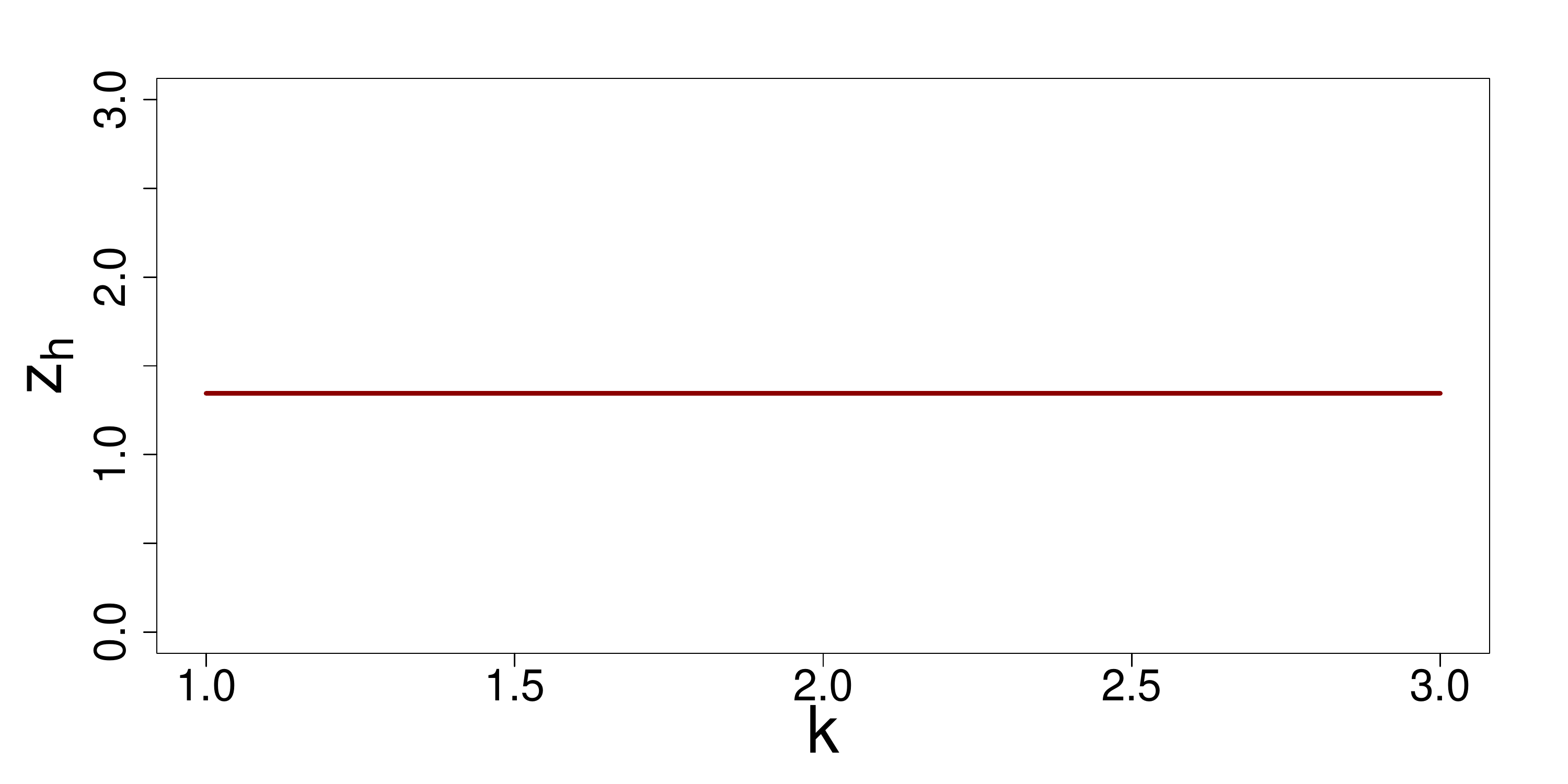} \\
\end{tabular}
    \begin{tablenotes}
      \footnotesize
      \item Notes. This design varies the distribution of the receiver type ($x$) by changing $k$. Recall the matching function $x = k z^q$. We keep the same distribution on senders' type. The lower bound of an interval delegation is always zero ($t_l = 0$). The lower bound of a well-behaved equilibrium is also zero ($z_l=0$).
    \end{tablenotes}
\end{threeparttable}
\end{figure}

\begin{figure}[p]
\begin{threeparttable}
\caption{Design 3. Receiver Distribution Change by $q$}\label{fg:D3}
\centering
\begin{tabular}{c c}
\multicolumn{2}{c}{\underline{Beta(1,1)}}   \\
\includegraphics[scale=0.25]{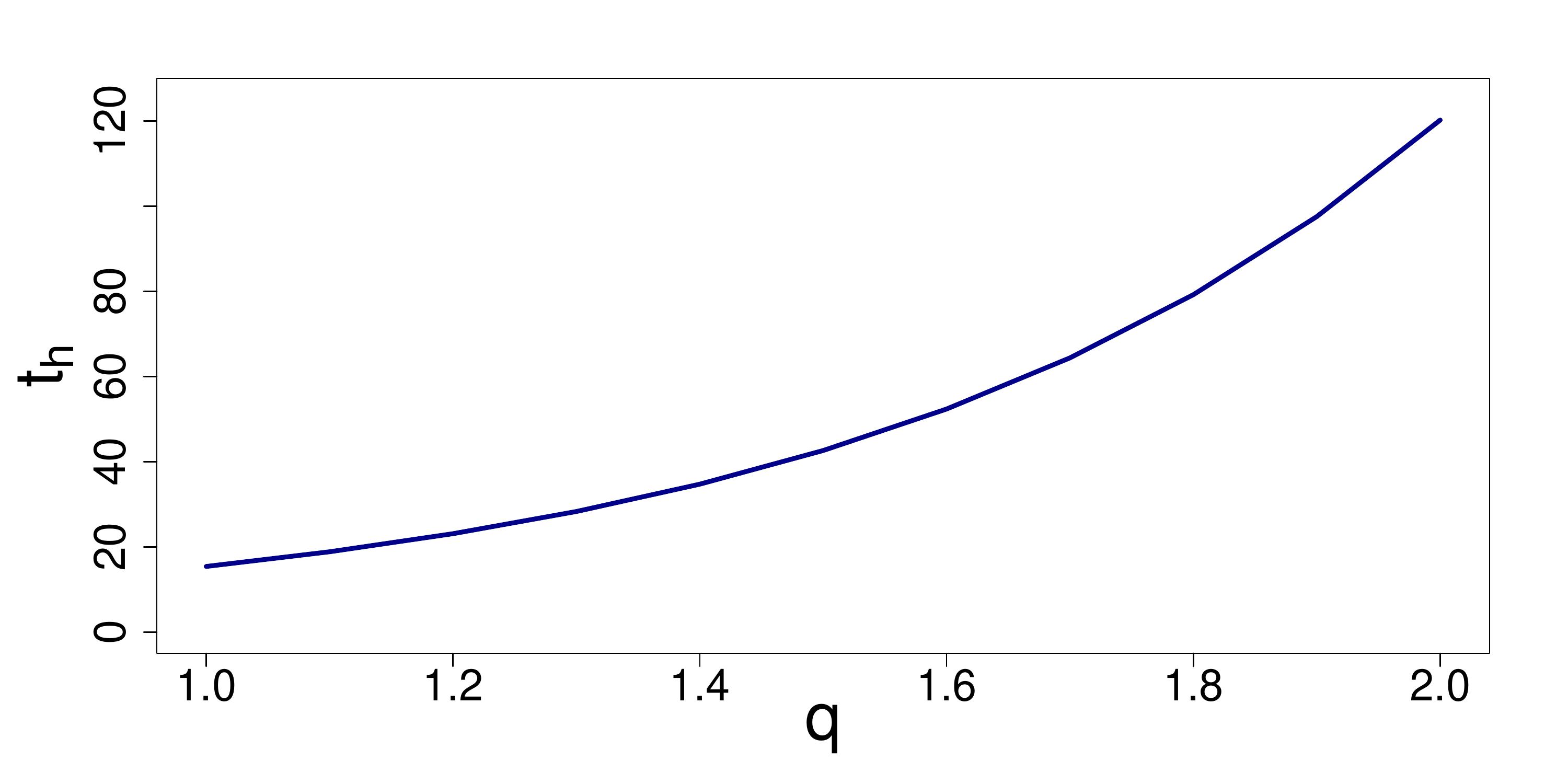} &
\includegraphics[scale=0.25]{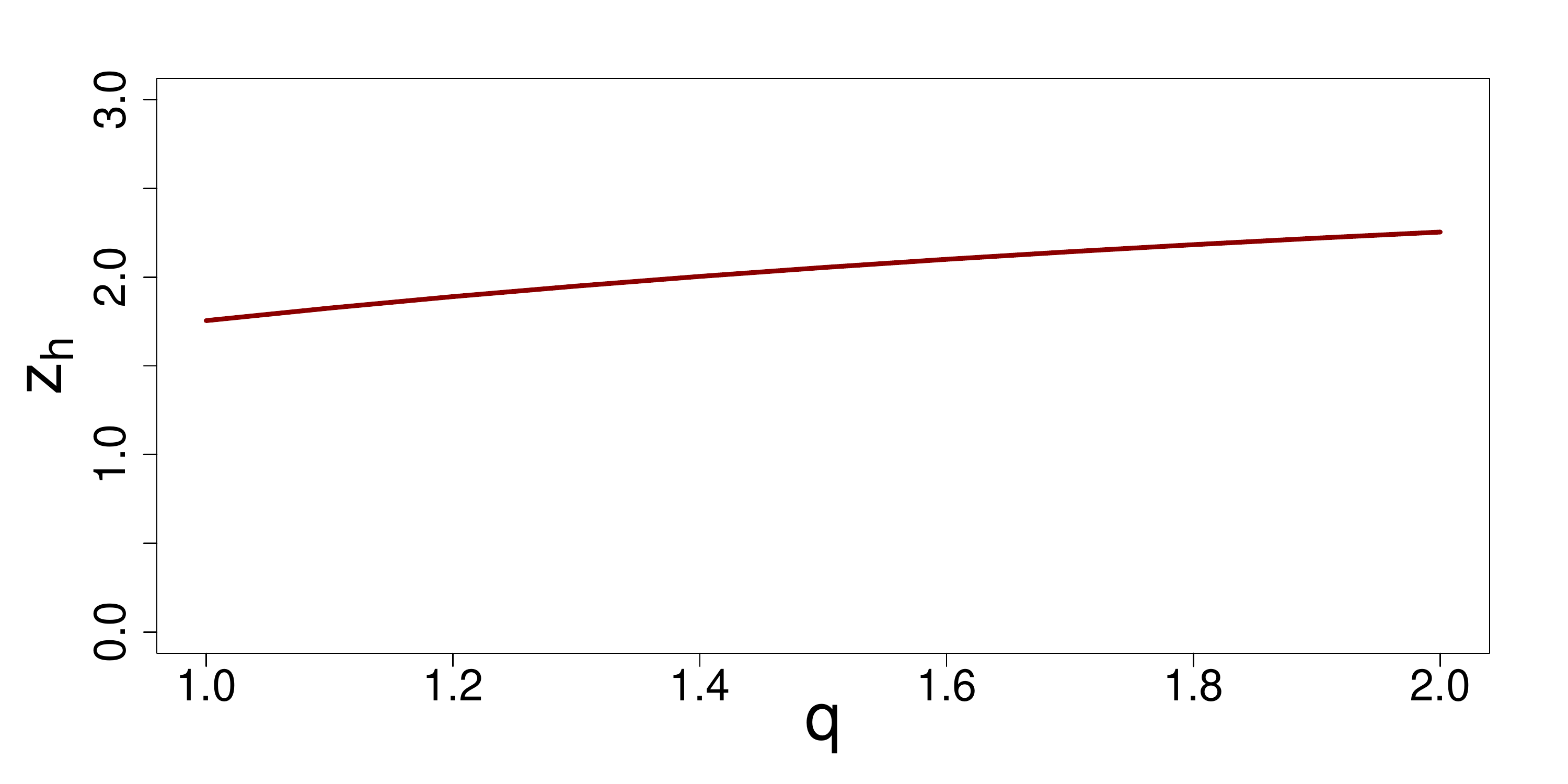} \\
\multicolumn{2}{c}{\underline{Beta(3,5)}}   \\
\includegraphics[scale=0.25]{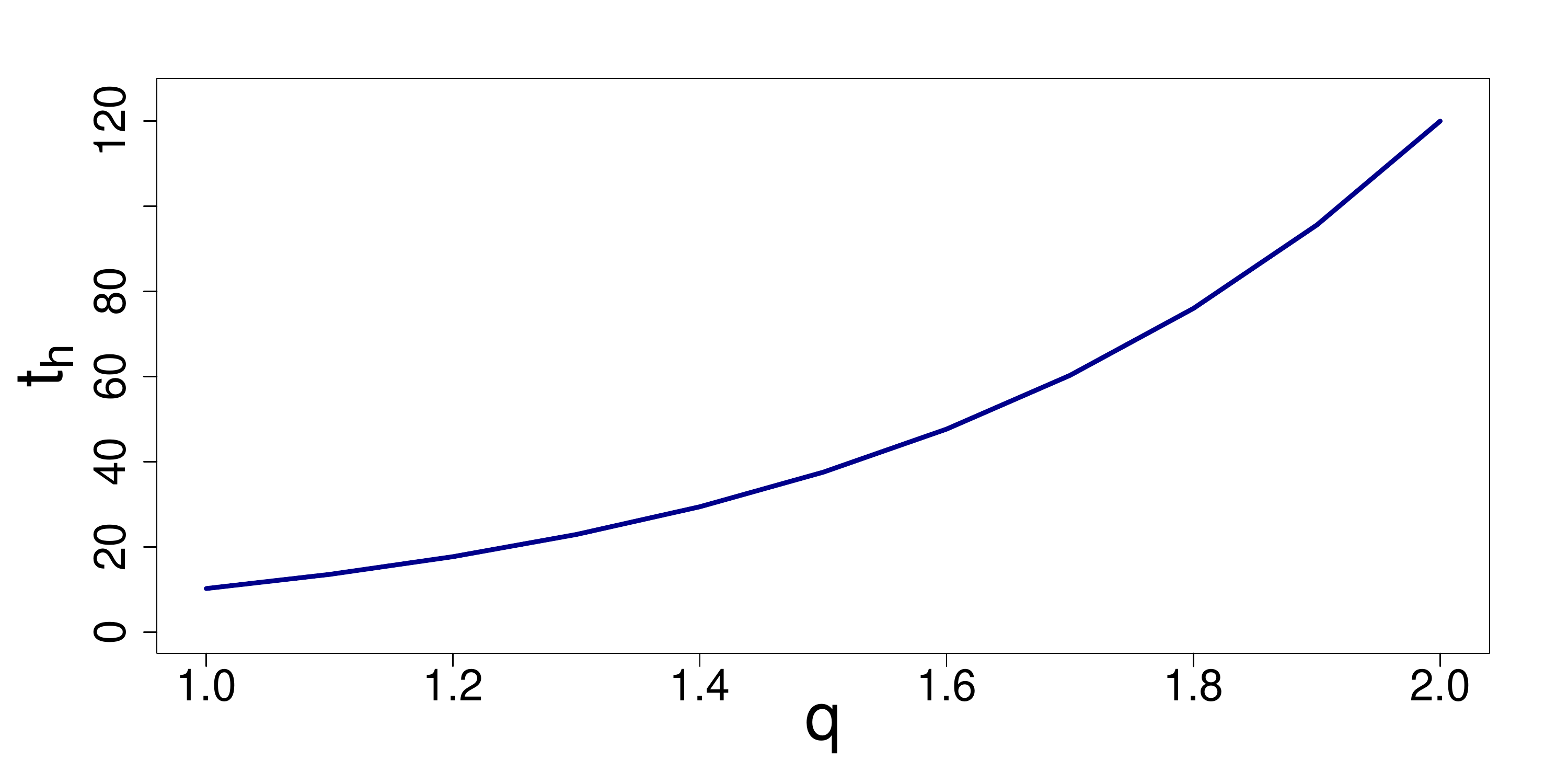} &
\includegraphics[scale=0.25]{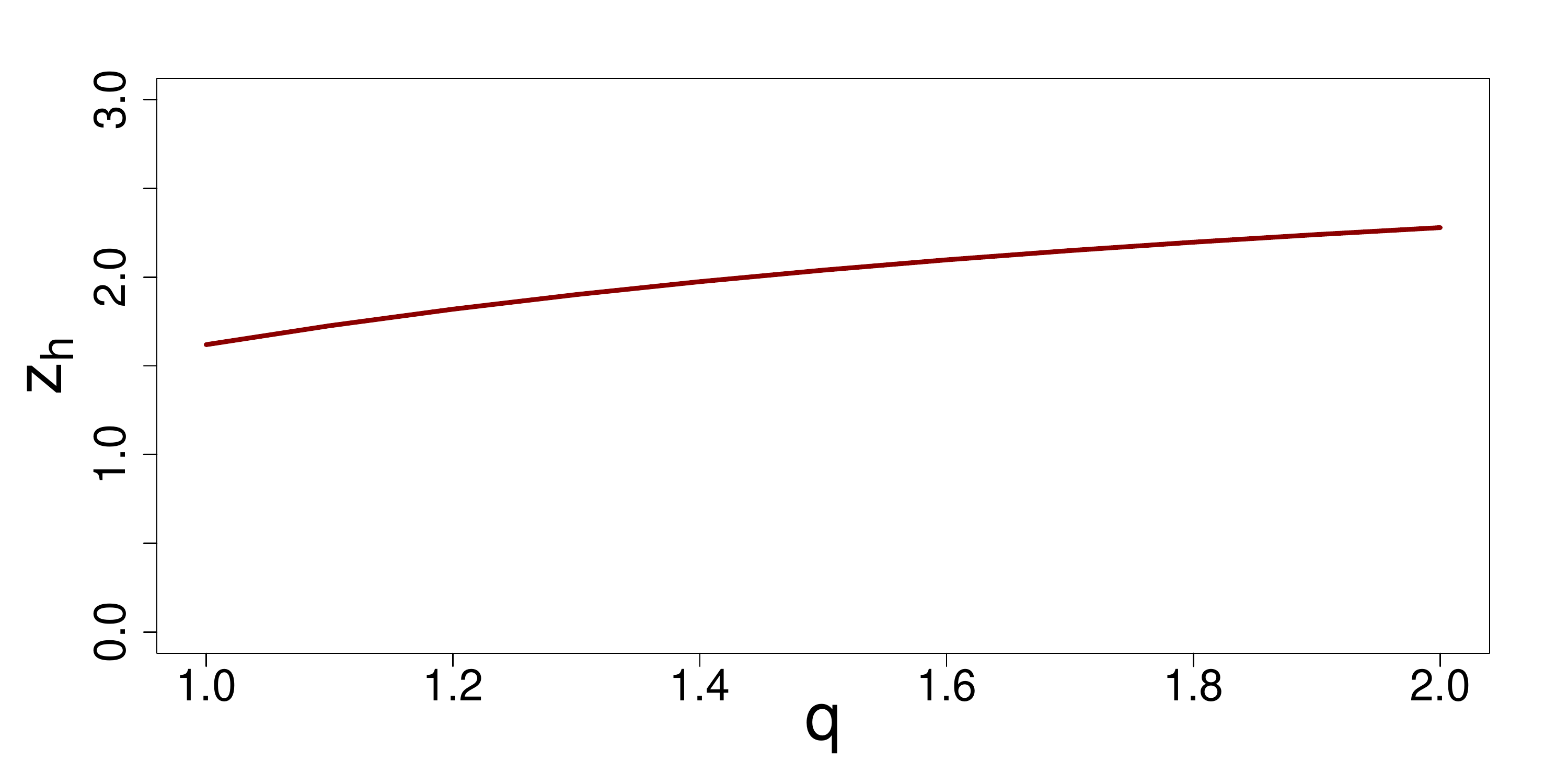} \\
\multicolumn{2}{c}{\underline{Beta(5,5)}}   \\
\includegraphics[scale=0.25]{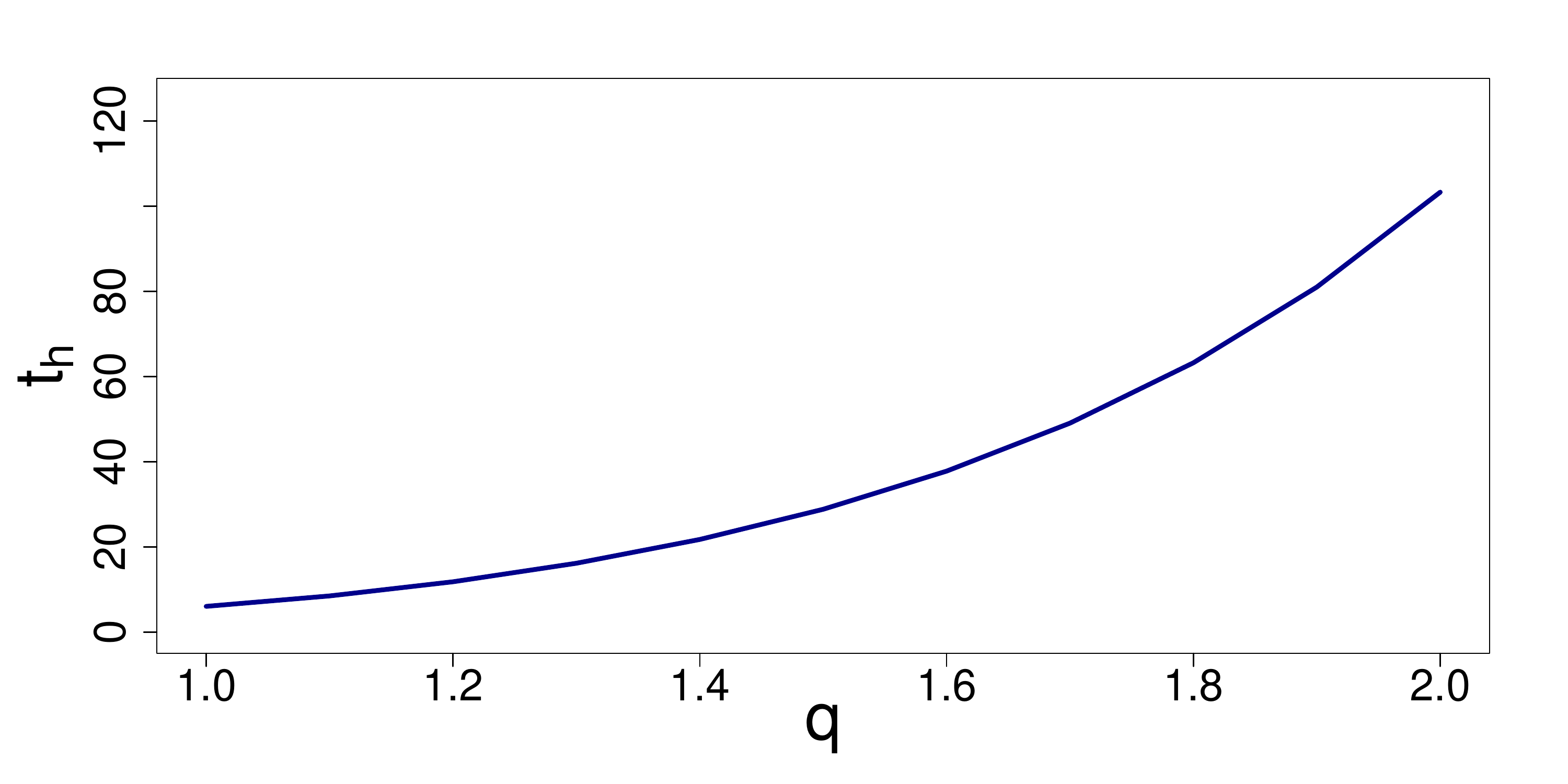} &
\includegraphics[scale=0.25]{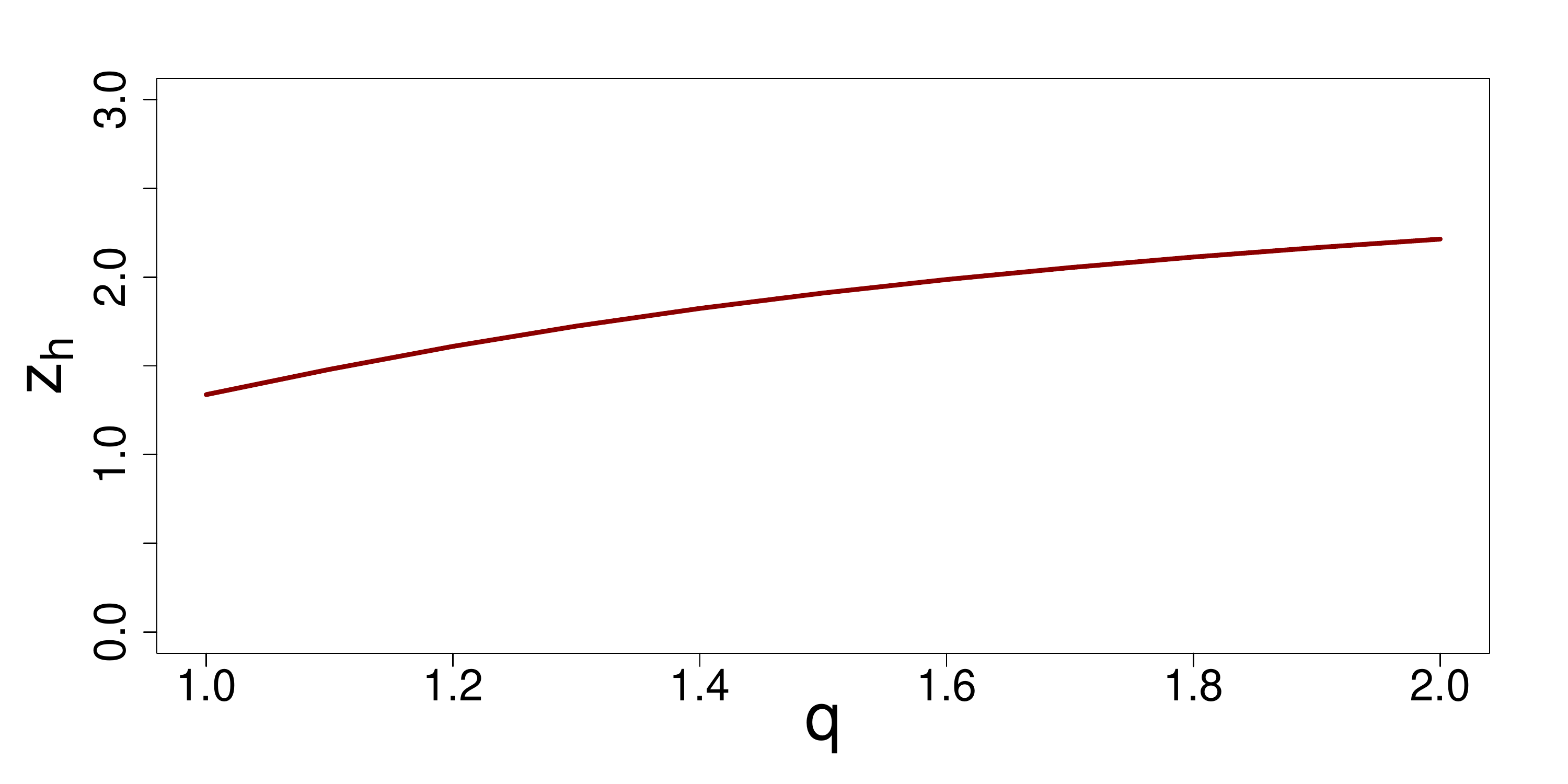} \\
\multicolumn{2}{c}{\underline{Beta(5,3)}}   \\
\includegraphics[scale=0.25]{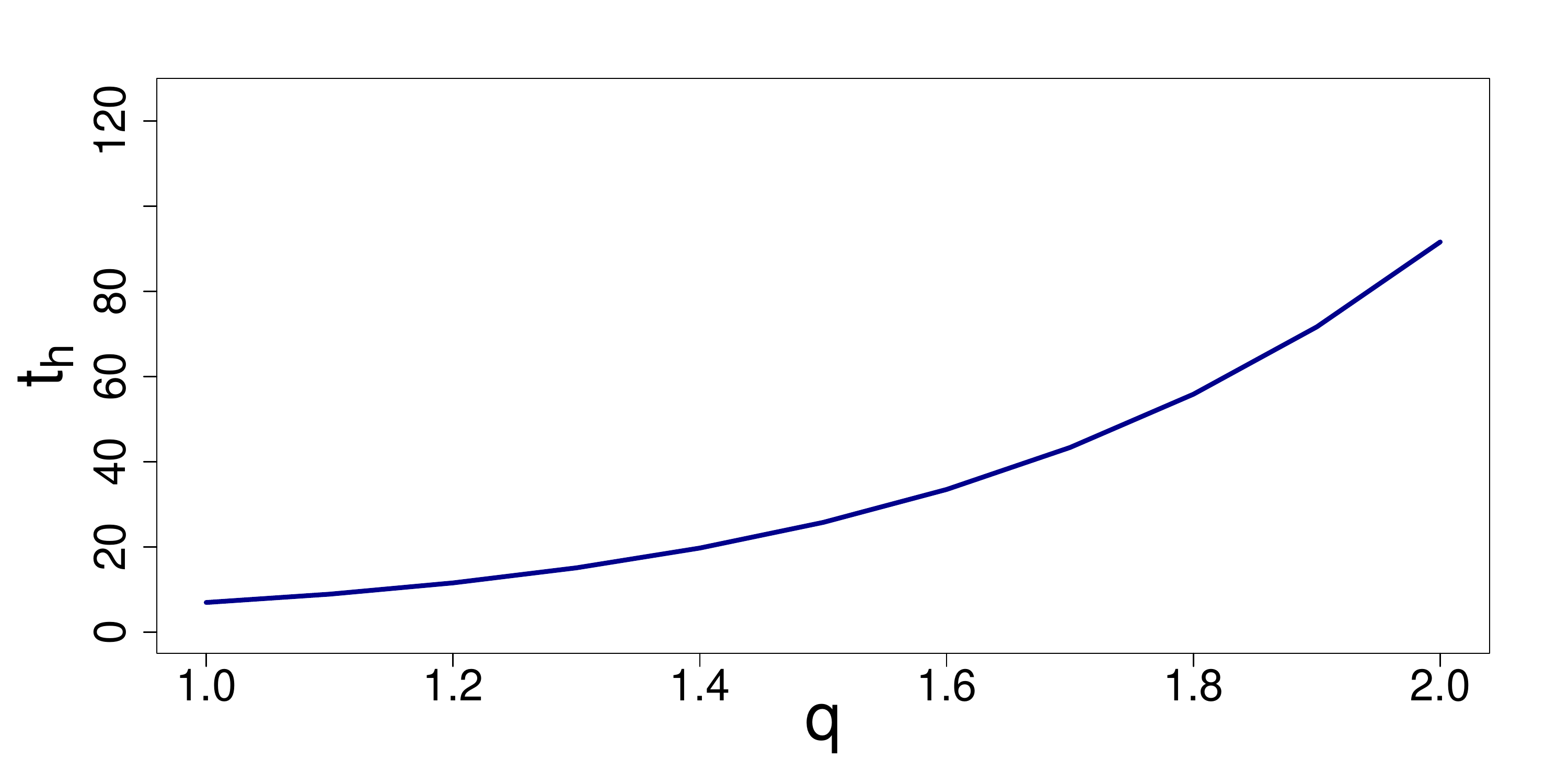} &
\includegraphics[scale=0.25]{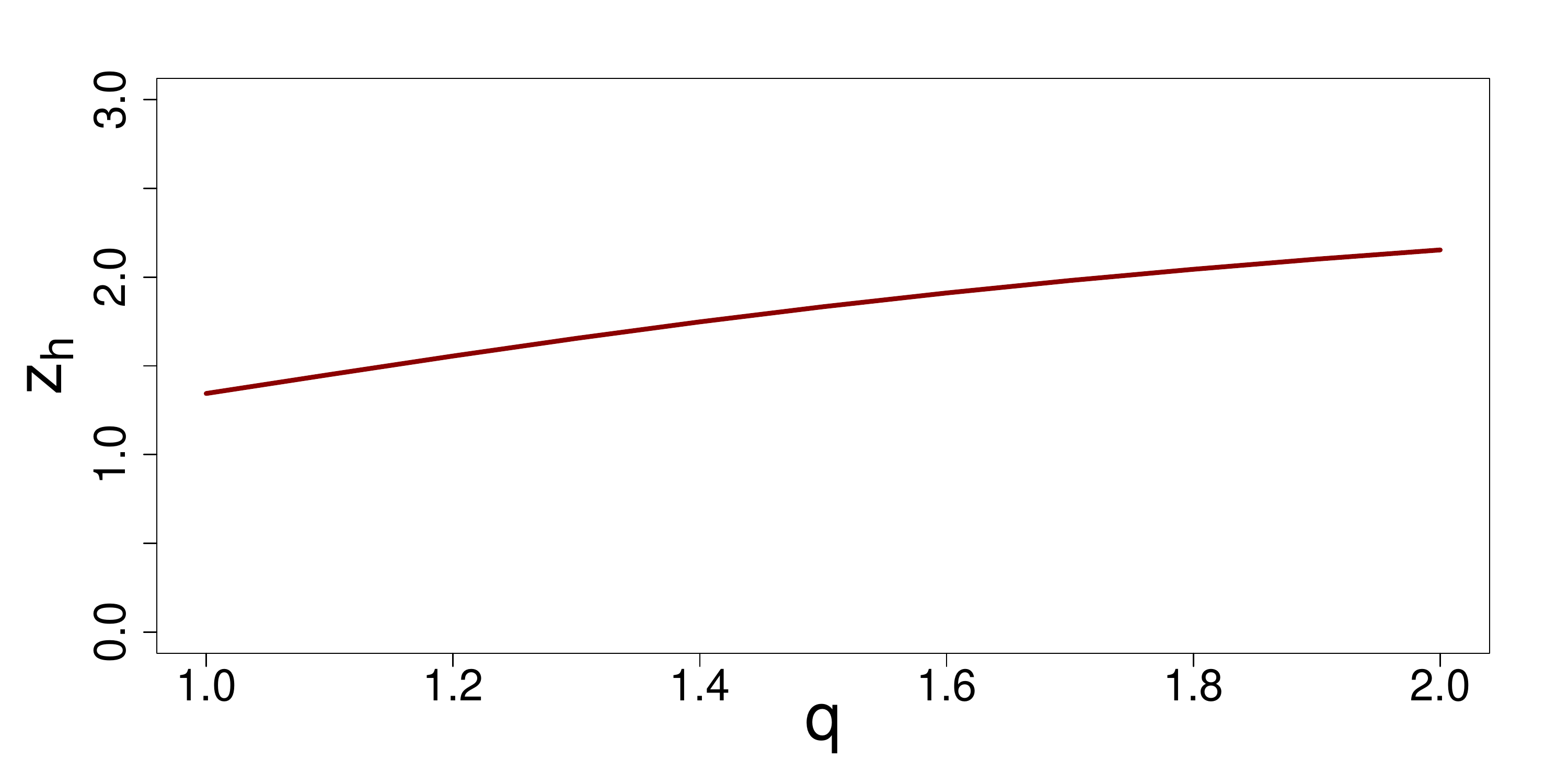} \\
\end{tabular}
    \begin{tablenotes}
      \footnotesize
      \item Notes. This design varies the distribution of the receiver type ($x$) by changing $q$. Recall the matching function $x = k z^q$. We keep the same distribution on senders' type. The lower bound of an interval delegation is always zero ($t_l = 0$). The lower bound of a well-behaved equilibrium is also zero ($z_l=0$).
    \end{tablenotes}
\end{threeparttable}
\end{figure}

\begin{figure}[p]
\begin{threeparttable}
\caption{Design 4. Signal Contribution ($a$) and Receiver Type ($q$)}\label{fg:D4}
\centering
\begin{tabular}{c c}
\multicolumn{2}{c}{\underline{Beta(1,1)}}   \\
\includegraphics[scale=0.25]{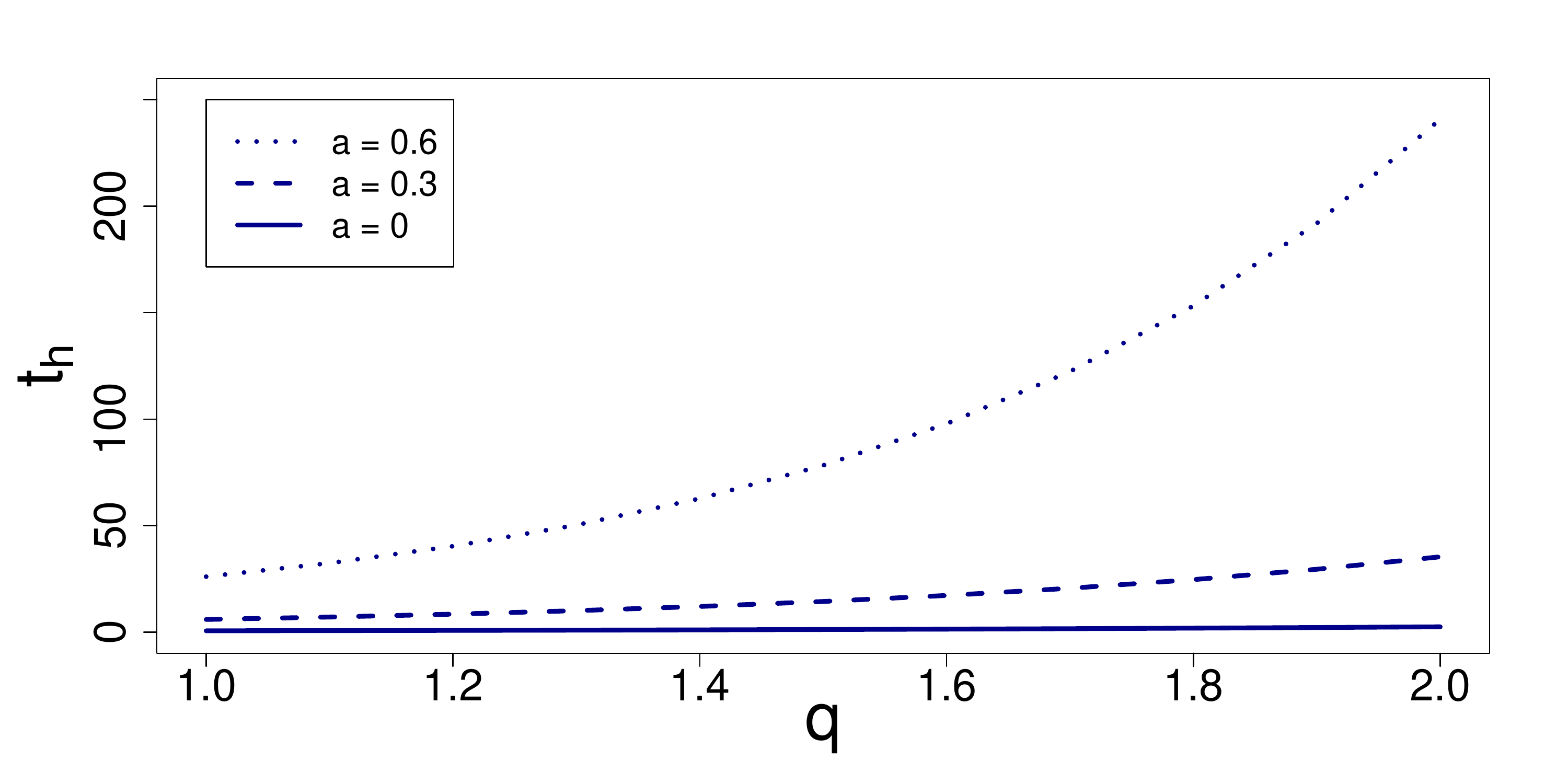} &
\includegraphics[scale=0.25]{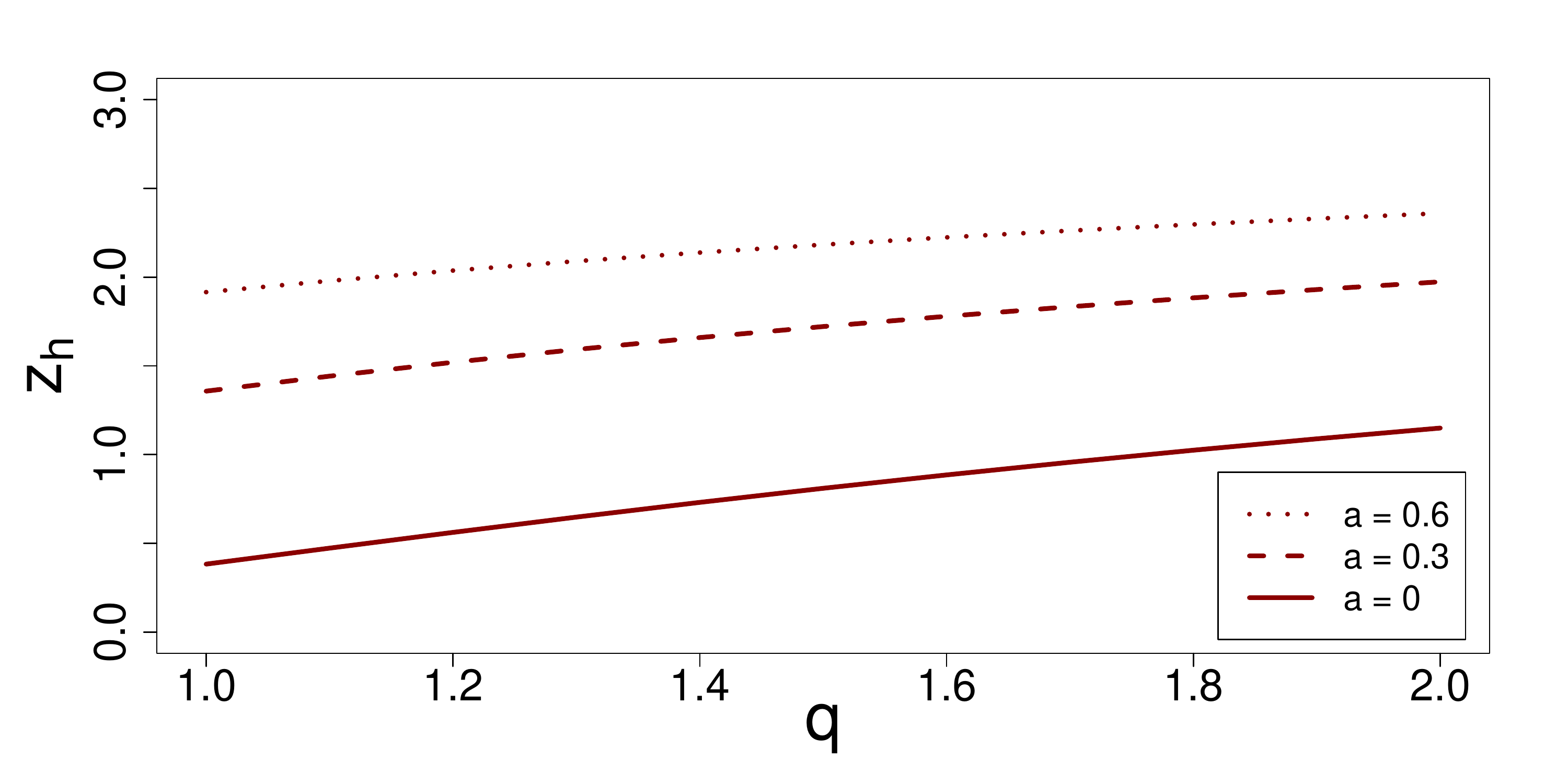} \\
\multicolumn{2}{c}{\underline{Beta(3,5)}}   \\
\includegraphics[scale=0.25]{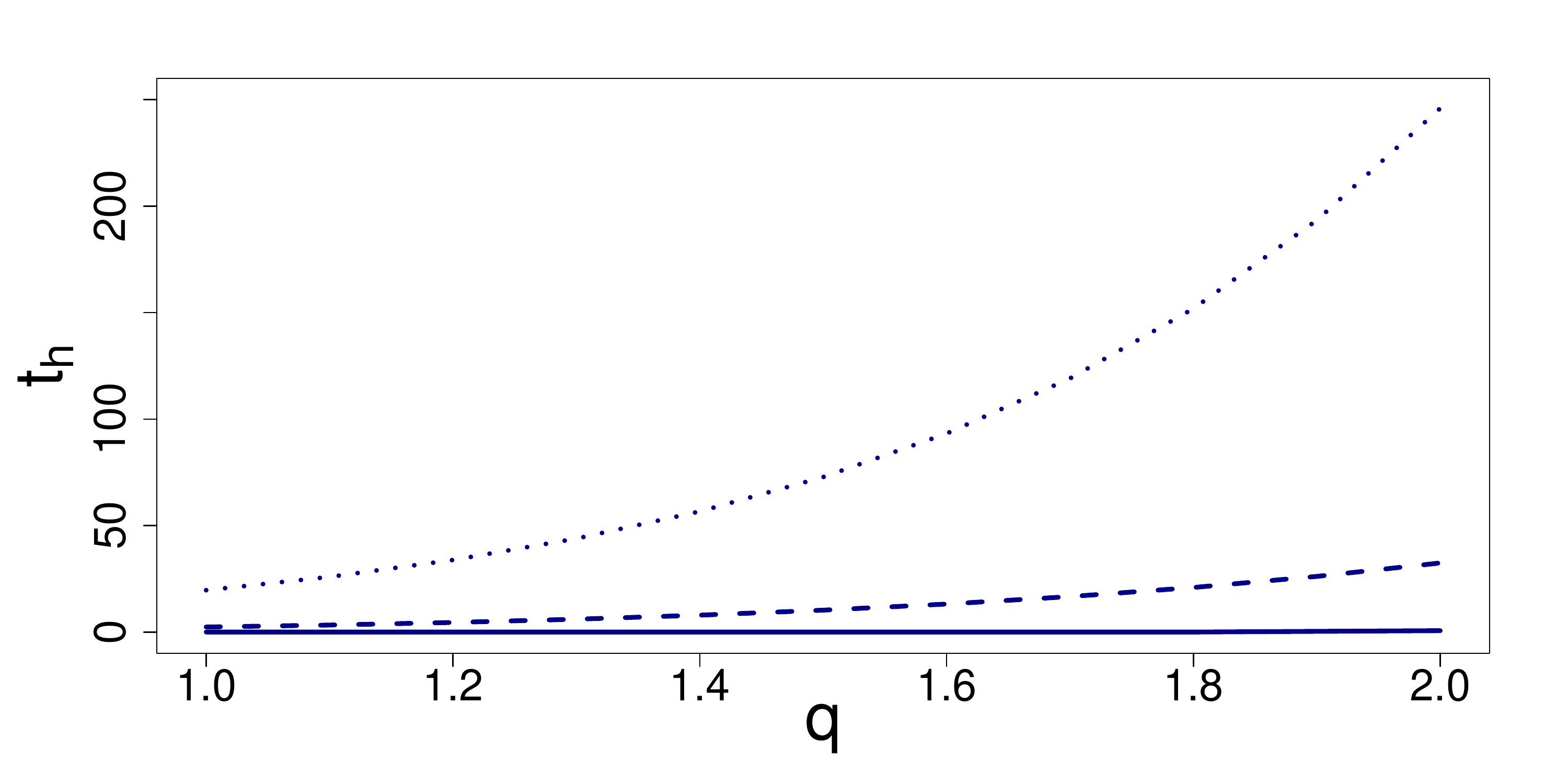} &
\includegraphics[scale=0.25]{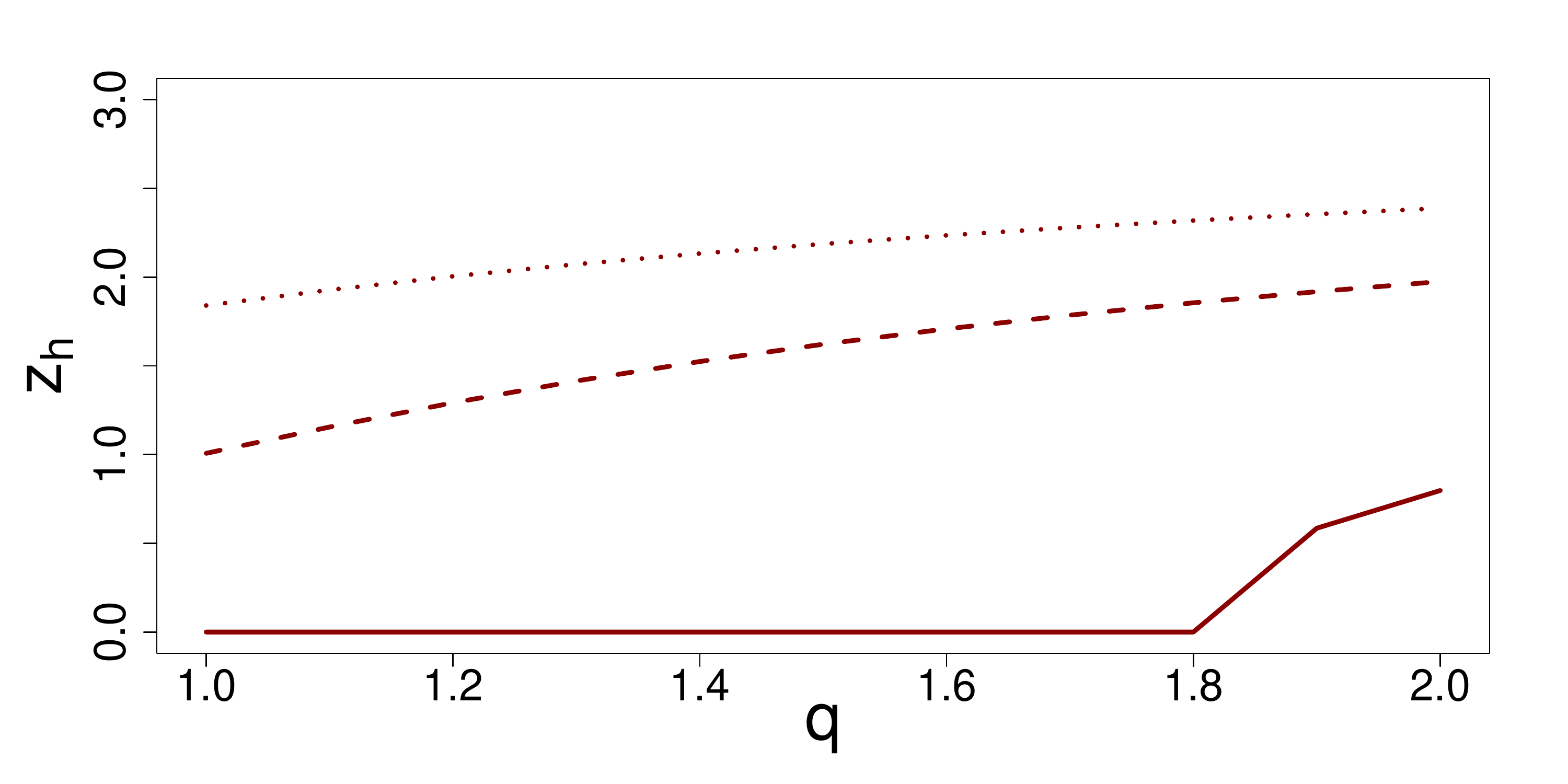} \\
\multicolumn{2}{c}{\underline{Beta(5,5)}}   \\
\includegraphics[scale=0.25]{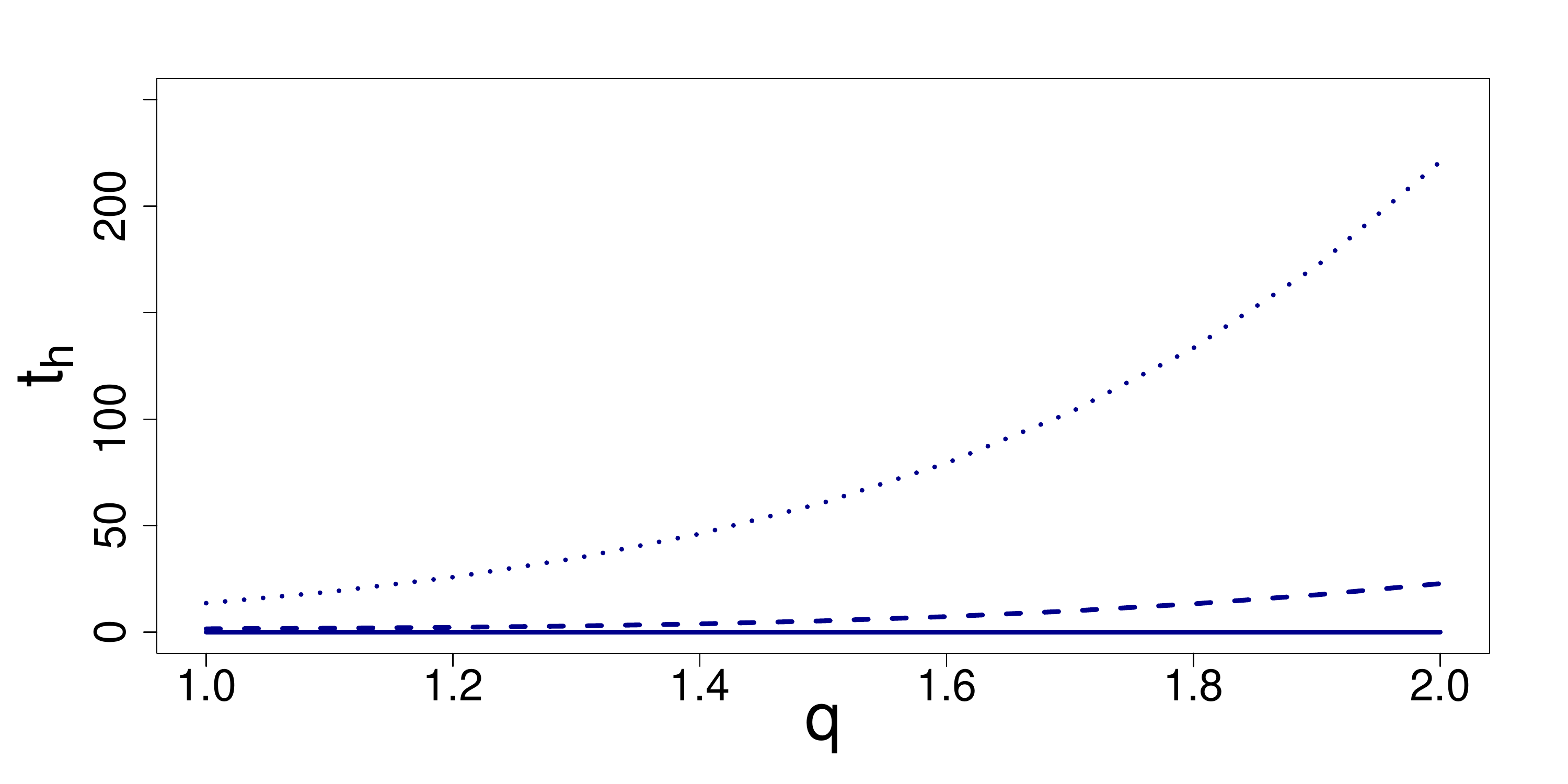} &
\includegraphics[scale=0.25]{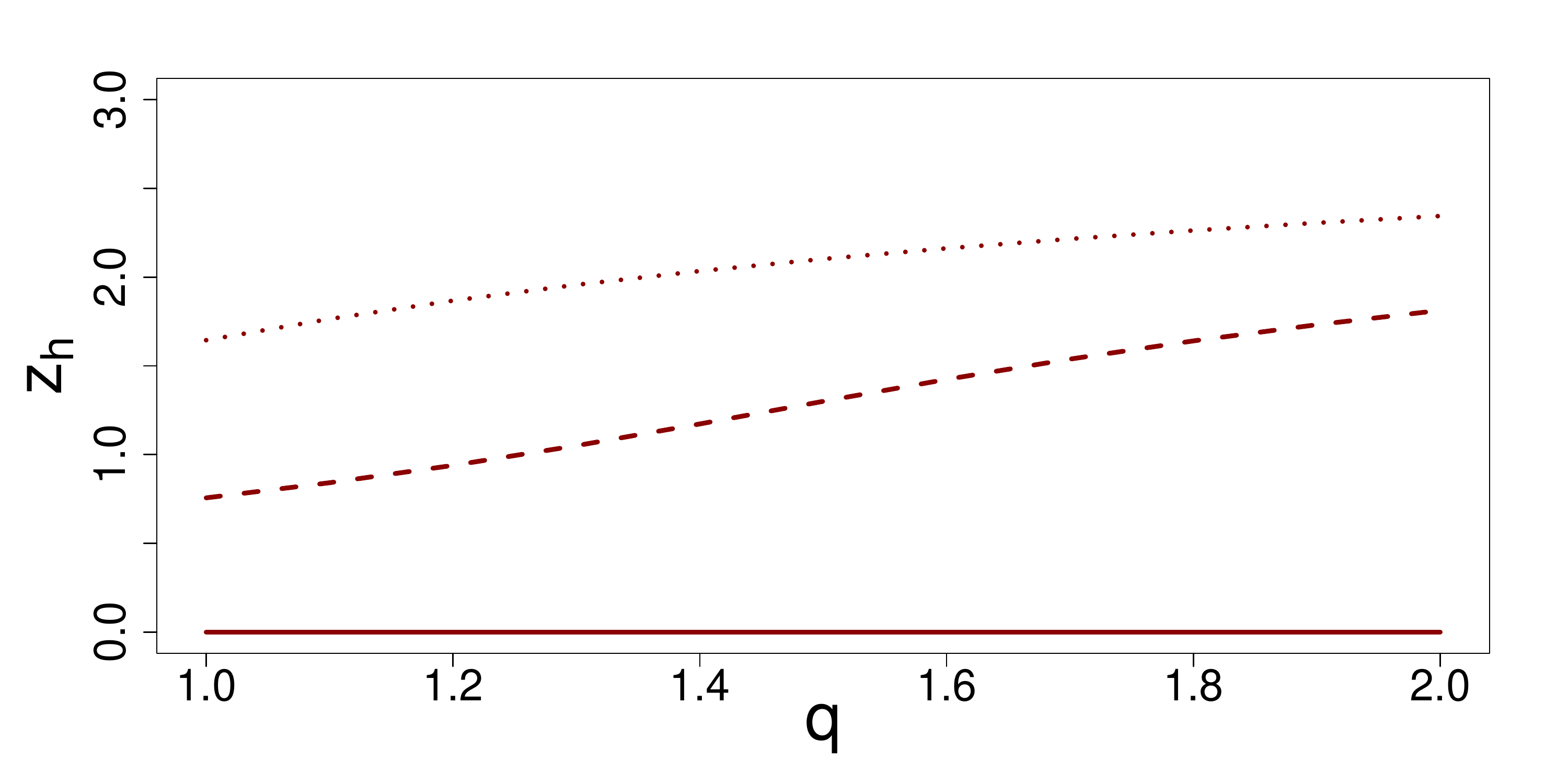} \\
\multicolumn{2}{c}{\underline{Beta(5,3)}}   \\
\includegraphics[scale=0.25]{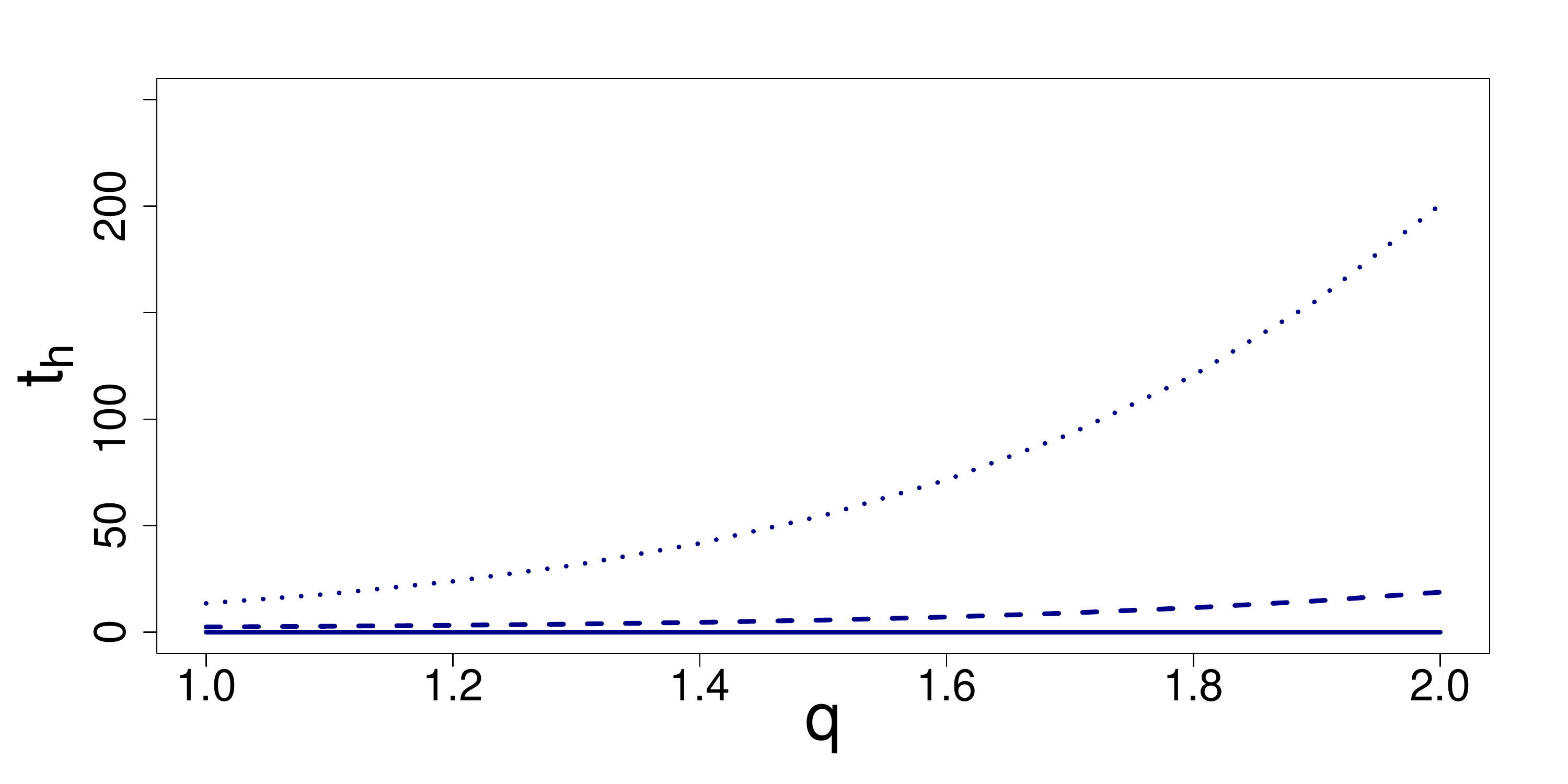} &
\includegraphics[scale=0.25]{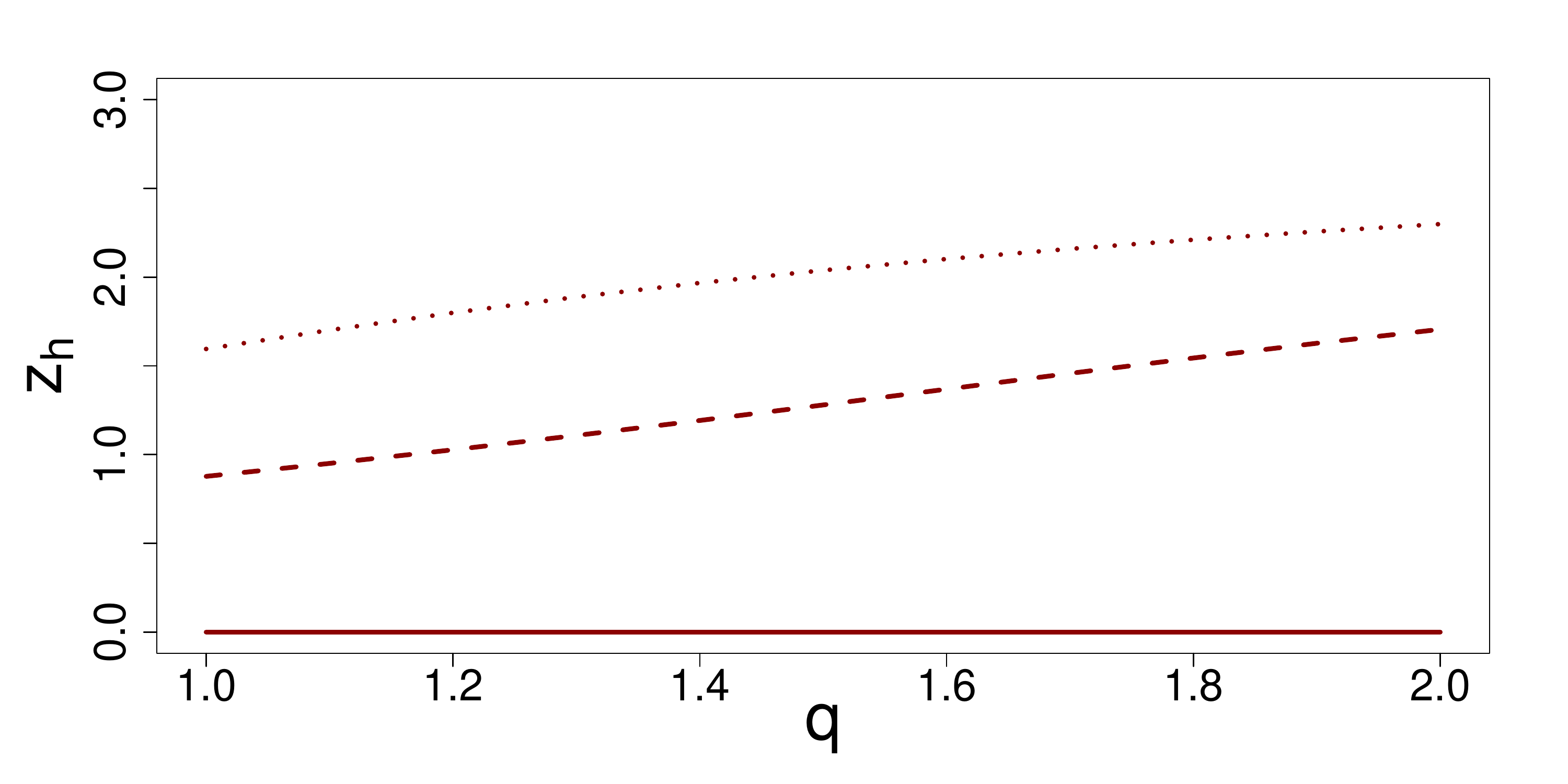} \\
\end{tabular}
    \begin{tablenotes}
      \footnotesize
      \item Notes. This design varies both the signal contribution parameter $a$ and the distribution of the receiver type by $q$. Recall the matching function $x = k z^q$ and the match surplus function $v(x,s,z)=As^axz$. We keep the same distribution on senders' type. The lower bound of an interval delegation is always zero ($t_l = 0$). The lower bound of a well-behaved equilibrium is also zero ($z_l=0$). Graphs with $a=0.9$ are relegated to the appendix because of the scale difference.
    \end{tablenotes}
\end{threeparttable}
\end{figure}

Other parameters in the baseline design are set as follows: $A=1,$ $\beta
=0.5,$ $a=0.5,$ $k=1$ and $q=1.$ Note that, given $k=q=1,$ we have that $%
n(z)=x$ in (\ref{matching_function}). In each design, we depart from the
baseline design by changing some parameter values above. In Design 1,
senders and receivers have the same type distribution (i.e.\ a symmetric
structure) by keeping $k=1$ and $q=1$. However, we change the support of
type distributions. Specifically, we vary the upper bound of the support
between 1 and 3. Recall that the upper bound of the baseline design is 3.
Then, we check the optimal reaction interval and the corresponding stronger
monotone equilibrium. In Design 2, we set the sender type distribution back
to the baseline design, i.e.\ $z\in \lbrack 0,3]$ and change the scale
parameter $k$. Design 3 varies only the relative spacing parameter $q$ while
holding all other conditions of the baseline design. Finally, Design 4
varies the signal contribution parameter $a$ as well as the relative spacing
parameter $q$.

Due to Theorem \ref{thm_optimal_design}, we can first derive the optimal $%
z_{\ell }$ and $z_{h}$ given the type distributions and other parameter
values. This determines the type of the optimal stronger monotone CSE. We
can then retrieve the corresponding $t_{\ell }$ and $t_{h},$ the lower and
upper bounds of the optimal delegation interval. In all cases, $z_{\ell }$
turns out zero so that it is optimal not to exclude anyone in the market and
hence $t_{\ell }=0.$ Therefore, we report only the optimal solution paths
for $t_{h}$ and $z_{h}$ in Figures \ref{fg:D1}--\ref{fg:D4}. We also report
the solution paths for other variables ($x_{h}$ and $s_{h}$) as well as the
support bounds ($\bar{x}$ and $\bar{z}$) in Tables \ref{tb:D1_Beta(1,1)}--%
\ref{tb:D4_Beta(5,3)} that are relegated to the appendix.

Design 1 restricts our attention to the symmetric structure between the
sender type and the receiver type, changing only the upper bound of the type
distribution support ($\bar{z}=\overline{x}$). Regardless of the value of $%
\bar{z},$ $z_{h}$ is less than $\overline{z}$ and greater than $z_{\ell }=0.$
Therefore, the optimal stronger monotone CSE is \emph{strictly well behaved}
so that it includes a pooling part on the top. This suggests that the
savings in the cost associated with the pooled action chosen by sender types
above $z_{h}$ outweighs the inefficiency associated with random matching in
the pooling part. As we increase the type distribution support from $[0,1]$
to $[0,3]$, Figure \ref{fg:D1} shows that both $t_{h}$ and $z_{h}$ increase
monotonically. Therefore, as the upper bound of the type distribution
support increases, the planner gives more discretion to receivers with a
higher upper bound of the delegation interval, increasing the threshold
sender type for pooling. Note that $z_{h}$ has a linear relationship with $%
\bar{z}$ while $t_{h}$ increases exponentially In fact, the percentile of
the threshold sender type $z_{h}$ that starts choosing the pooled action
does not change even though $z_{h}$ itself increases as the upper bound of
type distribution changes.\footnote{%
The percentiles of $z_{h}$ are 58.5\% in Beta (1,1), 83.4\% in Beta (3,5),
36.9\% in Beta (5,5), and 15.1\% in Beta (5,1).} These results hold across
different Beta distributions.%

Design 2 departs from the symmetric design by changing the scale parameter $%
k\in \lbrack 1,3]$ in $n(z)=kz^{q}$. All other parameters are set to the
baseline values. As $k$ increases from 1, it increases every receiver type $%
x $ by the \emph{same scale}.\footnote{%
Suppose that $x$ is the $y$ $\%$ percentile of the receiver type when $k=1.$
Then, the $y$ $\%$ precentile of the receiver type is simply $kx$ when $k>1$
for all $y\in \lbrack 0,100].$} Regardless of the value of $k$, the optimal
stronger monotone CSE is again strictly well behaved. Note that $z_{h}$ does
not vary over different $k$ values. Because the sender type distribution is
fixed, this implies that the percentile of $z_{h}$ does not change (See the
second column in Figure \ref{fg:D2}). This makes sense because $k$ has only
the scaling effect on the receiver type. However, $t_{h}$ increases in $k$
as the threshold receiver type $\bar{x}$ for pooling increases in $k$ (See
the first column in Figure \ref{fg:D2}). Therefore, we confirm that the
planner gives more discretion to receivers when the support of firms gets
larger though the scale parameter.

In Design 3, we fix $k=1$ and increase the value of $q$ from 1. Because the
initial support of the receiver type is $[0,3]$, the whole distribution of
receiver types changes in a way that the receiver type in $[0,1]$ decreases
and the receiver type in $[1,3]$ increases by the \emph{same factor} as $q$
increases. This change results in an increase in $z_{h}$. Since the sender
type distribution is fixed in each case, it implies that the percentile of $%
z_{h}$ increases as $q$ increases. This seems to suggest that the spreading
effect on the receiver type in $[1,3]$ of the initial distribution dominates
the shrinking effect on the receiver type in $[0,1]$. Subsequently, the
planner gives more discretion to receivers by increasing $t_{h}$.

A change in $k$ or a change in $q$ each results in an increase in $t_{h}.$
However, the percentile of $z_{h}$ remains the same with a change in $k$,
whereas it increases in $q$. Therefore, an increase in $t_{h}$ should be
sufficiently high to satisfy the utility indifference conditions, (\ref%
{jumping_sellers}) and (\ref{jumping_buyers}), at a higher percentile of the
threshold type for pooling when $q$ increases. Figure \ref{fg:D3} shows that
both $t_{h}$ and $z_{h}$ increase monotonically as $q$ increases. However,
the solution paths of $t_{h}$ become \emph{more convex} since the nonlinear
transformation moves more weights on higher types.

Designs 2 and 3 also reflect the empirical findings in Poschke (2018) such
that the mean and variance of the firm size distribution ($x$) are larger in
rich countries and have increased over time for U.S. firms, where the firm
size is measured by the number of workers employed. In fact, an increase in $%
k$ or $q$ each results in increases in both the mean and variance of the
receiver type distribution. The results in Designs 2 and 3 show that as the
mean and variance of the firm size distribution increase, the planner gives
more discretion to receivers. However, when increases in the mean and
variance of the receiver type distribution come from an increase in $q$
rather than $k$, the planner gives a lot more discretion to receivers due to
more convex solution paths of $t_{h}$ with respect to $q$.%

The first three designs provide comparative statics as the underlying type
distributions change given a positive value of the productivity paramter of
the sender's action ($a=0.5$). In Design 4, we change the value of $a$ along
with $q$. We vary $a\in \{0,0.3,0.6,0.9\}$, but we report the results up to $%
a\leq 0.6$ because of the scale difference. The whole results including $%
a=0.9$ can be found in the appendix. Figure \ref{fg:D4} reports the results.

Notably, when the sender's action has only a pure signaling effect (i.e., $%
a=0$) and $q$ is small, the optimal sender threshold type for pooling is $%
z_{h}=0$.\footnote{%
We present the results for $q\geq 1$ and $a\in \{0,0.3,0.6,0.9\}$ here and
in Appendix. Note that in Beta (1,1), the optimal stronger monotone CSE is
strictly well-behaved in these sets of values for $q$ and $a$. However, in
Beta (0,0), the optimal stronger monotone CSE is pooling (i.e., $%
z_{h}=t_{h}=0 $) when $q$ is close to zero and $a=0.$} This implies that the
optimal stronger monotone CSE is pooling with no separating part.
Subsequently, the optimal upper bound of reactions is zero ($t_{h}=0$).
Therefore, it is optimal for the planner to impose a single reaction ($%
t_{\ell }=t_{h}=0$) to all receivers instead of delegating reaction choices
to receivers.

For any given positive value of $a$ and any given $q,$ delegation is better
than imposing a single reaction because the optimal stronger monotone CSE is
strictly well-behaved. As $a$ increases, both $t_{h}$ and $z_{h}$ increase.
Because we fix the sender type distribution in each case of Design 4, an
increase in $z_{h}$ implies an increase in the percentile of the threshold
type for pooling in the optimal stronger monotone CSE.

\begin{table}[hp]
\centering
\caption{Design 5} \label{tb:fsd}
\begin{tabular}{rrrrrrrrr}
  \hline
$q$ & $k$ & Beta & $\bar{z}$ & $\bar{x}$ & $t_h$ & $z_h$ & $x_h$ & $s_h$ \\ 
  \hline
  1 & 1 & (3,5) & 3 & 3 & 10.211 & 1.620 & 1.620 & 3.367 \\ 
  1 & 1 & (5,5) & 3 & 3 & 6.030  & 1.338 & 1.338 & 2.686 \\ 
  1 & 1 & (5,3) & 3 & 3 & 6.940  & 1.344 & 1.344 & 3.089 \\ 
   \hline
\end{tabular}
\end{table}

\begin{figure}[ph]
\caption{Cumulative Distribution Functions}
\label{fg:beta-cdf}\centering
\par
{} \includegraphics[scale=0.25]{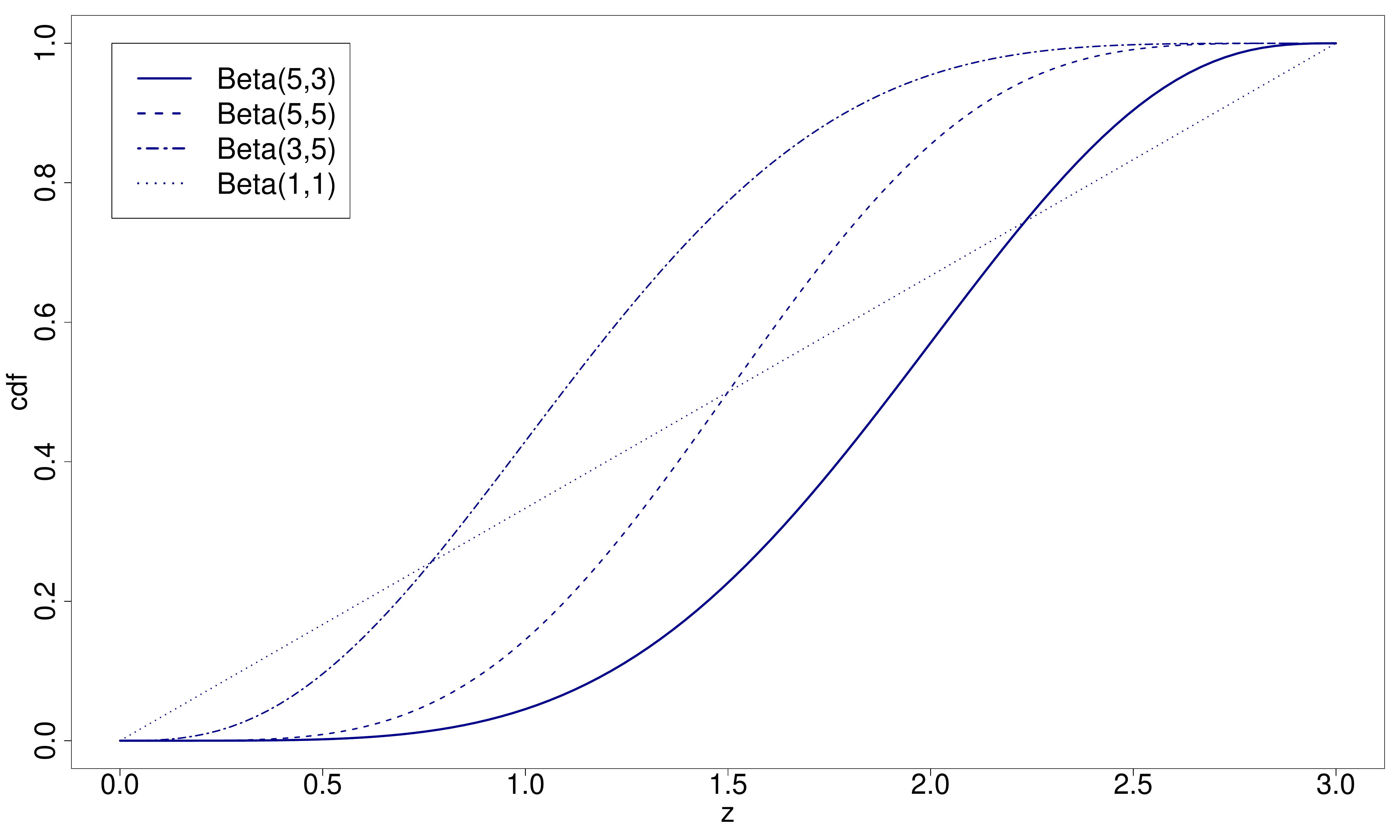}
\end{figure}

Finally, we remark that the optimal solution paths are quite nonlinear and
they do not have a monotone relationship with underlying parameters in
general. For example, we report the optimal delegation and other solutions
over the first order stochastic dominant distributions in Table \ref{tb:fsd}%
. Both sides of the market have the same Beta distribution and we change it
according to the first-order stochastic dominance given $A=k=q=1$ and $%
a=\beta =0.5$. As we can see from Figure \ref{fg:beta-cdf}, the Beta
distributions in the table satisfy the first order stochastic dominance
relationship: Beta (5,3) first order stochastically dominates Beta (5,5),
which first order stochastically dominates Beta (3,5). As the same type
distribution on both sides of the market moves to a first-order
stochastically dominant distribution, one may expect that $z_{h}$ and $t_{h}$
would increase because there are relatively more high types with a
first-order stochastically dominant distribution. However, that's not the
case according to Table \ref{tb:fsd}.\textbf{\ }The utility indifference
conditions, (\ref{jumping_sellers}) and (\ref{jumping_buyers}), play the
crucial role in determining the upper bound of the reaction interval, but
the whole type distributions affect the determination of the upper bound in
a non-trivial way through the indifference conditions. Therefore, one cannot
say that the planner gives more discretion to receivers as the type
distributions on both sides of the market move to first-order stochastically
dominant ones.

\section{Conclusion}

Our studies on optimal delegation provides a noble insight into the
principal's willingness to delegate and the determination of optimal
delegation in markets for matching with signaling. We first introduce a new
method that shows how to derive an optimal reaction interval for the full
implementation of the full implementation of a stronger monotone CSE, the
notion of equilibrium adopted from HSS (2023). Choosing an reaction interval
is equivalent to choosing the two threshold sender types because we can
retrieve the lower and upper bounds of the corresponding interval.
Furthermore, the two threshold sender types provide a sharp characterization
of a unique stronger monotone CSE.

The relative heterogeneity of receiver types and the productivity of the
sender' signal are crucial in deriving an optimal interval in the presence
of the trade-off between matching efficiency and signaling costs. Optimal
delegation studied in our paper assumes that there are no transfers between
the planner and receivers or between the planner and senders. Optimal
delegation can be studied with transfers. One prominent example is
delegation with taxation. It would be interesting but perhaps challenging to
examine how the presence of taxation alters the planner's incentives for
delegation and her design of optimal delegation.

\addcontentsline{toc}{section}{Appendices}

\renewcommand{\thesection}{\Alph{section}} \setcounter{section}{0} %
\setcounter{equation}{0} \renewcommand{\theequation}{A\arabic{equation}}

\clearpage

\begin{center}
{\LARGE \textbf{Appendix}}
\end{center}

\section{Proof of Lemma \protect\ref{lem_unified_well_behaved}}

It is clear that if $z_{h}\rightarrow \bar{z},$ $\left\{ \hat{\sigma},\hat{%
\mu},\hat{\tau},\hat{m}\right\} $ converges to the stronger monotone
separating equilibrium with the same lower threshold sender type $z_{\ell }$
as the pooling part vanishes. Note that as $z_{h}\rightarrow z_{\ell }$, $%
\lim_{z_{h}\rightarrow z_{\ell }}t_{h}=t_{\ell }$ and $\lim_{z_{h}%
\rightarrow z_{\ell }}s_{h}=\tilde{\sigma}(z_{\ell })=s_{\ell }$ because (%
\ref{jumping_sellers}) and (\ref{jumping_buyers}) are satisfied only when $%
(t_{h},s_{h})=(t_{\ell },\tilde{\sigma}(z_{\ell }))$. Therefore, combining (%
\ref{lem1}) and (\ref{lem2}) with (\ref{jumping_sellers}) and (\ref%
{jumping_buyers}) yields that
\begin{gather*}
\lim_{z_{h}\rightarrow z_{\ell }}\left[ t_{h}-c\left( s_{h},z_{h}\right) %
\right] =t_{\ell }-c\left( s_{\ell },z_{\ell }\right) =0, \\
\lim_{z_{h}\rightarrow z_{\ell }}\left[ \mathbb{E}[v(n\left( z_{h}\right)
,s_{h},z^{\prime })|z^{\prime }\geq z_{h}]-t_{h}\right] =\mathbb{E}%
[v(n\left( z_{\ell }\right) ,s_{\ell },z^{\prime })|z^{\prime }\geq z_{\ell
}]-t_{\ell }\geq 0,
\end{gather*}%
where the inequality holds with equality if $z_{\ell }>\underline{z}.$ The
second equality of the first line and the inequality of the second line are
a consequence of Theorem 6 in HSS (2023). These imply that $\left\{ \hat{%
\sigma},\hat{\mu},\hat{\tau},\hat{m}\right\} $ converges the stronger
monotone pooling equilibrium where $t_{\ell }$ is the single feasible
reaction, $z_{\ell }$ is the threshold sender type for market entry and $%
s_{\ell }$ is the pooled action for senders in the market. Note that when $%
z_{\ell }=\underline{z},$ we have $s_{\ell }=0$ and $t_{\ell }=0$ in $%
\left\{ \hat{\sigma},\hat{\mu},\hat{\tau},\hat{m}\right\} $ due to Theorem 6
in HSS (2023).

\section{\noindent Proof of Proposition \protect\ref{prop_unbounded_design}}

When $z_{\ell }=\underline{z}$, $s_{\ell }=\zeta (\underline{x},\underline{z}%
)=0$ and $t_{\ell }=c\left( \zeta (\underline{x},\underline{z}),\underline{z}%
\right) =0$. When $t_{\ell }=0,$ $z_{\ell }=\underline{z}$ and $s_{\ell
}=\zeta (\underline{x},\underline{z})=0$. Now consider the case with $%
z_{\ell }\in (\underline{z},\overline{z}).$ First consider case (i) in
Assumption 7 in HSS (2023). For any given $z_{\ell }\in (\underline{z},%
\overline{z}]$, (\ref{lem1}) and (\ref{lem2}) induce the equation:
\begin{equation}
v\left( n\left( z_{\ell }\right) ,s,z_{\ell }\right) -c\left( s,z_{\ell
}\right) =0.  \label{z_l_determination}
\end{equation}

If $s=0,$ then the left-hand side of (\ref{z_l_determination}) is positive.
As $s\rightarrow \infty ,$ the left-hand side approaches $-\infty $ because
of Assumption 4 in HSS (2023). Given Assumption 7.(i) in HSS (2023), $%
v\left( n\left( z_{\ell }\right) ,s,z_{\ell }\right) $ is positive and it is
independent of $s.$ Because $c$ is continuous in $s$ (Assumption 4 in HSS
(2023)), it means that there exists a unique $s_{\ell }$ satisfying ((\ref%
{z_l_determination}). Then, a unique $t_{\ell }$ is determined by either (%
\ref{lem1}) or (\ref{lem2}) given $s_{\ell }$ and $z_{\ell }$.

Now consider case (ii) in Assumption 7 in HSS (2023). For any given $z_{\ell
}\in (\underline{z},\overline{z})$, the left hand side of (\ref%
{z_l_determination}) is zero at $s=0.$ However, we cannot have $s_{\ell }=0.$
If $s_{\ell }=0,$ then $t_{\ell }$ must be zero. Then every seller's utility
is zero by entering the market. This implies that every sender will enter
the market so $z_{\ell }$ cannot be greater than $\underline{z}$ given our
assumption that everyone enters the market if she is indifferent between
entering the market and staying out of it.

Because of Assumptions 4 and 5 in HSS (2023), $v-c$ is strictly concave and
the left hand side of (\ref{z_l_determination}) approaches $-\infty $ as $%
s\rightarrow \infty $. Since the left hand side of (\ref{z_l_determination})
is zero at $s=0$, this implies that there exists a unique positive $s_{\ell
} $ satisfying (\ref{z_l_determination}). Then, a unique $t_{\ell }$ is
determined by either (\ref{lem1}) or (\ref{lem2}) given $s_{\ell }$ and $%
z_{\ell }$.

Suppose that the planner chooses $t_{\ell }$, a part of the unique solution $%
(t_{\ell },s_{\ell })$ that solves (\ref{lem1}) and (\ref{lem2}) given $%
z_{\ell }.$ Then, $(z_{\ell },s_{\ell })$ is a unique solution that solves (%
\ref{lem1}) and (\ref{lem2}) because of Lemma 7 in HSS (2023).

\section{\noindent Proof of Proposition \protect\ref{prop_bounded_design}}

First consider case (i) in Assumption 7 in HSS (2023). For any given $%
z_{h}\in \left( z_{\ell },\overline{z}\right) $, (\ref{jumping_sellers}) and
(\ref{jumping_buyers}) induces the equation:
\begin{equation}
c\left( s,z_{h}\right) -c\left( \tilde{\sigma}\left( z_{h}\right)
,z_{h}\right) =\mathbb{E}[v(n\left( z_{h}\right) ,s,z^{\prime })|z^{\prime
}\geq z_{h}]-v\left( n\left( z_{h}\right) ,\tilde{\sigma}\left( z_{h}\right)
,z_{h}\right) .  \label{jumping_both}
\end{equation}

The right hand side of (\ref{jumping_both}) is positive because it is
independent of $s$ and $\tilde{\sigma}\left( z_{h}\right) $ and $z^{\prime
}\geq z_{h}$ for $z_{h}<\overline{z}.$ The left-hand side is continuous and
increasing in $s$. Because of Assumptions 1.(i) and 4, the left hand side is
increasing in $s$ with $\lim_{s\searrow 0}\left[ c\left( s,z_{h}\right)
-c\left( \tilde{\sigma}\left( z_{h}\right) ,z_{h}\right) \right] <0$ and $%
\lim_{s\nearrow \infty }\left[ c\left( s,z_{h}\right) -c\left( \tilde{\sigma}%
\left( z_{h}\right) ,z_{h}\right) \right] =\infty .$ Therefore, we have a
unique solution for $s_{h}$ that solves (\ref{jumping_both}). Then, $t_{h}$
can be uniquely derived from either (\ref{jumping_sellers}) and (\ref%
{jumping_buyers}). Therefore, for any given $z_{h}\in \left( z_{\ell },%
\overline{z}\right) ,$ there exists a unique $(t_{h},s_{h})$ that satisfies (%
\ref{jumping_sellers}) and (\ref{jumping_buyers}).

Now consider case (ii) in Assumption 7 in HSS (2023). For any given $%
z_{h}\in \left( z_{\ell },\overline{z}\right) ,$ (\ref{jumping_sellers}) and
(\ref{jumping_buyers}) induce the equation:
\begin{equation}
\mathbb{E}[v(n\left( z_{h}\right) ,s,z^{\prime })|z^{\prime }\geq
z_{h}]-c\left( s,z_{h}\right) =v\left( n\left( z_{h}\right) ,\tilde{\sigma}%
\left( z_{h}\right) ,z_{h}\right) -c\left( \tilde{\sigma}\left( z_{h}\right)
,z_{h}\right) .  \label{jumping_both1}
\end{equation}%
The right hand side of (\ref{jumping_both1}) is positive because it is the
sum of the equilibrium utilities for the sender type $z_{h}$ and the
receiver type $n\left( z_{h}\right) $ for $z_{h}\in \left( z_{\ell },%
\overline{z}\right) $ in the stronger monotone separating equilibrium and
both equilibrium utilities for senders and receivers are increasing in types
in the separating equilibrium. Because of Assumption 4 ($c\left( 0,z\right)
=0$ for all $z$) and case (ii) in Assumption 7 ($v(x,0,z)=0$ for all $x$ and
$z$) in HSS (2023), we have that%
\begin{equation}
0=\mathbb{E}[v(n\left( z_{h}\right) ,0,z^{\prime })|z^{\prime }\geq
z_{h}]-c\left( 0,z_{h}\right) <v\left( n\left( z_{h}\right) ,\tilde{\sigma}%
\left( z_{h}\right) ,z_{h}\right) -c\left( \tilde{\sigma}\left( z_{h}\right)
,z_{h}\right)  \label{jumping_both2}
\end{equation}%
Because of Assumption 5 ($\lim_{s\rightarrow \infty }v_{s}(x,s,z)=0$ given
any $x$ and $z$ and $\lim_{s\rightarrow \infty }$ $c_{s}(s,z)=\infty $ given
any $z$) in HSS (2023), we have that
\begin{equation}
-\infty =\lim_{s\rightarrow \infty }\left[ \mathbb{E}[v(n\left( z_{h}\right)
,s,z^{\prime })|z^{\prime }\geq z_{h}]-c\left( s,z_{h}\right) \right]
<v\left( n\left( z_{h}\right) ,\tilde{\sigma}\left( z_{h}\right)
,z_{h}\right) -c\left( \tilde{\sigma}\left( z_{h}\right) ,z_{h}\right)
\label{jumping_both3}
\end{equation}%
When $s=\tilde{\sigma}\left( z_{h}\right) $, we have that
\begin{equation}
\mathbb{E}[v(n\left( z_{h}\right) ,\tilde{\sigma}\left( z_{h}\right)
,z^{\prime })|z^{\prime }\geq z_{h}]-c\left( \tilde{\sigma}\left(
z_{h}\right) ,z_{h}\right) >v\left( n\left( z_{h}\right) ,\tilde{\sigma}%
\left( z_{h}\right) ,z_{h}\right) -c\left( \tilde{\sigma}\left( z_{h}\right)
,z_{h}\right) .  \label{jumping_both4}
\end{equation}%
because $\mathbb{E}[v(n\left( z_{h}\right) ,\tilde{\sigma}\left(
z_{h}\right) ,z^{\prime })|z^{\prime }\geq z_{h}]>v\left( n\left(
z_{h}\right) ,\tilde{\sigma}\left( z_{h}\right) ,z_{h}\right) .$

Because $\mathbb{E}[v(n\left( z_{h}\right) ,s,z^{\prime })|z^{\prime }\geq
z_{h}]-c\left( s,z_{h}\right) $ is strictly concave in $s$ due to Assumption
5 in HSS (2023), (\ref{jumping_both2}), (\ref{jumping_both3}), and (\ref%
{jumping_both4}) imply that there are two values, $s^{0}$ and $s^{1}$ that
satisfy (\ref{jumping_both1}) with $s^{0}<\tilde{\sigma}\left( z_{h}\right)
<s^{1}$. Because $\tilde{\sigma}\left( z_{h}\right) <s_{h}$ from Lemma 13 in
HSS (2023), $s_{h}$ is equal to $s^{1}$. Because $t_{h}$ can be uniquely
derived from either (\ref{jumping_sellers}) and (\ref{jumping_buyers})$,$
there exists a unique $(t_{h},s_{h})$ that satisfies (\ref{jumping_sellers})
and (\ref{jumping_buyers}) for any given $z_{h}\in \left( z_{\ell },%
\overline{z}\right) .$ Therefore, in both cases in Assumption 7 in HSS
(2023), there exists a unique $t_{h}$ and a unique $s_{h}$ that satisfies (%
\ref{jumping_sellers}) and (\ref{jumping_buyers}) for any given $z_{h}\in
\left( z_{\ell },\overline{z}\right) $. Suppose that the planner chooses $%
t_{h}$, a part of the unique solution $(t_{h},s_{h})$ that solves (\ref%
{jumping_sellers}) and (\ref{jumping_buyers}) given $z_{h}\in \left( z_{\ell
},\overline{z}\right) $. Then, $(z_{h},s_{h})$ is a unique solution that
solves (\ref{jumping_sellers}) and (\ref{jumping_buyers}) because of Lemma
12.

Because all functions in (\ref{jumping_sellers}) and (\ref{jumping_buyers})
are continuous (Assumptions 3.(ii), 4.(i), 6.(ii) and Theorem 4 in HSS
(2023)), it is clear the solution $(t_{h},s_{h})$ is continuous in $z_{h}$
and that $z_{h}$, the part of the solution ($z_{h},s_{h})$ is also
continuous in $t_{h}.$ Finally, $\lim_{z_{h}\rightarrow z_{\ell
}}t_{h}=t_{\ell }$ and $\lim_{z_{h}\rightarrow z_{\ell }}s_{h}=\tilde{\sigma}%
\left( z_{\ell }\right) $ because (\ref{jumping_sellers}) and (\ref%
{jumping_buyers}) are satisfied only when $(t_{h},s)=(t_{\ell },\tilde{\sigma%
}\left( z_{\ell }\right) )$ as $z_{h}\rightarrow \underline{z}$.

\section{Proof of Proposition \protect\ref{prop_diff_eq}}

Consider the initial value problem:
\begin{equation*}
\left\{
\begin{array}{cc}
v_{s}(n(\mu (s)),s,\mu (s))+v_{z}(n(\mu (s)),s,\mu (s))\mu ^{\prime
}(s)-c_{s}(s,\mu (s))=0 &  \\
\mu (s_{\ell })=z_{\ell } &
\end{array}%
\right. \text{ }.
\end{equation*}

Given $v(x,s,z)=As^{a}xz$,$~$ $c(s,z)=\beta \dfrac{s^{2}}{z}$, $~$and $%
n(z)=kz^{q}$, we have that $v_{z}(x,s,z)=As^{a}x$,$~$ $%
v_{s}(x,s,z)=aAs^{a-1}xz$,$~$and $c_{s}(s,z)=2\beta \dfrac{s}{z}$.
Therefore, the above IVP becomes
\begin{equation}
aAs^{a-1}k\mu (s)^{q}\mu (s)+As^{a}k\mu (s)^{q}\mu ^{\prime }(s)-\dfrac{%
2\beta s}{\mu (s)}=0.  \label{eq:1}
\end{equation}%
Rewriting (\ref{eq:1}) gives

\begin{equation}
\dfrac{Ak}{2\beta }\mu (s)^{1+q}\mu ^{\prime }(s)+\dfrac{aAk}{2\beta s}\mu
(s)^{2+q}=s^{1-a}.  \label{eq:4}
\end{equation}%
Let $D=:\dfrac{Ak}{2\beta }$. Thus, (\ref{eq:4}) becomes

\begin{equation}
D\mu ^{1+q}\mu ^{\prime }+\dfrac{aD}{s}\mu ^{2+q}=s^{1-a}  \label{eq:5}
\end{equation}%
where we denote $\mu =:\mu (s)$ for simplicity. Let $v=\mu ^{2+q}$. Then $%
v^{\prime }=(2+q)\mu ^{1+q}\mu ^{\prime }$ and (\ref{eq:5}) becomes%
\begin{equation}
v^{\prime }+\dfrac{a(2+q)}{s}v=\left( \dfrac{2+q}{D}\right) s^{1-a}
\label{eq:6}
\end{equation}%
which is a first order linear differential equation with integrating factor $%
\mathcal{I}(s)=s^{a(2+q)}$. Therefore, (\ref{eq:6}) is equivalent to $\frac{d%
}{ds}\left\{ vs^{a(2+q)}\right\} =\left( \dfrac{2+q}{D}\right) s^{1+a+aq},$
which implies
\begin{equation}
vs^{a(2+q)}=\left( \dfrac{2+q}{D}\right) \int s^{1+a+aq}ds+\kappa ,
\label{eq:7}
\end{equation}%
where $\kappa $ is some integration constant. Equation (\ref{eq:7}) implies%
\begin{equation}
\mu (s)^{2+q}=\left( \dfrac{2+q}{D}\right) \left( \dfrac{s^{2-a}}{2+a+aq}%
\right) +\dfrac{\kappa }{s^{a(2+q)}}.  \label{eq:9}
\end{equation}%
given that $v=\mu ^{2+q}$. By using the initial condition $\mu (s_{\ell
})=z_{\ell }$, we compute $\kappa $ as follows:
\begin{equation}
z_{\ell }^{2+q}=\left( \dfrac{2+q}{D}\right) \left( \dfrac{s_{\ell }^{2-a}}{%
2+a+aq}\right) +\dfrac{\kappa }{s_{\ell }^{a(2+q)}}  \label{eq:10}
\end{equation}%
which gives
\begin{equation}
\kappa =\dfrac{s_{\ell }^{a(2+q)}\left[ D(2+a+aq)z_{\ell
}^{2+q}-(2+q)s_{\ell }^{(2-a)}\right] }{D(2+a+aq)}.  \label{eq:12}
\end{equation}%
Plugging (\ref{eq:12}) into (\ref{eq:9}) gives%
\begin{equation}
\tilde{\mu}(s)=\left[ \left( \dfrac{2\beta (2+q)}{Ak}\right) \dfrac{s^{2-a}}{%
2+a+aq}+\left( \dfrac{s_{\ell }}{s}\right) ^{a(2+q)}\dfrac{\left[
Ak(2+a+aq)z_{\ell }^{2+q}-2\beta (2+q)s_{\ell }^{(2-a)}\right] }{Ak(2+a+aq)}%
\right] ^{\dfrac{1}{2+q}},  \label{eq:14}
\end{equation}%
where $s_{\ell }$ is determined by (\ref{s_l_determination}) and it is $%
s_{\ell }(z_{\ell })=\left( \frac{Ak}{\beta }z_{\ell }^{q+2}\right) ^{\frac{1%
}{2-a}}.$

\section{Proof of Proposition \protect\ref{thm_eq_design}}

First, we construct a (unique) stronger monotone pooling equilibrium with a
positive single feasible reaction $t^{\ast }>0$. Because (\ref%
{pooling_sender1}) and (\ref{pooling_receiver1}) must hold with equality by
Theorem 6 in HSS (2023) with $z^{\ast }>\underline{z}$, the pooled action $%
s^{\ast }(z^{\ast },q,a)$ solves $As^{\ast }{}^{a}kz^{\ast }{}^{q}\mathbb{E}%
\left[ z|z\geq z^{\ast }\right] -\beta \frac{s^{\ast }{}^{2}}{z^{\ast }}=0$
and hence we have that $s^{\ast }(z^{\ast },q,a)=\left( \frac{z^{\ast q+1}Ak%
\mathbb{E}\left[ z|z\geq z^{\ast }\right] }{\beta }\right) ^{\frac{1}{2-a}}$%
. Therefore, we have that $\lim_{q,a\rightarrow 0}s^{\ast }(t,q,a)=\sqrt{%
z^{\ast }Ak\mathbb{E}\left[ z|z\geq z^{\ast }\right] /\beta }.$

The aggregate net surplus in the stronger monotone pooling equilibrium is
\begin{equation*}
\Pi _{p}(z^{\ast },q,a,G):=\int_{z^{\ast }}^{\bar{z}}As^{\ast }(z^{\ast
},q,a)^{a}kz^{\ast q}\mathbb{E}\left[ z|z\geq z^{\ast }\right] dG(z)-\beta
s^{\ast }(z^{\ast },q,a)^{2}\int_{z^{\ast }}^{\bar{z}}\frac{1}{z}dG(z).
\end{equation*}%
This implies that
\begin{gather*}
\lim_{q,a\rightarrow 0}\Pi _{p}(z^{\ast },q,a,G)=\int_{z^{\ast }}^{\bar{z}}Ak%
\mathbb{E}\left[ z|z\geq z^{\ast }\right] dG(z)-z^{\ast }Ak\mathbb{E}\left[
z|z\geq z^{\ast }\right] \int_{z^{\ast }}^{\bar{z}}\frac{1}{z}dG(z), \\
\lim_{z^{\ast }\rightarrow 0}\left[ \lim_{q,a\rightarrow 0}\Pi _{p}(z^{\ast
},q,a,G)\right] =\int_{0}^{\bar{z}}Ak\mu _{z}dG(z)=Ak\mu _{z}
\end{gather*}%
where $\mu _{z}$ is the unconditional mean of the sender type $z.$ Because $%
\lim_{q,a\rightarrow 0}\Pi ^{\ast }(q,a,G)=\frac{Ak\mu _{z}}{2}$, we have
that
\begin{equation}
\lim_{z^{\ast }\rightarrow 0}\left[ \lim_{q,a\rightarrow 0}\Pi _{p}(z^{\ast
},q,a,G)\right] -\lim_{q,a\rightarrow 0}\Pi _{\ast }(q,a,G)=\frac{Ak\mu _{z}%
}{2}>0.  \label{belief_design6}
\end{equation}

Because $\Pi _{p}(z^{\ast },q,a,G)$ and $\Pi _{\ast }(q,a,G)$ are
continuous, there exists $\tilde{q}>0$, and $\tilde{a}>0$ and $z^{\ast }(%
\tilde{q},\tilde{a})\in $ Int $Z$ such that for every $(q,a)\in \lbrack 0,%
\tilde{q}]\times \lbrack 0,\tilde{a}]$ and every $z^{\ast }\in (0,z^{\ast }(%
\tilde{q},\tilde{a})]$, $\Pi _{p}(z^{\ast },q,a,G)>\Pi _{s}(q,a,G)$. We can
retrieve $t^{\ast }>0$ given $z^{\ast }\in (0,z^{\ast }(\tilde{q},\tilde{a}%
)] $.

\section{Additional Figures and Tables}

\subsection{Figures}

\begin{figure}[p]
\begin{threeparttable}
\caption{Design 4. Change of Signal Contribution ($a$) and Receiver Type ($k$)}\label{fg:D4_long}
\centering
\begin{tabular}{c c}
\multicolumn{2}{c}{\underline{Beta(1,1)}}   \\
\includegraphics[scale=0.25]{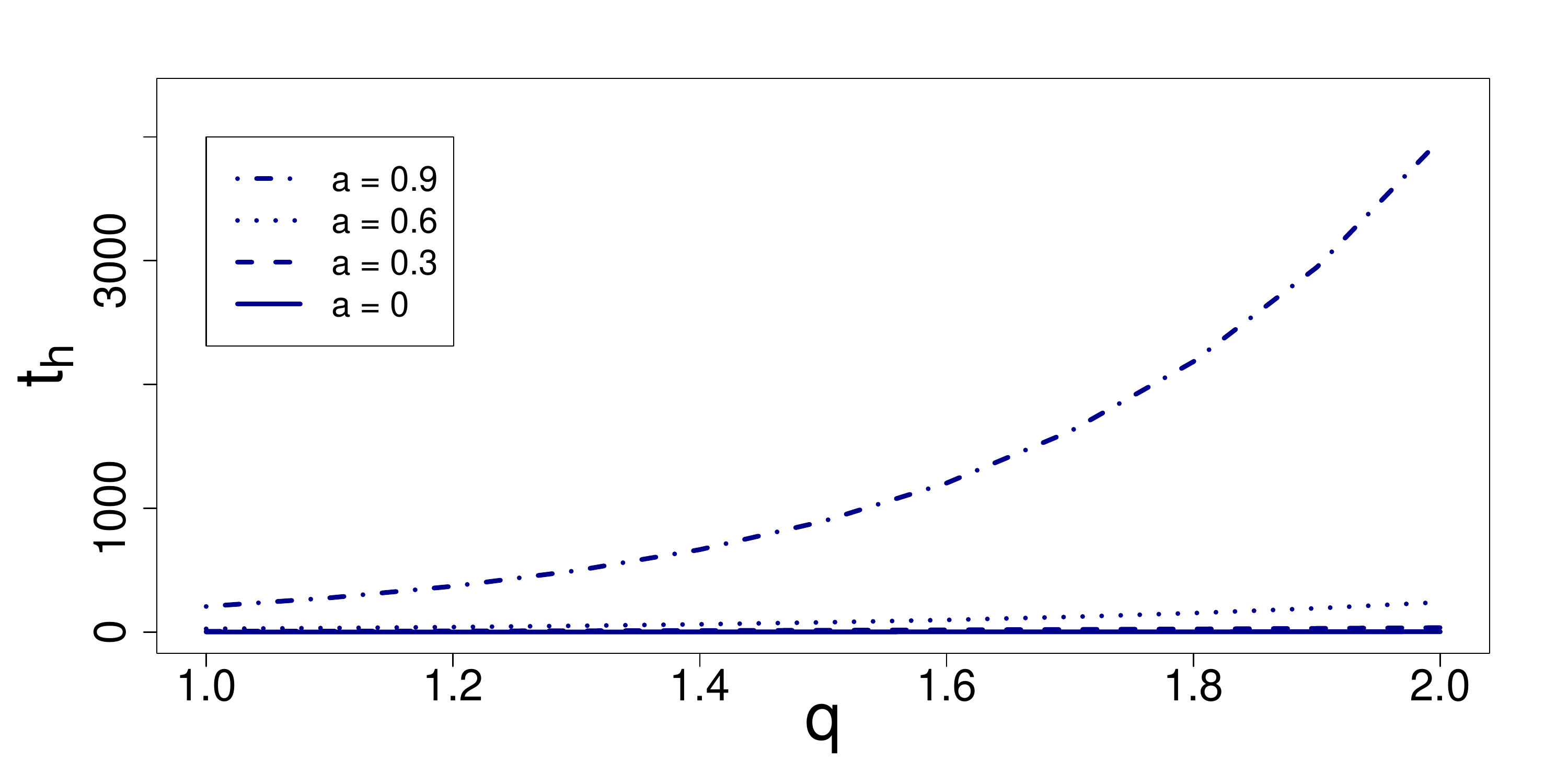} &
\includegraphics[scale=0.25]{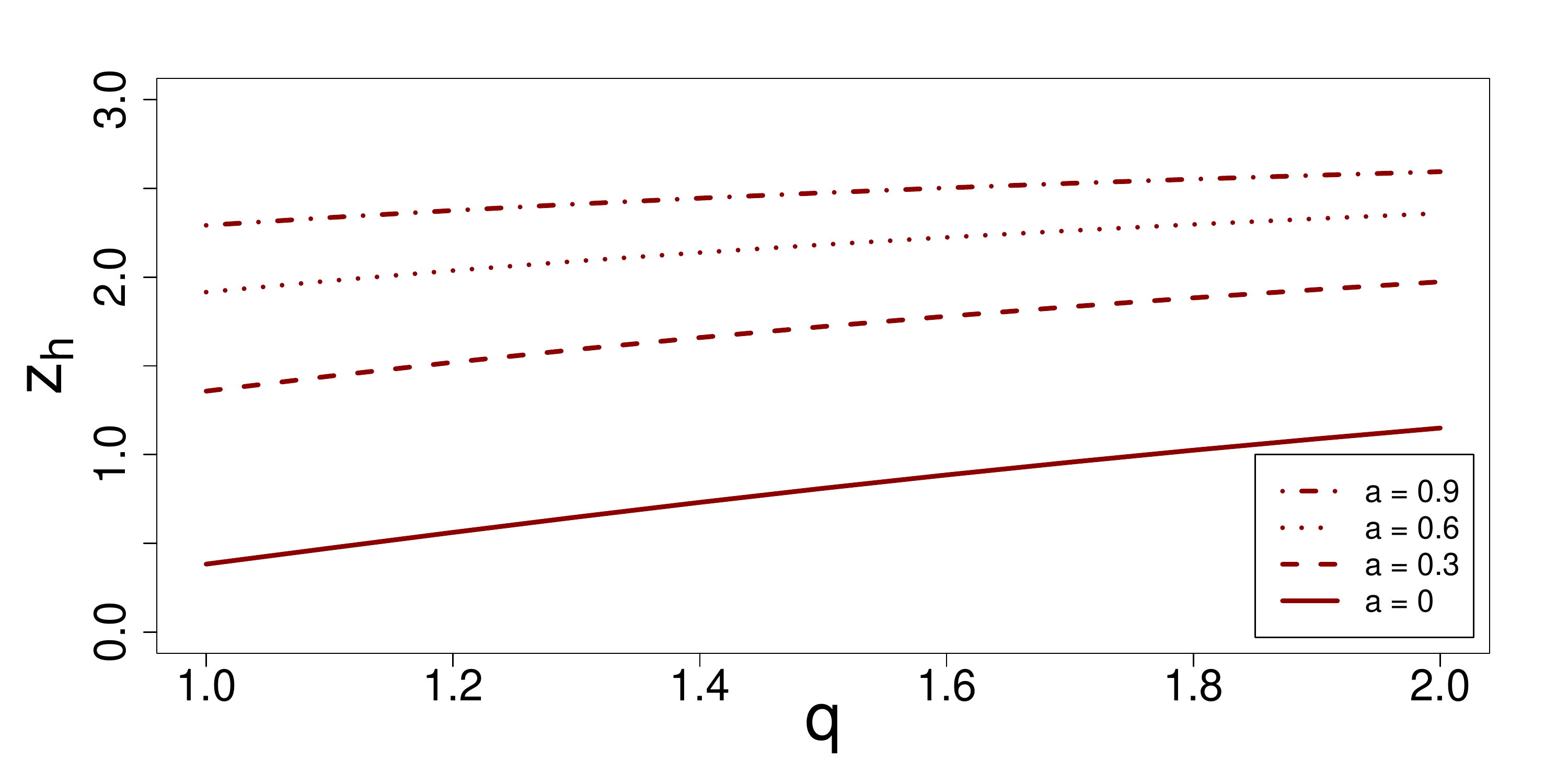} \\
\multicolumn{2}{c}{\underline{Beta(3,5)}}   \\
\includegraphics[scale=0.25]{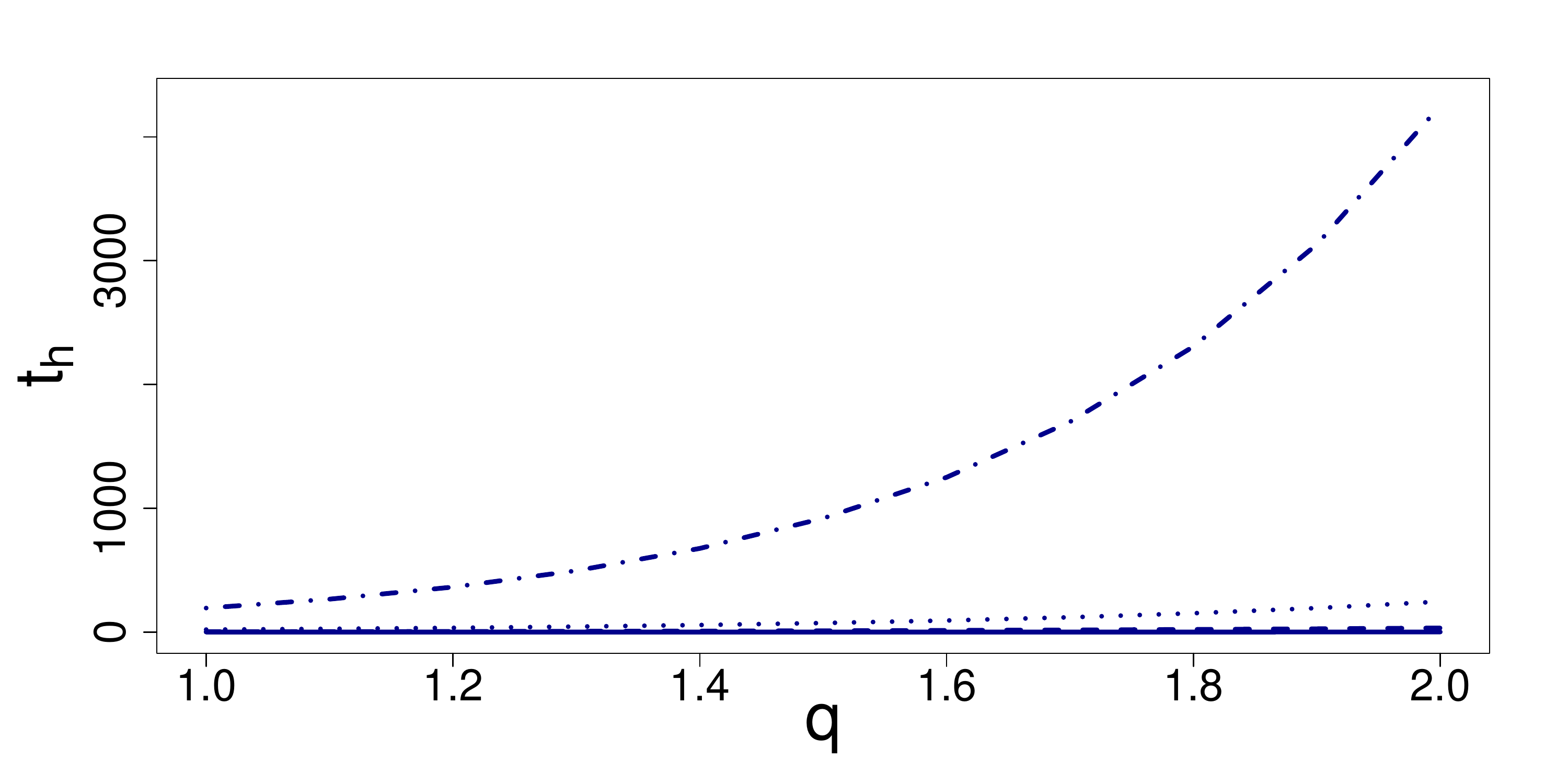} &
\includegraphics[scale=0.25]{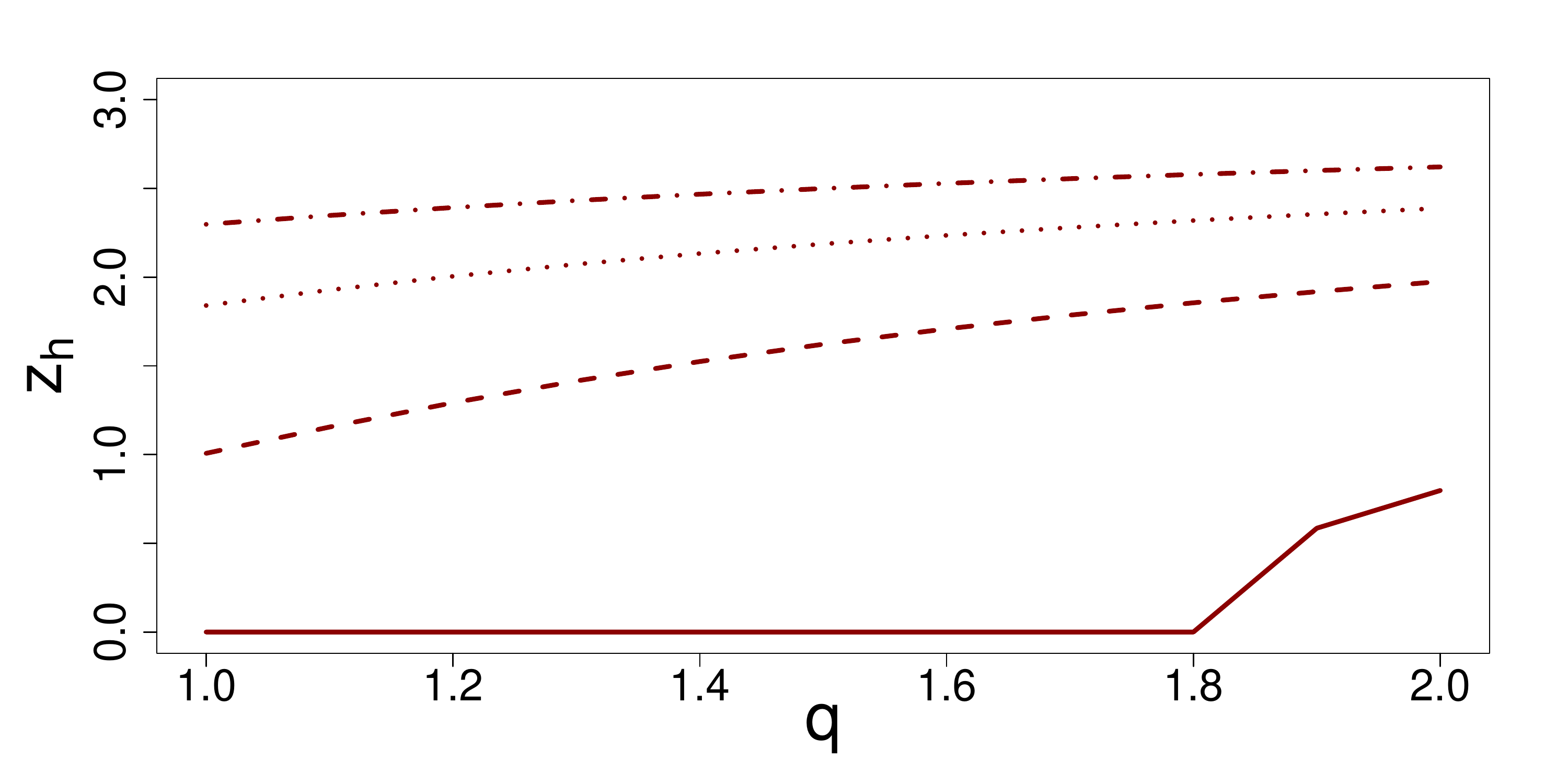} \\
\multicolumn{2}{c}{\underline{Beta(5,5)}}   \\
\includegraphics[scale=0.25]{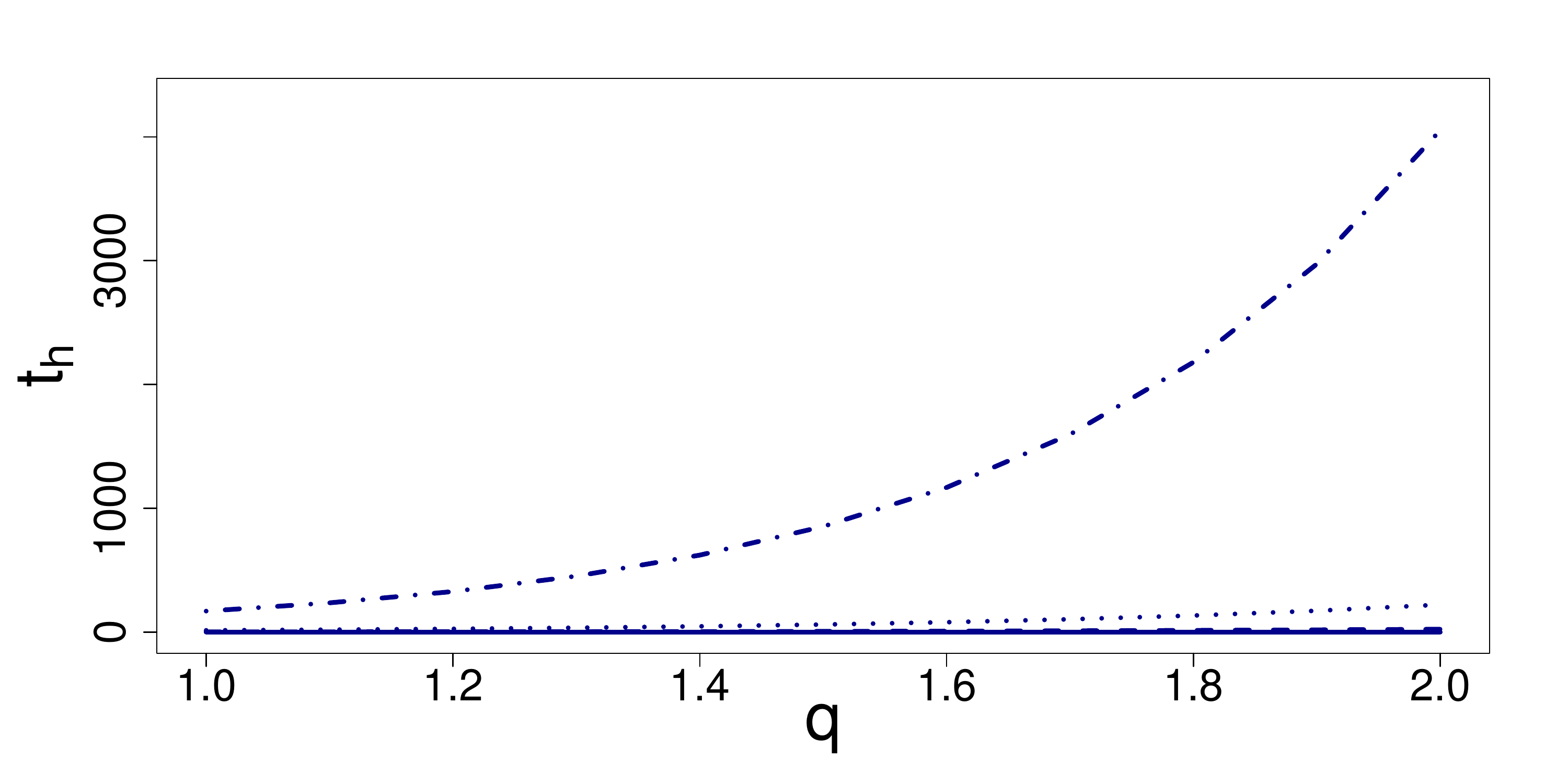} &
\includegraphics[scale=0.25]{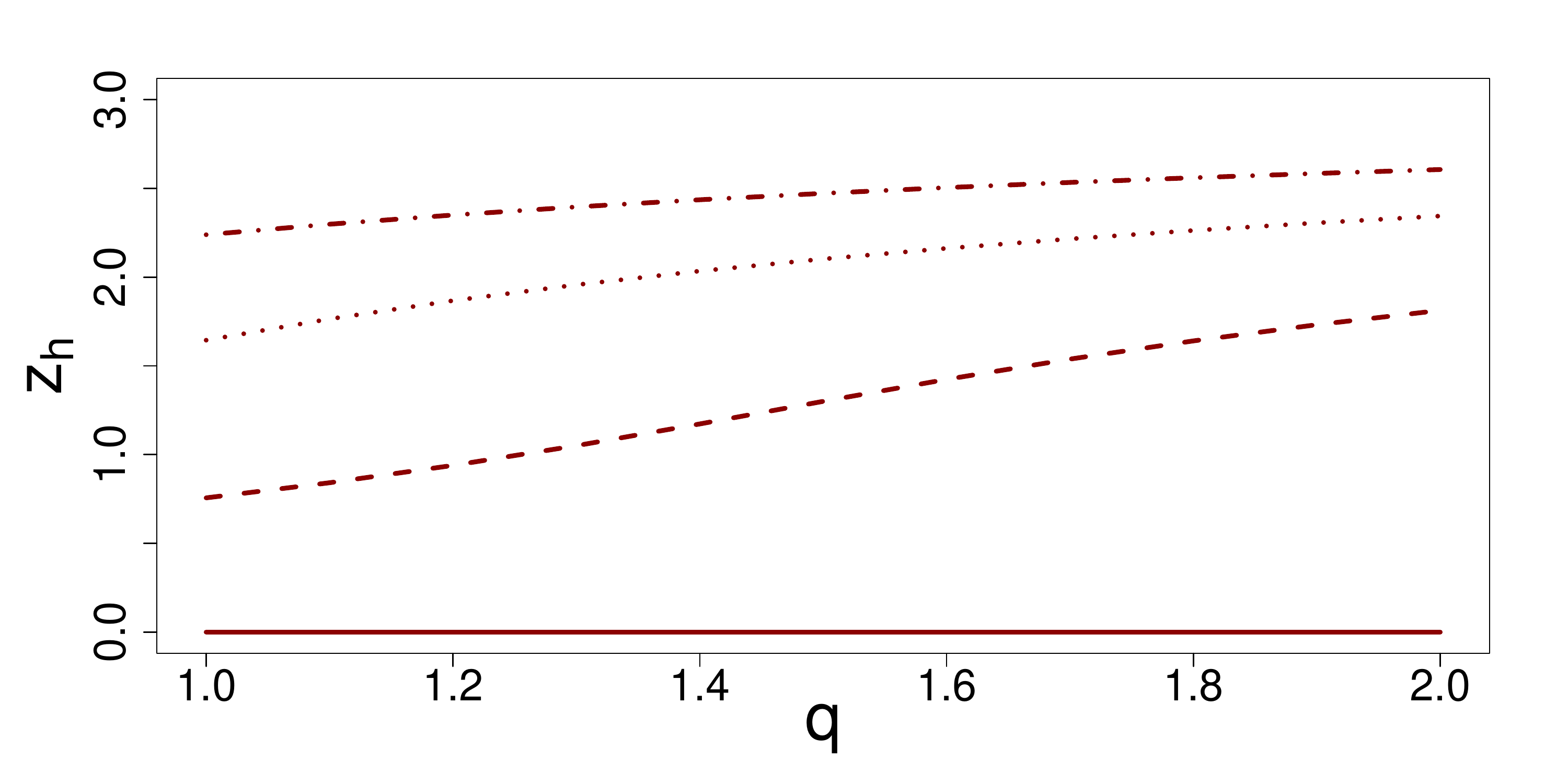} \\
\multicolumn{2}{c}{\underline{Beta(5,3)}}   \\
\includegraphics[scale=0.25]{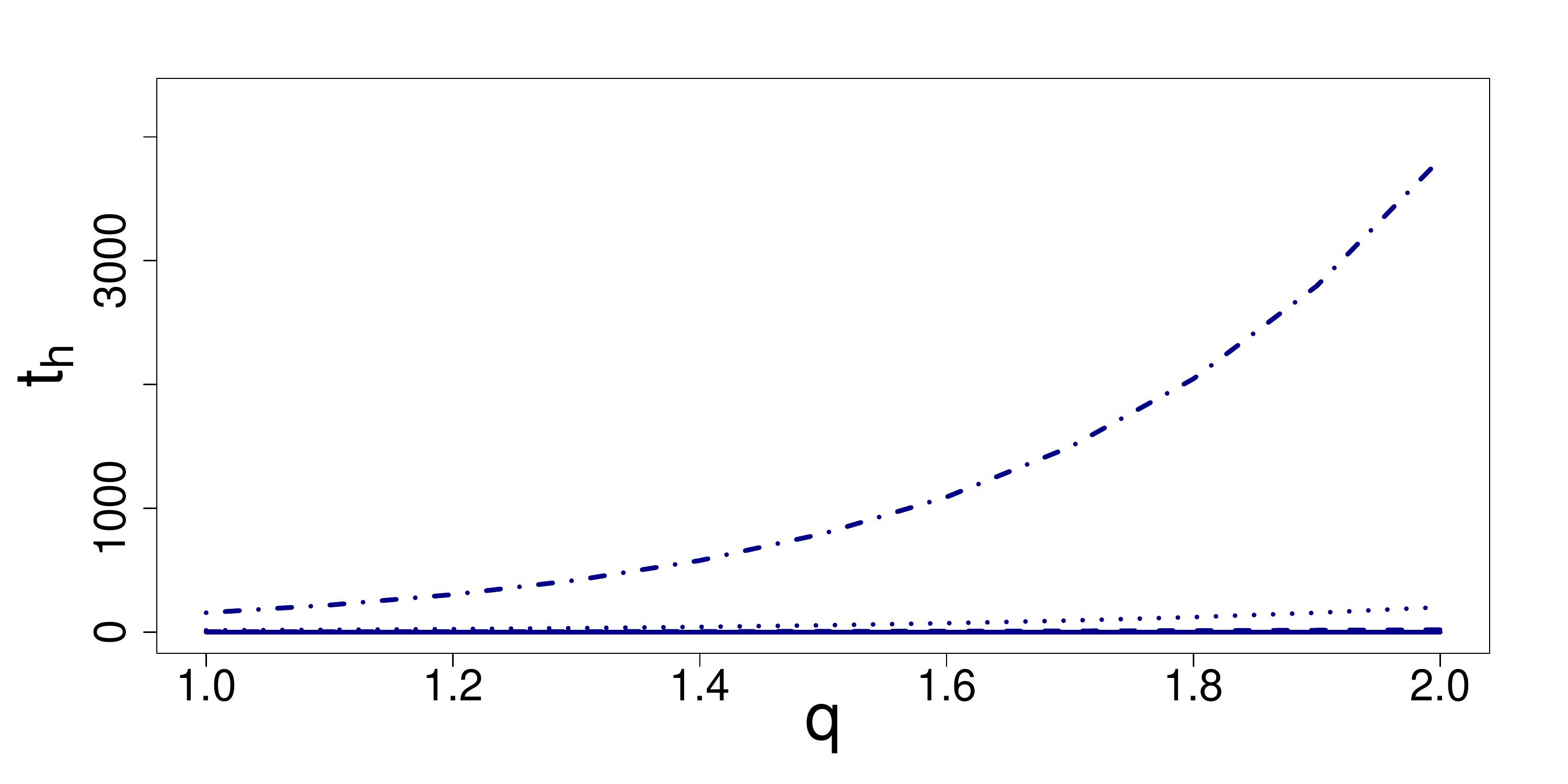} &
\includegraphics[scale=0.25]{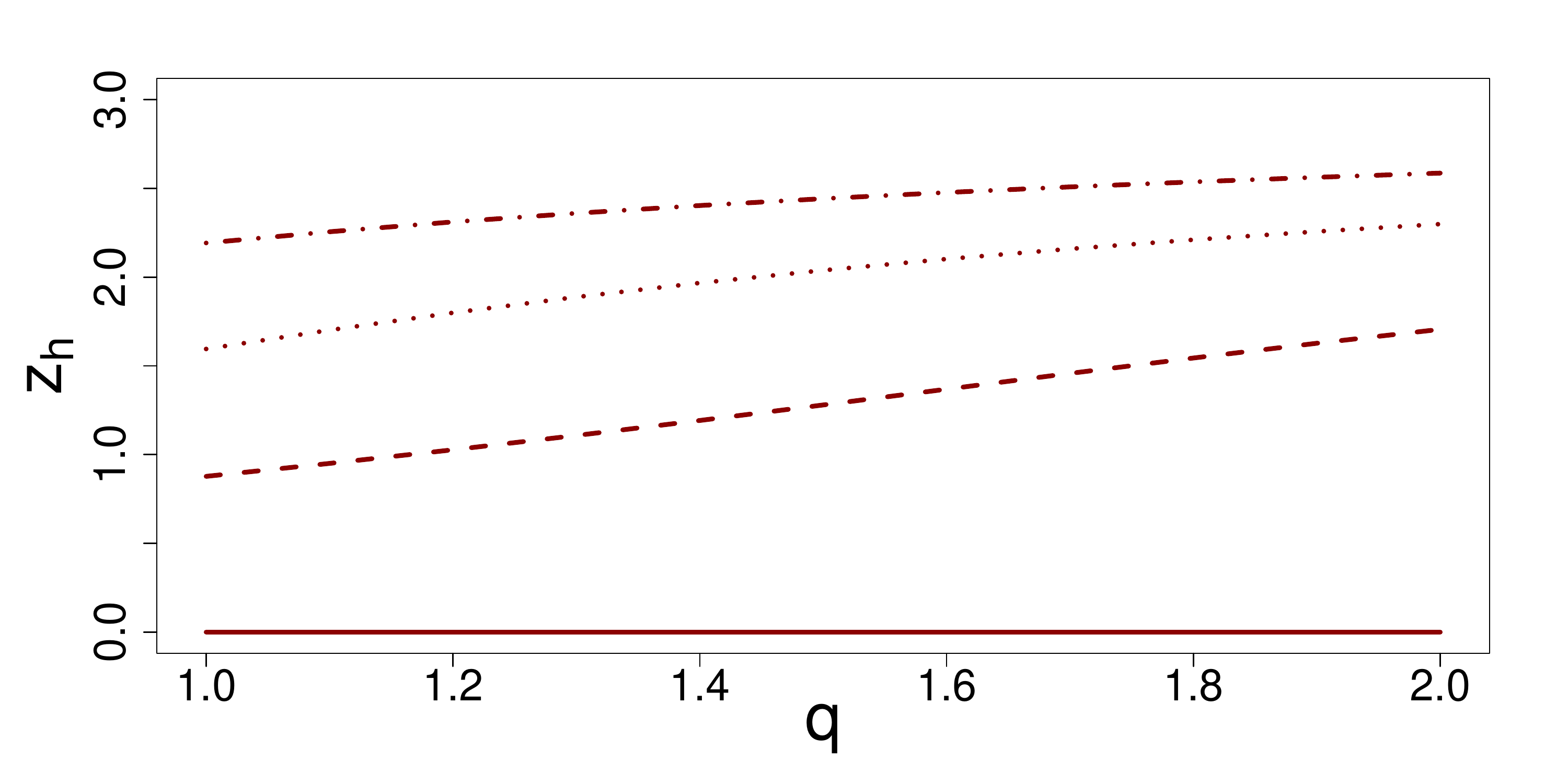} \\
\end{tabular}
    \begin{tablenotes}
      \footnotesize
      \item Notes. This design varies both the signal contribution parameter $a$ and the distribution of the receiver type by $q$. Recall the matching function $x = k z^q$ and the match surplus function $v(x,s,z)=As^axz$. We keep the same distribution on senders' type. The lower bound of an interval delegation is always zero ($t_l = 0$). The lower bound of a well-behaved equilibrium is also zero ($z_l=0$).
    \end{tablenotes}
\end{threeparttable}
\end{figure}

\clearpage

\subsection{Tables}

\begin{table}[hp]\label{tb:D1_Beta(1,1)}
\centering
\caption{Design 1 with Beta(1,1)} \label{tb:D1_Beta(1,1)}
\begin{tabular}{rrrrrrrr}
  \hline
$q$ & $k$ & $\bar{z}$ & $\bar{x}$ & $t_h$ & $z_h$ & $x_h$ & $s_h$ \\ 
  \hline
1.00 & 1.00 & 1.00 & 1.00 & 0.41 & 0.58 & 0.58 & 0.53 \\ 
  1.00 & 1.00 & 1.20 & 1.20 & 0.74 & 0.70 & 0.70 & 0.76 \\ 
  1.00 & 1.00 & 1.40 & 1.40 & 1.22 & 0.82 & 0.82 & 1.03 \\ 
  1.00 & 1.00 & 1.60 & 1.60 & 1.89 & 0.94 & 0.94 & 1.34 \\ 
  1.00 & 1.00 & 1.80 & 1.80 & 2.78 & 1.05 & 1.05 & 1.70 \\ 
  1.00 & 1.00 & 2.00 & 2.00 & 3.94 & 1.17 & 1.17 & 2.10 \\ 
  1.00 & 1.00 & 2.20 & 2.20 & 5.41 & 1.29 & 1.29 & 2.54 \\ 
  1.00 & 1.00 & 2.40 & 2.40 & 7.24 & 1.40 & 1.40 & 3.03 \\ 
  1.00 & 1.00 & 2.60 & 2.60 & 9.47 & 1.52 & 1.52 & 3.55 \\ 
  1.00 & 1.00 & 2.80 & 2.80 & 12.18 & 1.64 & 1.64 & 4.12 \\ 
  1.00 & 1.00 & 3.00 & 3.00 & 15.40 & 1.75 & 1.75 & 4.73 \\ 
   \hline
\end{tabular}
\end{table}
 %
\begin{table}[hp]
\centering
\caption{Design 1 with Beta(5,5)} 
\begin{tabular}{rrrrrrrr}
  \hline
$q$ & $k$ & $\bar{z}$ & $\bar{x}$ & $t_h$ & $z_h$ & $x_h$ & $s_h$ \\ 
  \hline
1.00 & 1.00 & 1.00 & 1.00 & 0.17 & 0.45 & 0.45 & 0.30 \\ 
  1.00 & 1.00 & 1.20 & 1.20 & 0.30 & 0.54 & 0.54 & 0.43 \\ 
  1.00 & 1.00 & 1.40 & 1.40 & 0.50 & 0.62 & 0.62 & 0.58 \\ 
  1.00 & 1.00 & 1.60 & 1.60 & 0.76 & 0.71 & 0.71 & 0.76 \\ 
  1.00 & 1.00 & 1.80 & 1.80 & 1.12 & 0.80 & 0.80 & 0.97 \\ 
  1.00 & 1.00 & 2.00 & 2.00 & 1.58 & 0.89 & 0.89 & 1.19 \\ 
  1.00 & 1.00 & 2.20 & 2.20 & 2.15 & 0.98 & 0.98 & 1.45 \\ 
  1.00 & 1.00 & 2.40 & 2.40 & 2.87 & 1.07 & 1.07 & 1.72 \\ 
  1.00 & 1.00 & 2.60 & 2.60 & 3.74 & 1.16 & 1.16 & 2.02 \\ 
  1.00 & 1.00 & 2.80 & 2.80 & 4.79 & 1.25 & 1.25 & 2.34 \\ 
  1.00 & 1.00 & 3.00 & 3.00 & 6.03 & 1.34 & 1.34 & 2.69 \\ 
   \hline
\end{tabular}
\end{table}
 %
\begin{table}[hp]
\centering
\caption{Design 1 with Beta(3,5)} 
\begin{tabular}{rrrrrrrr}
  \hline
$q$ & $k$ & $\bar{z}$ & $\bar{x}$ & $t_h$ & $z_h$ & $x_h$ & $s_h$ \\ 
  \hline
1.00 & 1.00 & 1.00 & 1.00 & 0.27 & 0.54 & 0.54 & 0.37 \\ 
  1.00 & 1.00 & 1.20 & 1.20 & 0.48 & 0.65 & 0.65 & 0.54 \\ 
  1.00 & 1.00 & 1.40 & 1.40 & 0.79 & 0.76 & 0.76 & 0.73 \\ 
  1.00 & 1.00 & 1.60 & 1.60 & 1.22 & 0.86 & 0.86 & 0.96 \\ 
  1.00 & 1.00 & 1.80 & 1.80 & 1.81 & 0.97 & 0.97 & 1.21 \\ 
  1.00 & 1.00 & 2.00 & 2.00 & 2.57 & 1.08 & 1.08 & 1.50 \\ 
  1.00 & 1.00 & 2.20 & 2.20 & 3.54 & 1.19 & 1.19 & 1.81 \\ 
  1.00 & 1.00 & 2.40 & 2.40 & 4.76 & 1.30 & 1.30 & 2.15 \\ 
  1.00 & 1.00 & 2.60 & 2.60 & 6.25 & 1.40 & 1.40 & 2.53 \\ 
  1.00 & 1.00 & 2.80 & 2.80 & 8.05 & 1.51 & 1.51 & 2.93 \\ 
  1.00 & 1.00 & 3.00 & 3.00 & 10.21 & 1.62 & 1.62 & 3.37 \\ 
   \hline
\end{tabular}
\end{table}
 %
\begin{table}[hp]
\centering
\caption{Design 1 with Beta(5,3)} 
\begin{tabular}{rrrrrrrr}
  \hline
$q$ & $k$ & $\bar{z}$ & $\bar{x}$ & $t_h$ & $z_h$ & $x_h$ & $s_h$ \\ 
  \hline
1.00 & 1.00 & 1.00 & 1.00 & 0.20 & 0.45 & 0.45 & 0.34 \\ 
  1.00 & 1.00 & 1.20 & 1.20 & 0.36 & 0.54 & 0.54 & 0.49 \\ 
  1.00 & 1.00 & 1.40 & 1.40 & 0.59 & 0.63 & 0.63 & 0.67 \\ 
  1.00 & 1.00 & 1.60 & 1.60 & 0.90 & 0.72 & 0.72 & 0.88 \\ 
  1.00 & 1.00 & 1.80 & 1.80 & 1.31 & 0.81 & 0.81 & 1.11 \\ 
  1.00 & 1.00 & 2.00 & 2.00 & 1.84 & 0.90 & 0.90 & 1.37 \\ 
  1.00 & 1.00 & 2.20 & 2.20 & 2.51 & 0.99 & 0.99 & 1.66 \\ 
  1.00 & 1.00 & 2.40 & 2.40 & 3.33 & 1.08 & 1.08 & 1.98 \\ 
  1.00 & 1.00 & 2.60 & 2.60 & 4.33 & 1.17 & 1.17 & 2.32 \\ 
  1.00 & 1.00 & 2.80 & 2.80 & 5.53 & 1.25 & 1.25 & 2.69 \\ 
  1.00 & 1.00 & 3.00 & 3.00 & 6.94 & 1.34 & 1.34 & 3.09 \\ 
   \hline
\end{tabular}
\end{table}
\begin{table}[hp]
\centering
\caption{Design 2 with Beta(1,1)} 
\begin{tabular}{rrrrrrrr}
  \hline
$q$ & $k$ & $\bar{z}$ & $\bar{x}$ & $t_h$ & $z_h$ & $x_h$ & $s_h$ \\ 
  \hline
1.00 & 1.00 & 3.00 & 3.00 & 15.40 & 1.75 & 1.75 & 4.73 \\ 
  1.00 & 1.20 & 3.00 & 3.60 & 20.68 & 1.75 & 2.11 & 5.34 \\ 
  1.00 & 1.40 & 3.00 & 4.20 & 26.61 & 1.75 & 2.46 & 5.92 \\ 
  1.00 & 1.60 & 3.00 & 4.80 & 33.17 & 1.75 & 2.81 & 6.47 \\ 
  1.00 & 1.80 & 3.00 & 5.40 & 40.36 & 1.75 & 3.16 & 7.00 \\ 
  1.00 & 2.00 & 3.00 & 6.00 & 48.16 & 1.75 & 3.51 & 7.51 \\ 
  1.00 & 2.20 & 3.00 & 6.60 & 56.57 & 1.75 & 3.86 & 8.00 \\ 
  1.00 & 2.40 & 3.00 & 7.20 & 65.58 & 1.75 & 4.21 & 8.48 \\ 
  1.00 & 2.60 & 3.00 & 7.80 & 75.19 & 1.75 & 4.56 & 8.94 \\ 
  1.00 & 2.80 & 3.00 & 8.40 & 85.39 & 1.75 & 4.91 & 9.39 \\ 
  1.00 & 3.00 & 3.00 & 9.00 & 96.18 & 1.75 & 5.26 & 9.84 \\ 
   \hline
\end{tabular}
\end{table}
 %
\begin{table}[hp]
\centering
\caption{Design 2 with Beta(5,5)} 
\begin{tabular}{rrrrrrrr}
  \hline
$q$ & $k$ & $\bar{z}$ & $\bar{x}$ & $t_h$ & $z_h$ & $x_h$ & $s_h$ \\ 
  \hline
1.00 & 1.00 & 3.00 & 3.00 & 6.03 & 1.34 & 1.34 & 2.69 \\ 
  1.00 & 1.20 & 3.00 & 3.60 & 8.04 & 1.34 & 1.61 & 3.03 \\ 
  1.00 & 1.40 & 3.00 & 4.20 & 10.29 & 1.34 & 1.87 & 3.36 \\ 
  1.00 & 1.60 & 3.00 & 4.80 & 12.76 & 1.34 & 2.14 & 3.67 \\ 
  1.00 & 1.80 & 3.00 & 5.40 & 15.45 & 1.34 & 2.41 & 3.98 \\ 
  1.00 & 2.00 & 3.00 & 6.00 & 18.36 & 1.34 & 2.68 & 4.26 \\ 
  1.00 & 2.20 & 3.00 & 6.60 & 21.48 & 1.34 & 2.94 & 4.54 \\ 
  1.00 & 2.40 & 3.00 & 7.20 & 24.82 & 1.34 & 3.21 & 4.82 \\ 
  1.00 & 2.60 & 3.00 & 7.80 & 28.36 & 1.34 & 3.48 & 5.08 \\ 
  1.00 & 2.80 & 3.00 & 8.40 & 32.12 & 1.34 & 3.75 & 5.34 \\ 
  1.00 & 3.00 & 3.00 & 9.00 & 36.08 & 1.34 & 4.01 & 5.59 \\ 
   \hline
\end{tabular}
\end{table}
 %
\begin{table}[hp]
\centering
\caption{Design 2 with Beta(3,5)} 
\begin{tabular}{rrrrrrrr}
  \hline
$q$ & $k$ & $\bar{z}$ & $\bar{x}$ & $t_h$ & $z_h$ & $x_h$ & $s_h$ \\ 
  \hline
1.00 & 1.00 & 3.00 & 3.00 & 10.21 & 1.62 & 1.62 & 3.37 \\ 
  1.00 & 1.20 & 3.00 & 3.60 & 13.78 & 1.62 & 1.94 & 3.80 \\ 
  1.00 & 1.40 & 3.00 & 4.20 & 17.80 & 1.62 & 2.27 & 4.21 \\ 
  1.00 & 1.60 & 3.00 & 4.80 & 22.27 & 1.62 & 2.59 & 4.61 \\ 
  1.00 & 1.80 & 3.00 & 5.40 & 27.18 & 1.62 & 2.92 & 4.98 \\ 
  1.00 & 2.00 & 3.00 & 6.00 & 32.52 & 1.62 & 3.24 & 5.34 \\ 
  1.00 & 2.20 & 3.00 & 6.60 & 38.29 & 1.62 & 3.56 & 5.70 \\ 
  1.00 & 2.40 & 3.00 & 7.20 & 44.49 & 1.62 & 3.89 & 6.04 \\ 
  1.00 & 2.60 & 3.00 & 7.80 & 51.12 & 1.62 & 4.21 & 6.37 \\ 
  1.00 & 2.80 & 3.00 & 8.40 & 58.16 & 1.62 & 4.54 & 6.69 \\ 
  1.00 & 3.00 & 3.00 & 9.00 & 65.62 & 1.62 & 4.86 & 7.00 \\ 
   \hline
\end{tabular}
\end{table}
 %
\begin{table}[hp]
\centering
\caption{Design 2 with Beta(5,3)} 
\begin{tabular}{rrrrrrrr}
  \hline
$q$ & $k$ & $\bar{z}$ & $\bar{x}$ & $t_h$ & $z_h$ & $x_h$ & $s_h$ \\ 
  \hline
1.00 & 1.00 & 3.00 & 3.00 & 6.94 & 1.34 & 1.34 & 3.09 \\ 
  1.00 & 1.20 & 3.00 & 3.60 & 9.21 & 1.34 & 1.61 & 3.49 \\ 
  1.00 & 1.40 & 3.00 & 4.20 & 11.73 & 1.34 & 1.88 & 3.87 \\ 
  1.00 & 1.60 & 3.00 & 4.80 & 14.49 & 1.34 & 2.15 & 4.22 \\ 
  1.00 & 1.80 & 3.00 & 5.40 & 17.48 & 1.34 & 2.42 & 4.57 \\ 
  1.00 & 2.00 & 3.00 & 6.00 & 20.71 & 1.34 & 2.69 & 4.90 \\ 
  1.00 & 2.20 & 3.00 & 6.60 & 24.17 & 1.34 & 2.96 & 5.22 \\ 
  1.00 & 2.40 & 3.00 & 7.20 & 27.85 & 1.34 & 3.23 & 5.54 \\ 
  1.00 & 2.60 & 3.00 & 7.80 & 31.75 & 1.34 & 3.50 & 5.84 \\ 
  1.00 & 2.80 & 3.00 & 8.40 & 35.87 & 1.34 & 3.76 & 6.14 \\ 
  1.00 & 3.00 & 3.00 & 9.00 & 40.21 & 1.34 & 4.03 & 6.42 \\ 
   \hline
\end{tabular}
\end{table}
\begin{table}[hp]
\centering
\caption{Design 3 with Beta(1,1)} 
\begin{tabular}{rrrrrrrr}
  \hline
$q$ & $k$ & $\bar{z}$ & $\bar{x}$ & $t_h$ & $z_h$ & $x_h$ & $s_h$ \\ 
  \hline
1.00 & 1.00 & 3.00 & 3.00 & 15.40 & 1.75 & 1.75 & 4.73 \\ 
  1.10 & 1.00 & 3.00 & 3.35 & 18.85 & 1.83 & 1.94 & 5.16 \\ 
  1.20 & 1.00 & 3.00 & 3.74 & 23.09 & 1.89 & 2.15 & 5.62 \\ 
  1.30 & 1.00 & 3.00 & 4.17 & 28.30 & 1.95 & 2.38 & 6.12 \\ 
  1.40 & 1.00 & 3.00 & 4.66 & 34.72 & 2.00 & 2.65 & 6.65 \\ 
  1.50 & 1.00 & 3.00 & 5.20 & 42.63 & 2.05 & 2.94 & 7.21 \\ 
  1.60 & 1.00 & 3.00 & 5.80 & 52.37 & 2.10 & 3.28 & 7.82 \\ 
  1.70 & 1.00 & 3.00 & 6.47 & 64.39 & 2.14 & 3.65 & 8.47 \\ 
  1.80 & 1.00 & 3.00 & 7.22 & 79.22 & 2.18 & 4.08 & 9.17 \\ 
  1.90 & 1.00 & 3.00 & 8.06 & 97.56 & 2.22 & 4.55 & 9.92 \\ 
  2.00 & 1.00 & 3.00 & 9.00 & 120.23 & 2.25 & 5.08 & 10.72 \\ 
   \hline
\end{tabular}
\end{table}
 %
\begin{table}[hp]
\centering
\caption{Design 3 with Beta(5,5)} 
\begin{tabular}{rrrrrrrr}
  \hline
$q$ & $k$ & $\bar{z}$ & $\bar{x}$ & $t_h$ & $z_h$ & $x_h$ & $s_h$ \\ 
  \hline
1.00 & 1.00 & 3.00 & 3.00 & 6.03 & 1.34 & 1.34 & 2.69 \\ 
  1.10 & 1.00 & 3.00 & 3.35 & 8.48 & 1.48 & 1.54 & 3.15 \\ 
  1.20 & 1.00 & 3.00 & 3.74 & 11.80 & 1.61 & 1.77 & 3.65 \\ 
  1.30 & 1.00 & 3.00 & 4.17 & 16.16 & 1.72 & 2.03 & 4.18 \\ 
  1.40 & 1.00 & 3.00 & 4.66 & 21.74 & 1.82 & 2.32 & 4.75 \\ 
  1.50 & 1.00 & 3.00 & 5.20 & 28.83 & 1.91 & 2.64 & 5.34 \\ 
  1.60 & 1.00 & 3.00 & 5.80 & 37.78 & 1.99 & 3.00 & 5.96 \\ 
  1.70 & 1.00 & 3.00 & 6.47 & 49.06 & 2.05 & 3.40 & 6.62 \\ 
  1.80 & 1.00 & 3.00 & 7.22 & 63.23 & 2.11 & 3.85 & 7.32 \\ 
  1.90 & 1.00 & 3.00 & 8.06 & 81.00 & 2.17 & 4.34 & 8.06 \\ 
  2.00 & 1.00 & 3.00 & 9.00 & 103.29 & 2.21 & 4.90 & 8.85 \\ 
   \hline
\end{tabular}
\end{table}
 %
\begin{table}[hp]
\centering
\caption{Design 3 with Beta(3,5)} 
\begin{tabular}{rrrrrrrr}
  \hline
$q$ & $k$ & $\bar{z}$ & $\bar{x}$ & $t_h$ & $z_h$ & $x_h$ & $s_h$ \\ 
  \hline
1.00 & 1.00 & 3.00 & 3.00 & 10.21 & 1.62 & 1.62 & 3.37 \\ 
  1.10 & 1.00 & 3.00 & 3.35 & 13.53 & 1.73 & 1.82 & 3.83 \\ 
  1.20 & 1.00 & 3.00 & 3.74 & 17.70 & 1.82 & 2.05 & 4.31 \\ 
  1.30 & 1.00 & 3.00 & 4.17 & 22.91 & 1.90 & 2.31 & 4.81 \\ 
  1.40 & 1.00 & 3.00 & 4.66 & 29.41 & 1.98 & 2.59 & 5.35 \\ 
  1.50 & 1.00 & 3.00 & 5.20 & 37.53 & 2.04 & 2.91 & 5.91 \\ 
  1.60 & 1.00 & 3.00 & 5.80 & 47.66 & 2.10 & 3.27 & 6.51 \\ 
  1.70 & 1.00 & 3.00 & 6.47 & 60.27 & 2.15 & 3.67 & 7.15 \\ 
  1.80 & 1.00 & 3.00 & 7.22 & 75.99 & 2.20 & 4.12 & 7.83 \\ 
  1.90 & 1.00 & 3.00 & 8.06 & 95.57 & 2.24 & 4.63 & 8.56 \\ 
  2.00 & 1.00 & 3.00 & 9.00 & 119.98 & 2.28 & 5.20 & 9.33 \\ 
   \hline
\end{tabular}
\end{table}
 %
\begin{table}[hp]
\centering
\caption{Design 3 with Beta(5,3)} 
\begin{tabular}{rrrrrrrr}
  \hline
$q$ & $k$ & $\bar{z}$ & $\bar{x}$ & $t_h$ & $z_h$ & $x_h$ & $s_h$ \\ 
  \hline
1.00 & 1.00 & 3.00 & 3.00 & 6.94 & 1.34 & 1.34 & 3.09 \\ 
  1.10 & 1.00 & 3.00 & 3.35 & 8.89 & 1.45 & 1.51 & 3.46 \\ 
  1.20 & 1.00 & 3.00 & 3.74 & 11.54 & 1.55 & 1.70 & 3.89 \\ 
  1.30 & 1.00 & 3.00 & 4.17 & 15.08 & 1.66 & 1.93 & 4.37 \\ 
  1.40 & 1.00 & 3.00 & 4.66 & 19.72 & 1.75 & 2.19 & 4.89 \\ 
  1.50 & 1.00 & 3.00 & 5.20 & 25.74 & 1.83 & 2.48 & 5.45 \\ 
  1.60 & 1.00 & 3.00 & 5.80 & 33.48 & 1.91 & 2.82 & 6.05 \\ 
  1.70 & 1.00 & 3.00 & 6.47 & 43.33 & 1.98 & 3.20 & 6.70 \\ 
  1.80 & 1.00 & 3.00 & 7.22 & 55.84 & 2.04 & 3.62 & 7.39 \\ 
  1.90 & 1.00 & 3.00 & 8.06 & 71.66 & 2.10 & 4.10 & 8.13 \\ 
  2.00 & 1.00 & 3.00 & 9.00 & 91.60 & 2.15 & 4.64 & 8.93 \\ 
   \hline
\end{tabular}
\end{table}
\begin{table}[hp]
\centering
\caption{Design 4 with Beta(1,1)} 
\begingroup\footnotesize
\begin{tabular}{rrrrrrrrr}
  \hline
$q$ & $k$ & $a$ & $\bar{z}$ & $\bar{x}$ & $t_h$ & $z_h$ & $x_h$ & $s_h$ \\ 
  \hline
1.00 & 1.00 & 0.00 & 3.00 & 3.00 & 0.57 & 0.38 & 0.38 & 0.65 \\ 
  1.10 & 1.00 & 0.00 & 3.00 & 3.35 & 0.66 & 0.47 & 0.44 & 0.77 \\ 
  1.20 & 1.00 & 0.00 & 3.00 & 3.74 & 0.77 & 0.56 & 0.50 & 0.89 \\ 
  1.30 & 1.00 & 0.00 & 3.00 & 4.17 & 0.89 & 0.65 & 0.57 & 1.01 \\ 
  1.40 & 1.00 & 0.00 & 3.00 & 4.66 & 1.02 & 0.73 & 0.64 & 1.13 \\ 
  1.50 & 1.00 & 0.00 & 3.00 & 5.20 & 1.18 & 0.81 & 0.73 & 1.25 \\ 
  1.60 & 1.00 & 0.00 & 3.00 & 5.80 & 1.36 & 0.89 & 0.82 & 1.38 \\ 
  1.70 & 1.00 & 0.00 & 3.00 & 6.47 & 1.58 & 0.96 & 0.93 & 1.51 \\ 
  1.80 & 1.00 & 0.00 & 3.00 & 7.22 & 1.83 & 1.02 & 1.04 & 1.64 \\ 
  1.90 & 1.00 & 0.00 & 3.00 & 8.06 & 2.12 & 1.09 & 1.18 & 1.78 \\ 
  2.00 & 1.00 & 0.00 & 3.00 & 9.00 & 2.46 & 1.15 & 1.32 & 1.92 \\ 
  1.00 & 1.00 & 0.30 & 3.00 & 3.00 & 5.90 & 1.36 & 1.36 & 2.77 \\ 
  1.10 & 1.00 & 0.30 & 3.00 & 3.35 & 7.04 & 1.44 & 1.50 & 3.01 \\ 
  1.20 & 1.00 & 0.30 & 3.00 & 3.74 & 8.41 & 1.52 & 1.65 & 3.28 \\ 
  1.30 & 1.00 & 0.30 & 3.00 & 4.17 & 10.05 & 1.59 & 1.83 & 3.55 \\ 
  1.40 & 1.00 & 0.30 & 3.00 & 4.66 & 12.01 & 1.66 & 2.03 & 3.85 \\ 
  1.50 & 1.00 & 0.30 & 3.00 & 5.20 & 14.36 & 1.72 & 2.26 & 4.15 \\ 
  1.60 & 1.00 & 0.30 & 3.00 & 5.80 & 17.18 & 1.78 & 2.51 & 4.48 \\ 
  1.70 & 1.00 & 0.30 & 3.00 & 6.47 & 20.56 & 1.83 & 2.80 & 4.82 \\ 
  1.80 & 1.00 & 0.30 & 3.00 & 7.22 & 24.62 & 1.88 & 3.13 & 5.19 \\ 
  1.90 & 1.00 & 0.30 & 3.00 & 8.06 & 29.48 & 1.93 & 3.49 & 5.58 \\ 
  2.00 & 1.00 & 0.30 & 3.00 & 9.00 & 35.33 & 1.97 & 3.90 & 5.99 \\ 
  1.00 & 1.00 & 0.60 & 3.00 & 3.00 & 26.02 & 1.92 & 1.92 & 6.19 \\ 
  1.10 & 1.00 & 0.60 & 3.00 & 3.35 & 32.37 & 1.98 & 2.12 & 6.77 \\ 
  1.20 & 1.00 & 0.60 & 3.00 & 3.74 & 40.31 & 2.04 & 2.35 & 7.40 \\ 
  1.30 & 1.00 & 0.60 & 3.00 & 4.17 & 50.24 & 2.09 & 2.61 & 8.08 \\ 
  1.40 & 1.00 & 0.60 & 3.00 & 4.66 & 62.68 & 2.14 & 2.90 & 8.81 \\ 
  1.50 & 1.00 & 0.60 & 3.00 & 5.20 & 78.26 & 2.18 & 3.23 & 9.60 \\ 
  1.60 & 1.00 & 0.60 & 3.00 & 5.80 & 97.80 & 2.22 & 3.59 & 10.45 \\ 
  1.70 & 1.00 & 0.60 & 3.00 & 6.47 & 122.34 & 2.26 & 4.01 & 11.37 \\ 
  1.80 & 1.00 & 0.60 & 3.00 & 7.22 & 153.17 & 2.30 & 4.47 & 12.36 \\ 
  1.90 & 1.00 & 0.60 & 3.00 & 8.06 & 191.95 & 2.33 & 4.99 & 13.44 \\ 
  2.00 & 1.00 & 0.60 & 3.00 & 9.00 & 240.76 & 2.36 & 5.57 & 14.59 \\ 
  1.00 & 1.00 & 0.90 & 3.00 & 3.00 & 206.64 & 2.29 & 2.29 & 15.92 \\ 
  1.10 & 1.00 & 0.90 & 3.00 & 3.35 & 276.32 & 2.34 & 2.54 & 17.74 \\ 
  1.20 & 1.00 & 0.90 & 3.00 & 3.74 & 369.97 & 2.38 & 2.82 & 19.73 \\ 
  1.30 & 1.00 & 0.90 & 3.00 & 4.17 & 495.99 & 2.41 & 3.14 & 21.94 \\ 
  1.40 & 1.00 & 0.90 & 3.00 & 4.66 & 665.73 & 2.45 & 3.50 & 24.38 \\ 
  1.50 & 1.00 & 0.90 & 3.00 & 5.20 & 894.60 & 2.48 & 3.89 & 27.07 \\ 
  1.60 & 1.00 & 0.90 & 3.00 & 5.80 & 1203.39 & 2.50 & 4.34 & 30.05 \\ 
  1.70 & 1.00 & 0.90 & 3.00 & 6.47 & 1620.49 & 2.53 & 4.84 & 33.34 \\ 
  1.80 & 1.00 & 0.90 & 3.00 & 7.22 & 2184.06 & 2.55 & 5.40 & 36.97 \\ 
  1.90 & 1.00 & 0.90 & 3.00 & 8.06 & 2946.38 & 2.57 & 6.03 & 40.99 \\ 
  2.00 & 1.00 & 0.90 & 3.00 & 9.00 & 3977.63 & 2.59 & 6.73 & 45.44 \\ 
   \hline
\end{tabular}
\endgroup
\end{table}
 %
\begin{table}[hp]
\centering
\caption{Design 4 with Beta(5,5)} 
\begingroup\footnotesize
\begin{tabular}{rrrrrrrrr}
  \hline
$q$ & $k$ & $a$ & $\bar{z}$ & $\bar{x}$ & $t_h$ & $z_h$ & $x_h$ & $s_h$ \\ 
  \hline
1.00 & 1.00 & 0.00 & 3.00 & 3.00 & 0.00 & 0.00 & 0.00 & 0.00 \\ 
  1.10 & 1.00 & 0.00 & 3.00 & 3.35 & 0.00 & 0.00 & 0.00 & 0.00 \\ 
  1.20 & 1.00 & 0.00 & 3.00 & 3.74 & 0.00 & 0.00 & 0.00 & 0.00 \\ 
  1.30 & 1.00 & 0.00 & 3.00 & 4.17 & 0.00 & 0.00 & 0.00 & 0.00 \\ 
  1.40 & 1.00 & 0.00 & 3.00 & 4.66 & 0.00 & 0.00 & 0.00 & 0.00 \\ 
  1.50 & 1.00 & 0.00 & 3.00 & 5.20 & 0.00 & 0.00 & 0.00 & 0.00 \\ 
  1.60 & 1.00 & 0.00 & 3.00 & 5.80 & 0.00 & 0.00 & 0.00 & 0.00 \\ 
  1.70 & 1.00 & 0.00 & 3.00 & 6.47 & 0.00 & 0.00 & 0.00 & 0.00 \\ 
  1.80 & 1.00 & 0.00 & 3.00 & 7.22 & 0.00 & 0.00 & 0.00 & 0.00 \\ 
  1.90 & 1.00 & 0.00 & 3.00 & 8.06 & 0.00 & 0.00 & 0.00 & 0.00 \\ 
  2.00 & 1.00 & 0.00 & 3.00 & 9.00 & 0.00 & 0.00 & 0.00 & 0.00 \\ 
  1.00 & 1.00 & 0.30 & 3.00 & 3.00 & 1.51 & 0.76 & 0.76 & 1.21 \\ 
  1.10 & 1.00 & 0.30 & 3.00 & 3.35 & 1.79 & 0.84 & 0.83 & 1.34 \\ 
  1.20 & 1.00 & 0.30 & 3.00 & 3.74 & 2.20 & 0.94 & 0.93 & 1.50 \\ 
  1.30 & 1.00 & 0.30 & 3.00 & 4.17 & 2.85 & 1.05 & 1.07 & 1.71 \\ 
  1.40 & 1.00 & 0.30 & 3.00 & 4.66 & 3.82 & 1.17 & 1.25 & 1.98 \\ 
  1.50 & 1.00 & 0.30 & 3.00 & 5.20 & 5.25 & 1.30 & 1.48 & 2.29 \\ 
  1.60 & 1.00 & 0.30 & 3.00 & 5.80 & 7.24 & 1.42 & 1.76 & 2.65 \\ 
  1.70 & 1.00 & 0.30 & 3.00 & 6.47 & 9.87 & 1.54 & 2.08 & 3.03 \\ 
  1.80 & 1.00 & 0.30 & 3.00 & 7.22 & 13.24 & 1.64 & 2.44 & 3.43 \\ 
  1.90 & 1.00 & 0.30 & 3.00 & 8.06 & 17.48 & 1.73 & 2.84 & 3.86 \\ 
  2.00 & 1.00 & 0.30 & 3.00 & 9.00 & 22.74 & 1.81 & 3.29 & 4.30 \\ 
  1.00 & 1.00 & 0.60 & 3.00 & 3.00 & 13.56 & 1.64 & 1.64 & 4.06 \\ 
  1.10 & 1.00 & 0.60 & 3.00 & 3.35 & 18.87 & 1.76 & 1.87 & 4.69 \\ 
  1.20 & 1.00 & 0.60 & 3.00 & 3.74 & 25.78 & 1.87 & 2.12 & 5.36 \\ 
  1.30 & 1.00 & 0.60 & 3.00 & 4.17 & 34.68 & 1.96 & 2.39 & 6.06 \\ 
  1.40 & 1.00 & 0.60 & 3.00 & 4.66 & 46.10 & 2.03 & 2.70 & 6.80 \\ 
  1.50 & 1.00 & 0.60 & 3.00 & 5.20 & 60.70 & 2.10 & 3.05 & 7.60 \\ 
  1.60 & 1.00 & 0.60 & 3.00 & 5.80 & 79.34 & 2.16 & 3.43 & 8.44 \\ 
  1.70 & 1.00 & 0.60 & 3.00 & 6.47 & 103.09 & 2.22 & 3.87 & 9.34 \\ 
  1.80 & 1.00 & 0.60 & 3.00 & 7.22 & 133.36 & 2.26 & 4.35 & 10.31 \\ 
  1.90 & 1.00 & 0.60 & 3.00 & 8.06 & 171.88 & 2.31 & 4.89 & 11.35 \\ 
  2.00 & 1.00 & 0.60 & 3.00 & 9.00 & 220.90 & 2.34 & 5.50 & 12.46 \\ 
  1.00 & 1.00 & 0.90 & 3.00 & 3.00 & 169.09 & 2.24 & 2.24 & 12.95 \\ 
  1.10 & 1.00 & 0.90 & 3.00 & 3.35 & 235.88 & 2.30 & 2.50 & 14.70 \\ 
  1.20 & 1.00 & 0.90 & 3.00 & 3.74 & 327.14 & 2.35 & 2.79 & 16.62 \\ 
  1.30 & 1.00 & 0.90 & 3.00 & 4.17 & 451.61 & 2.40 & 3.11 & 18.73 \\ 
  1.40 & 1.00 & 0.90 & 3.00 & 4.66 & 621.34 & 2.44 & 3.48 & 21.06 \\ 
  1.50 & 1.00 & 0.90 & 3.00 & 5.20 & 852.53 & 2.47 & 3.89 & 23.62 \\ 
  1.60 & 1.00 & 0.90 & 3.00 & 5.80 & 1167.55 & 2.50 & 4.34 & 26.45 \\ 
  1.70 & 1.00 & 0.90 & 3.00 & 6.47 & 1596.11 & 2.53 & 4.86 & 29.57 \\ 
  1.80 & 1.00 & 0.90 & 3.00 & 7.22 & 2179.49 & 2.56 & 5.43 & 33.02 \\ 
  1.90 & 1.00 & 0.90 & 3.00 & 8.06 & 2972.88 & 2.58 & 6.07 & 36.83 \\ 
  2.00 & 1.00 & 0.90 & 3.00 & 9.00 & 4052.01 & 2.61 & 6.79 & 41.04 \\ 
   \hline
\end{tabular}
\endgroup
\end{table}
 %
\begin{table}[hp]
\centering
\caption{Design 4 with Beta(3,5)} 
\begingroup\footnotesize
\begin{tabular}{rrrrrrrrr}
  \hline
$q$ & $k$ & $a$ & $\bar{z}$ & $\bar{x}$ & $t_h$ & $z_h$ & $x_h$ & $s_h$ \\ 
  \hline
1.00 & 1.00 & 0.00 & 3.00 & 3.00 & 0.00 & 0.00 & 0.00 & 0.00 \\ 
  1.10 & 1.00 & 0.00 & 3.00 & 3.35 & 0.00 & 0.00 & 0.00 & 0.00 \\ 
  1.20 & 1.00 & 0.00 & 3.00 & 3.74 & 0.00 & 0.00 & 0.00 & 0.00 \\ 
  1.30 & 1.00 & 0.00 & 3.00 & 4.17 & 0.00 & 0.00 & 0.00 & 0.00 \\ 
  1.40 & 1.00 & 0.00 & 3.00 & 4.66 & 0.00 & 0.00 & 0.00 & 0.00 \\ 
  1.50 & 1.00 & 0.00 & 3.00 & 5.20 & 0.00 & 0.00 & 0.00 & 0.00 \\ 
  1.60 & 1.00 & 0.00 & 3.00 & 5.80 & 0.00 & 0.00 & 0.00 & 0.00 \\ 
  1.70 & 1.00 & 0.00 & 3.00 & 6.47 & 0.00 & 0.00 & 0.00 & 0.00 \\ 
  1.80 & 1.00 & 0.00 & 3.00 & 7.22 & 0.00 & 0.00 & 0.00 & 0.00 \\ 
  1.90 & 1.00 & 0.00 & 3.00 & 8.06 & 0.33 & 0.59 & 0.36 & 0.58 \\ 
  2.00 & 1.00 & 0.00 & 3.00 & 9.00 & 0.63 & 0.80 & 0.64 & 0.87 \\ 
  1.00 & 1.00 & 0.30 & 3.00 & 3.00 & 2.37 & 1.01 & 1.01 & 1.49 \\ 
  1.10 & 1.00 & 0.30 & 3.00 & 3.35 & 3.29 & 1.16 & 1.17 & 1.77 \\ 
  1.20 & 1.00 & 0.30 & 3.00 & 3.74 & 4.50 & 1.29 & 1.36 & 2.06 \\ 
  1.30 & 1.00 & 0.30 & 3.00 & 4.17 & 6.03 & 1.42 & 1.57 & 2.37 \\ 
  1.40 & 1.00 & 0.30 & 3.00 & 4.66 & 7.93 & 1.52 & 1.80 & 2.70 \\ 
  1.50 & 1.00 & 0.30 & 3.00 & 5.20 & 10.27 & 1.62 & 2.07 & 3.03 \\ 
  1.60 & 1.00 & 0.30 & 3.00 & 5.80 & 13.15 & 1.71 & 2.36 & 3.37 \\ 
  1.70 & 1.00 & 0.30 & 3.00 & 6.47 & 16.65 & 1.79 & 2.68 & 3.73 \\ 
  1.80 & 1.00 & 0.30 & 3.00 & 7.22 & 20.92 & 1.86 & 3.04 & 4.11 \\ 
  1.90 & 1.00 & 0.30 & 3.00 & 8.06 & 26.12 & 1.92 & 3.45 & 4.50 \\ 
  2.00 & 1.00 & 0.30 & 3.00 & 9.00 & 32.45 & 1.98 & 3.90 & 4.92 \\ 
  1.00 & 1.00 & 0.60 & 3.00 & 3.00 & 19.65 & 1.84 & 1.84 & 4.73 \\ 
  1.10 & 1.00 & 0.60 & 3.00 & 3.35 & 25.90 & 1.93 & 2.06 & 5.33 \\ 
  1.20 & 1.00 & 0.60 & 3.00 & 3.74 & 33.83 & 2.00 & 2.30 & 5.96 \\ 
  1.30 & 1.00 & 0.60 & 3.00 & 4.17 & 43.89 & 2.07 & 2.58 & 6.63 \\ 
  1.40 & 1.00 & 0.60 & 3.00 & 4.66 & 56.64 & 2.13 & 2.89 & 7.36 \\ 
  1.50 & 1.00 & 0.60 & 3.00 & 5.20 & 72.78 & 2.19 & 3.23 & 8.13 \\ 
  1.60 & 1.00 & 0.60 & 3.00 & 5.80 & 93.22 & 2.24 & 3.62 & 8.95 \\ 
  1.70 & 1.00 & 0.60 & 3.00 & 6.47 & 119.11 & 2.28 & 4.06 & 9.84 \\ 
  1.80 & 1.00 & 0.60 & 3.00 & 7.22 & 151.89 & 2.32 & 4.54 & 10.80 \\ 
  1.90 & 1.00 & 0.60 & 3.00 & 8.06 & 193.38 & 2.36 & 5.09 & 11.83 \\ 
  2.00 & 1.00 & 0.60 & 3.00 & 9.00 & 245.96 & 2.39 & 5.70 & 12.93 \\ 
  1.00 & 1.00 & 0.90 & 3.00 & 3.00 & 193.76 & 2.30 & 2.30 & 13.59 \\ 
  1.10 & 1.00 & 0.90 & 3.00 & 3.35 & 265.69 & 2.35 & 2.56 & 15.32 \\ 
  1.20 & 1.00 & 0.90 & 3.00 & 3.74 & 363.34 & 2.39 & 2.85 & 17.23 \\ 
  1.30 & 1.00 & 0.90 & 3.00 & 4.17 & 495.79 & 2.43 & 3.17 & 19.33 \\ 
  1.40 & 1.00 & 0.90 & 3.00 & 4.66 & 675.43 & 2.47 & 3.54 & 21.65 \\ 
  1.50 & 1.00 & 0.90 & 3.00 & 5.20 & 918.77 & 2.50 & 3.95 & 24.21 \\ 
  1.60 & 1.00 & 0.90 & 3.00 & 5.80 & 1249.77 & 2.53 & 4.41 & 27.04 \\ 
  1.70 & 1.00 & 0.90 & 3.00 & 6.47 & 1697.94 & 2.55 & 4.93 & 30.16 \\ 
  1.80 & 1.00 & 0.90 & 3.00 & 7.22 & 2306.10 & 2.58 & 5.50 & 33.61 \\ 
  1.90 & 1.00 & 0.90 & 3.00 & 8.06 & 3130.75 & 2.60 & 6.15 & 37.43 \\ 
  2.00 & 1.00 & 0.90 & 3.00 & 9.00 & 4249.73 & 2.62 & 6.87 & 41.65 \\ 
   \hline
\end{tabular}
\endgroup
\end{table}
 %
\begin{table}[hp]
\centering
\caption{Design 4 with Beta(5,3)} \label{tb:D4_Beta(5,3)}
\begingroup\footnotesize
\begin{tabular}{rrrrrrrrr}
  \hline
$q$ & $k$ & $a$ & $\bar{z}$ & $\bar{x}$ & $t_h$ & $z_h$ & $x_h$ & $s_h$ \\ 
  \hline
1.00 & 1.00 & 0.00 & 3.00 & 3.00 & 0.00 & 0.00 & 0.00 & 0.00 \\ 
  1.10 & 1.00 & 0.00 & 3.00 & 3.35 & 0.00 & 0.00 & 0.00 & 0.00 \\ 
  1.20 & 1.00 & 0.00 & 3.00 & 3.74 & 0.00 & 0.00 & 0.00 & 0.00 \\ 
  1.30 & 1.00 & 0.00 & 3.00 & 4.17 & 0.00 & 0.00 & 0.00 & 0.00 \\ 
  1.40 & 1.00 & 0.00 & 3.00 & 4.66 & 0.00 & 0.00 & 0.00 & 0.00 \\ 
  1.50 & 1.00 & 0.00 & 3.00 & 5.20 & 0.00 & 0.00 & 0.00 & 0.00 \\ 
  1.60 & 1.00 & 0.00 & 3.00 & 5.80 & 0.00 & 0.00 & 0.00 & 0.00 \\ 
  1.70 & 1.00 & 0.00 & 3.00 & 6.47 & 0.00 & 0.00 & 0.00 & 0.00 \\ 
  1.80 & 1.00 & 0.00 & 3.00 & 7.22 & 0.00 & 0.00 & 0.00 & 0.00 \\ 
  1.90 & 1.00 & 0.00 & 3.00 & 8.06 & 0.00 & 0.00 & 0.00 & 0.00 \\ 
  2.00 & 1.00 & 0.00 & 3.00 & 9.00 & 0.00 & 0.00 & 0.00 & 0.00 \\ 
  1.00 & 1.00 & 0.30 & 3.00 & 3.00 & 2.39 & 0.88 & 0.88 & 1.65 \\ 
  1.10 & 1.00 & 0.30 & 3.00 & 3.35 & 2.72 & 0.95 & 0.95 & 1.78 \\ 
  1.20 & 1.00 & 0.30 & 3.00 & 3.74 & 3.17 & 1.03 & 1.03 & 1.93 \\ 
  1.30 & 1.00 & 0.30 & 3.00 & 4.17 & 3.76 & 1.11 & 1.14 & 2.11 \\ 
  1.40 & 1.00 & 0.30 & 3.00 & 4.66 & 4.55 & 1.19 & 1.28 & 2.32 \\ 
  1.50 & 1.00 & 0.30 & 3.00 & 5.20 & 5.62 & 1.28 & 1.45 & 2.57 \\ 
  1.60 & 1.00 & 0.30 & 3.00 & 5.80 & 7.05 & 1.37 & 1.65 & 2.85 \\ 
  1.70 & 1.00 & 0.30 & 3.00 & 6.47 & 8.94 & 1.46 & 1.90 & 3.16 \\ 
  1.80 & 1.00 & 0.30 & 3.00 & 7.22 & 11.42 & 1.54 & 2.19 & 3.50 \\ 
  1.90 & 1.00 & 0.30 & 3.00 & 8.06 & 14.63 & 1.63 & 2.52 & 3.88 \\ 
  2.00 & 1.00 & 0.30 & 3.00 & 9.00 & 18.72 & 1.71 & 2.91 & 4.29 \\ 
  1.00 & 1.00 & 0.60 & 3.00 & 3.00 & 13.47 & 1.60 & 1.60 & 4.35 \\ 
  1.10 & 1.00 & 0.60 & 3.00 & 3.35 & 17.90 & 1.70 & 1.79 & 4.91 \\ 
  1.20 & 1.00 & 0.60 & 3.00 & 3.74 & 23.80 & 1.80 & 2.02 & 5.54 \\ 
  1.30 & 1.00 & 0.60 & 3.00 & 4.17 & 31.56 & 1.89 & 2.28 & 6.21 \\ 
  1.40 & 1.00 & 0.60 & 3.00 & 4.66 & 41.64 & 1.97 & 2.58 & 6.94 \\ 
  1.50 & 1.00 & 0.60 & 3.00 & 5.20 & 54.67 & 2.04 & 2.91 & 7.72 \\ 
  1.60 & 1.00 & 0.60 & 3.00 & 5.80 & 71.44 & 2.10 & 3.28 & 8.57 \\ 
  1.70 & 1.00 & 0.60 & 3.00 & 6.47 & 92.95 & 2.16 & 3.70 & 9.47 \\ 
  1.80 & 1.00 & 0.60 & 3.00 & 7.22 & 120.49 & 2.21 & 4.17 & 10.44 \\ 
  1.90 & 1.00 & 0.60 & 3.00 & 8.06 & 155.69 & 2.26 & 4.70 & 11.49 \\ 
  2.00 & 1.00 & 0.60 & 3.00 & 9.00 & 200.64 & 2.30 & 5.29 & 12.61 \\ 
  1.00 & 1.00 & 0.90 & 3.00 & 3.00 & 156.12 & 2.19 & 2.19 & 13.24 \\ 
  1.10 & 1.00 & 0.90 & 3.00 & 3.35 & 217.95 & 2.26 & 2.45 & 15.00 \\ 
  1.20 & 1.00 & 0.90 & 3.00 & 3.74 & 302.74 & 2.31 & 2.73 & 16.94 \\ 
  1.30 & 1.00 & 0.90 & 3.00 & 4.17 & 418.87 & 2.36 & 3.05 & 19.08 \\ 
  1.40 & 1.00 & 0.90 & 3.00 & 4.66 & 577.62 & 2.40 & 3.41 & 21.43 \\ 
  1.50 & 1.00 & 0.90 & 3.00 & 5.20 & 794.56 & 2.44 & 3.82 & 24.03 \\ 
  1.60 & 1.00 & 0.90 & 3.00 & 5.80 & 1090.60 & 2.48 & 4.27 & 26.89 \\ 
  1.70 & 1.00 & 0.90 & 3.00 & 6.47 & 1494.52 & 2.51 & 4.78 & 30.05 \\ 
  1.80 & 1.00 & 0.90 & 3.00 & 7.22 & 2045.24 & 2.54 & 5.34 & 33.54 \\ 
  1.90 & 1.00 & 0.90 & 3.00 & 8.06 & 2795.97 & 2.56 & 5.98 & 37.40 \\ 
  2.00 & 1.00 & 0.90 & 3.00 & 9.00 & 3819.16 & 2.59 & 6.69 & 41.66 \\ 
   \hline
\end{tabular}
\endgroup
\end{table}

\end{document}